\documentclass[useAMS,usenatbib,aas_macros]{mn2e} 
\usepackage {graphicx,subfigure,times,aas_macros}

\newcommand{\equ}[1]{eq.~(\ref{eq:#1})}

\newcommand{\se}[1]{\S\ref{sec:#1}}
\newcommand{\fig}[1]{Fig.~\ref{fig:#1}}

\newcommand{\Fig}[1]{Figure~\ref{fig:#1}}

\newcommand{\tab}[1]{Table~\ref{tab:#1}}
\newcommand{\be}{\begin{equation}}
\newcommand{\ee}{\end{equation}}
\newcommand{\bea}{\begin{eqnarray}}
\newcommand{\eea}{\end{eqnarray}}

\newcommand{\msun}{{\rm M}_\odot}
\newcommand{\Msun}{M_\odot}

\newcommand{\ifm}[1]{\relax\ifmmode#1\else$\mathsurround=0pt #1$\fi}
\newcommand{\kms}{\ifmmode\,{\rm km}\,{\rm s}^{-1}\else km$\,$s$^{-1}$\fi}
\newcommand{\hmpc}{\,\ifm{h^{-1}}{\rm Mpc}}

\newcommand{\Mpc}{\,{\rm Mpc}}
\newcommand{\kpc}{\,{\rm kpc}}
\newcommand{\pc}{\,{\rm pc}}
\newcommand{\Gyr}{\,{\rm Gyr}}

\newcommand{\Myr}{\,{\rm Myr}}

\newcommand{\ltsima}{$\; \buildrel < \over \sim \;$}
\newcommand{\lsim}{\lower.5ex\hbox{\ltsima}}
\newcommand{\gtsima}{$\; \buildrel > \over \sim \;$}
\newcommand{\gsim}{\lower.5ex\hbox{\gtsima}}
\newcommand{\prop}{\propto}

\newcommand{\pa}{\partial}

\def\la{\langle}
\def\ra{\rangle}

\def\omm{\Omega_{\rm m}}
\def\oml{\Omega_{\Lambda}}
\def\omb{\Omega_{\rm b}}
\def\sy{\,M_\odot\, {\rm yr}^{-1}}

\def\cmc{\,{\rm cm}^{-3}}
\def\cms{\,{\rm cm}^{-2}}
\def\ergs{\,{\rm erg}\,{\rm s}^{-1}}

\def\Mv{M_{\rm v}}
\def\Rv{R_{\rm v}}
\def\Vv{V_{\rm v}}

\def\Md{M_{\rm d}}
\def\Mc{M_{\rm c}}
\def\Rc{R_{\rm c}}
\def\Rd{R_{\rm d}}

\def\Mt{M_{\rm tot}}

\def\Ms{M_*}

\def\sigc{\sigma_{\rm c}}
\def\sigs{\sigma_{\rm s}}
\def\sigg{\sigma_{\rm g}}

\def\rdi{r_{\rm d}}
\def\rt{{r}}
\def\rd0{r_{{\rm d}0}}
\def\Vdi0{V_{{\rm d}0}}
\def\Vc{V_{\rm rot}}
\def\Vphi{V_{\rm rot}}
\def\Vci{V_{\rm rot,i}}
\def\Omci{\omega_{\rm c,i}}
\def\Vcc{V_{\rm circ}}

\def\Rot{{\mathcal R}}
\def\Rci{R_{\rm ci}}
\def\sigd{\sigma_{\rm d}}
\def\sigc{\sigma_{\rm c}}
\def\rhoc{\rho_{\rm c}}
\def\Sigd{\Sigma_{\rm d}}
\def\Sigc{\Sigma_{\rm c}}
\def\Omd{\Omega_{\rm d}}
\def\gv{V_{\rm grad}}
\def\Rs{R_{\rm s}}
\def\Rsh{R_{\rm sh}}

\def\insitu{{\it in situ\ }}
\def\exsitu{{\it ex situ\ }}

\newcommand{\Halpha}{H${\alpha}$}
\newcommand{\sigmar}{\sigma_r}

\usepackage{color}

\begin{document} 

\large 

\title[Rotational Support of Giant Disc Clumps] 
{Rotational Support of Giant Clumps in High-z Disc Galaxies}

\author[Ceverino et al.] 
{\parbox[t]{\textwidth} 
{ 
Daniel Ceverino$^1$ \thanks{E-mail: ceverino@phys.huji.ac.il}, 
Avishai Dekel$^1$\thanks{E-mail: dekel@phys.huji.ac.il}, 
Nir Mandelker$^1$,  
Frederic Bournaud$^2$,  
\\ 
Andreas Burkert$^3$, 
Reinhard Genzel$^4$, 
Joel Primack$^5$ 
} 
\\ \\  
$^1$Racah Institute of Physics, The Hebrew University, Jerusalem 91904, 
Israel\\ 
$^2$Laboratoire AIM Paris-Saclay, CEA/IRFU/SAp - CNRS -  
Universite Paris Diderot, 91191 Gif-sur-Yvette Cedex, France\\ 
$^3$Universitats-Sternwarte Ludqig-Maximilians-Universitat, Scheinerstr. 1,  
Munchen, D-81679, Germany\\ 
$^4$Max-Planck-Institut fur Extraterrestische Physik, Giessenbachstr. 1, 
D-85748 Garching, Germany\\ 
$^5$Department of Physics, University of California, Santa Cruz, CA, 95064, USA 
} 
\date{} 
 
\pagerange{\pageref{firstpage}--\pageref{lastpage}} \pubyear{0000} 
 
\maketitle 
 
\label{firstpage}

\begin{abstract} 
We address the internal support against total free-fall collapse of the giant 
clumps that form by violent gravitational instability in high-$z$ disc 
galaxies.   
Guidance is provided by an analytic model, where the proto-clumps are cut from  
a rotating disc and collapse to equilibrium while preserving angular momentum.  
This model predicts prograde clump rotation, which dominates the support if  
the clump has contracted to a surface-density contrast $\gsim\!10$.   
This is confirmed in hydro-AMR zoom-in simulations of galaxies in a  
cosmological context.   
In most high-$z$ clumps, the centrifugal force dominates the support,  
$\Rot\!\equiv\!\Vc^2/\Vcc^2\!>\!0.5$,  where $\Vc$ is the rotation  
velocity and the circular velocity $\Vcc$ measures the potential well. 
The clump spin indeed tends to be in the sense of the global disc angular  
momentum, but substantial tilts are frequent, reflecting the highly warped  
nature of the high-$z$ discs.  
Most clumps are in Jeans equilibrium, with the rest of the support provided by  
turbulence, partly driven by the gravitational instability itself.  
The general agreement between model and simulations indicates that  
angular-momentum loss or gain in most clumps is limited to a factor of two.  
Simulations of isolated gas-rich  
discs that resolve the clump substructure  
reveal that the cosmological simulations may overestimate $\Rot$ by  
$\sim\!30\%$, but the dominance of rotational support at high $z$ 
is not a resolution artifact.  
In turn, isolated gas-poor disc simulations produce at $z=0$ smaller  
gaseous non-rotating transient clouds, indicating that the  
difference in rotational support is associated with the fraction of cold 
baryons in the disc. 
In our current cosmological simulations, 
the clump rotation velocity is typically more than twice the disc  
dispersion, $\Vc\!\sim\!100 \kms$, but when beam smearing of  
$\geq\!0.1$ arcsec is imposed, the rotation signal is reduced to a small  
gradient of $\leq 30 \kms\kpc^{-1}$ across the clump. 
The velocity dispersion in the simulated clumps is comparable to the  
disc dispersion so it is expected to leave only a marginal signal for any  
beam smearing. 
Retrograde minor-merging galaxies could lead to  
massive clumps that do not show rotation even when marginally resolved. 
Testable predictions of the scenario as simulated are that the mean stellar  
age of the clumps, and the stellar fraction, are declining linearly with 
distance from the disc center. 
 
\end{abstract} 
 

\section{Introduction} 
\label{sec:intro} 
 
More than half the massive star-forming galaxies  
observed in the redshift range $z \sim 1-3$  
\citep[e.g.][]{Steidel99,Adelberger04,Daddi04} 
turn out to be thick, turbulent, extended, rotating discs  
that are highly perturbed by transient elongated features and giant clumps 
\citep{Genzel06,Elmegreen06,Genzel08,Stark08,Law09,Forster09,Forster11a}.  
First dubbed ``clump-cluster" and ``chain" galaxies, based on their   
face-on or edge-on images  
\citep{Cowie95,Bergh96,Elmegreen04b,Elmegreen05,Elmegreen07}, 
a significant fraction of these galaxies ($>50\%$ for the more massive
ones) are confirmed by spectroscopic measurements to be rotating  
discs \citep{Genzel06,Shapiro08,Forster09}. 
The massive discs of $\sim 10^{11}\msun$ in baryons and radii $\sim 10\kpc$ 
tend to have high rotation-to-dispersion ratios of  
$\Vphi/\sigma \sim 2-7$ \citep{Cresci09}   
with high $\sigma \sim 20-80 \kms$ (one dimensional), 
while among the smaller galaxies there is a larger fraction of  
``dispersion-dominated" galaxies with $\Vphi/\sigma < 2$ 
\citep{Law07,Law09,Forster09}. 
 
These clumpy high-redshift discs are very different from the  
low-redshift disc galaxies of similar masses and sizes. 
In a typical $z \sim 2$ disc, up to half 
its restframe-UV light is emitted from a few giant clumps, 
each of a mass $\sim 10^9\msun$, a characteristic size $\sim 1\kpc$ 
that is resolved in HST imaging,  
and a star-formation rate (SFR) that could reach tens of $\!\sy$ 
\citep{Elmegreen04, ElmegreenE05, Forster06,Genzel08}. 
The non-negligible clump sizes, and the fact that the SFRs are not  
extremely high, indicate that the clumps are supported against catastrophic  
free-fall collapse. 
These clumps are much larger than the typical 
star-forming molecular clouds of $\sim 10^{5-6}\msun$ in local
galaxies\footnote{although there are also giant molecular associations of 
one to a few $10^7\msun$ in local galaxies 
\citep[e.g., in M51,][]{Rand90}.} 
This pronounced clumpy morphology is not a bandshift artifact in the sense  
that the restframe-UV images of low-redshift galaxies would  
not appear as clumpy if observed at high redshift with limited resolution and  
low signal-to-noise ratio \citep{Elmegreen09b},  
and indeed the high-redshift clumps are also seen in  
restframe-optical emission \citep{Genzel08,Forster09,Forster11a}. 
Estimates of the molecular gas to baryon fraction in the high-redshift discs,  
based on CO measurements, range from 0.2 to 0.8, with an average of  
$\sim 0.4-0.6$ at $z \sim 2$  
\citep{Daddi08,Daddi09,Tacconi10},   
systematically higher than the molecular gas fraction of $\sim 0.05-0.1$  
in today's discs \citep{saintonge11}. 
This gas may be in large molecular complexes, which give rise to  
star formation and the associated ionized gas that emits \Halpha. 
The typical ages of the stellar populations in these clumps  
are crudely estimated to range from one to several hundred Myr 
\citep{Elmegreen09,Forster09}, 
on the order of ten dynamical times, indicating that the clumps  
survive for such durations, but do not age unperturbed for Hubble times. 
The kinematic properties of $z \sim 2$ clumps are being studied 
by adaptive-optics \Halpha\ spectroscopy \citep{Genzel11}.  
 
The gravitational fragmentation of gas-rich and turbulent galactic discs  
into giant clumps has been addressed by simulations in the idealized   
context of an isolated galaxy  
\citep{Noguchi99, Immeli04a, Immeli04b,  
Elmegreen05, Bournaud07, Elmegreen08a, Bournaud09}, 
and in a cosmological context, using analytic theory 
\citep[][DSC09]{DSC} and cosmological simulations  
\citep{Agertz09b,CDB,Genel10}.  
According to the standard Toomre instability analysis \citep{toomre64}, 
a rotating disc becomes unstable to local gravitational  
collapse if its surface density is sufficiently high for its self-gravity to 
overcome the forces induced by rotation and velocity dispersion that resist  
the collapse.  
If the disc is maintained in a marginally unstable state with the maximum 
allowed velocity dispersion,  
and the cold fraction of the disc is about one third of the total mass 
within the disc radius, as obtained in a steady state (DSC09),  
the clumps are expected to be as massive as 
a few percent of the disc mass, with radii $\sim 10\%$ the disc radius.  
According to this robust theoretical understanding, most of the giant clumps  
in high-redshift discs were formed \insitu in the discs by gravitational  
instability of the gas and cold stars. 
The perturbed disc induces angular-momentum outflow and mass inflow  
into the disc centre, partly by clump migration through clump-clump  
interaction and dynamical friction. In the high-$z$ gas-rich discs, 
where the clumps are massive, this process operates on an  
timescale comparable to the orbital time at the disc edge. 
The gravitational energy gained by the inflow is driving turbulence 
that helps maintaining the disc in the marginally unstable state 
\citep{krumholz_burkert10,Cacciato11}.  
The mass inflow contributes to the growth of a central bulge, which in turn 
tends to stabilize the disc, but the continuous intense streaming of fresh  
cold gas from the cosmic web into the disc  
\citep{bd03,keres05,db06,ocvirk08,dekel09} 
maintains the high surface density 
and keeps the instability going for cosmological times (DSC09). 
 
Zoom-in cosmological simulations by \citet{Agertz09b} and  
\citet[][CDB10]{CDB}, which employed AMR hydrodynamics,  
confirmed the theoretical picture. 
With maximum resolution of $70\pc$ or less at all times,  
these simulations approached  
the physical conditions in star-forming regions, with densities well  
above $n \sim 10 \cmc$ and temperatures of 100 K. 
Indeed, the empirical values at high redshift are 
$n(H_2) \sim 10^{3-4} \cmc$ and $30-80$K \citep[e.g.,][]{Danielson11}. 
%
%
The simulations, zooming in on haloes of mass $\sim 5\times 10^{11}\msun$ 
at $z \sim 2.3$, typically revealed discs of a few $\times 10^{10} \msun$ 
and central bulges of comparable mass. 
The discs spend periods of $\sim 1 \Gyr$ in a marginally unstable state  
where they are continuously forming giant clumps, which account for $\sim 20\%$ 
of the disc mass at a given time and about half the total star formation  
rate. 
In these simulations, the individual clump masses indeed 
range from $10^8$ to a few times 10$^9 \msun$. 
The mean stellar ages within individual clumps 
range from $20$ to $700\Myr$, with a median at $150\Myr$ and
a systematic decline with distance from the disc center (\se{age}),
consistent with observations \citep{Elmegreen09,Genzel11,forster11b}  
and analytical estimates \citep[DSC09;][]{KrumholzDekel}.  
In these simulations,  
the clumps remain intact as they form stars and survive moderate  
supernova-driven outflows (but see other possibilities below). 
As expected, they migrate to the disc centre on an orbital timescale,  
where they coalesce with the bulge. 
External galaxies also come in with the streams --- the rare massive ones 
cause major mergers that help build the spheroid, while the more common 
small ones join the disc as mini-minor mergers, and can become a minor 
part of the clump population in or near the disc. 
 
It is still debatable
whether the \insitu giant clumps survive  
intact under stellar-feedback-driven outflows for durations longer than their  
migration timescale.  
There are observations of massive outflows from high-$z$ 
giant clumps \citep{Genzel11}, which, in extreme cases, are interpreted  
as potentially leading to disruption on a timescale comparable to the orbital  
time.  
Theoretical considerations argue that the typical high-$z$ giant clumps 
are supposed to survive stellar feedback, as long as the SFR efficiency  
(relative to star formation on a free-fall time) 
is obeying the local Kennicutt law between SFR and gas density.
In this case, energy-driven winds from standard supernovae 
do not provide enough power for un-binding the giant clumps  
\citep[DSC09, based on][]{DekelSilk86}.  
Momentum-driven outflows by stellar radiation and winds, which in principle 
could be more effective and may be responsible for clump disruption in 
local galaxies \citep{murray10}, are expected to remove only a fraction 
of the typical giant-clump mass at high redshift \citep{KrumholzDekel}. 
It has been argued, however, that the SFR efficiency could have been  
higher and then the feedback more effective, leading to disruption  
\citep{murray10}. Also, if the surface density is high enough, multiple 
scattering of photons may drive stronger outflows 
\citep[Dekel \& Krumholz, in preparation]{murray10,Hopkins11} 
This motivated \citet{Genel10} to simulate discs in which enhanced outflows 
do disrupt the clumps on a dynamical timescale. 
Even if the clumps are disrupted, the instability is associated with 
mass inflow within the disc toward bulge formation at its center,
as argued by theory and simulations \citep[DSC09,][]{Genel10}, 
as also indicated by observations \citep{Genzel08,Elmegreen09}. 
In this paper, we limit the analysis to five cases simulated as in CDB10, 
where the clumps survive intact for an orbital time $\sim 250 \Myr$ or more,  
such that the typical clump is caught in what seems to be 
dynamical equilibrium during its migration to the centre.

Our purpose in this paper is to find out to what extent the simulated 
giant clumps are in Jeans equilibrium  
between gravity, internal pressure forces and centrifugal forces, 
and to identify the mechanisms responsible for the support of the clumps 
against gravitational collapse. 
We estimate that the thermal pressure of the $\leq 10^4$K gas,
the pressure floor introduced in order to prevent artificial
fragmentation on the grid scale, and the supernova feedback as simulated,
each contributes energy per unit mass that amounts to only $\lsim (10\kms)^2$,
which falls short compared to the clump binding energy per unit mass
of $(40-200\kms)^2$.
The main source of pressure in the simulated clumps
is the macroscopic turbulent motions  
in the clumps, adding to a possible centrifugal force contribution  
if the clumps are rotating. 
Our main goal is thus to quantify the relative contributions of rotation and  
pressure by random motions to the support of the clumps against  
collapse under their own gravity.  
 
Using a simple model, we will quantify the expected rotation of the  
proto-clump in its own frame as induced by the global disc rotation,  
and the degree of rotational support after collapse by a given collapse  
factor under the assumption of conservation of angular momentum.  
An analysis of the simulated clumps with this simple model in mind will allow us 
to evaluate the validity of the simple model and to estimate the level of  
angular-momentum conservation during the clump collapse and evolution. 
We use five zoom-in cosmological simulations, extending the sample of three 
used in CDB10. The simulated discs produce giant clumps that form stars and  
survive for a couple of orbital times until they migrate to the centre and 
merge with the bulge. 
The cosmological simulations, with maximum resolution of 35-70 pc, 
only marginally resolve the clumps,  
so we complement our analysis with a convergence test using simulations of  
isolated discs with different resolutions reaching $\sim 1\pc$, which allow 
us to test the effects of clump substructure on the level of rotation support. 
The combination of cosmological simulations and the simulations of isolated 
discs with higher resolution enables coverage of a large dynamical range. 
The cosmological simulations form realistic unstable discs in steady state 
as they are continuously fed by streams from the cosmic web, while 
the simulations of isolated discs can follow in more detail the internal  
structure and dynamics within individual clumps.

This paper is organized as follows. 
In \se{theory} we describe a simple analytic model for clump support, 
that will guide our interpretation of the simulation results. 
In \se{sim} (and Appendix \se{art}) 
we describe the cosmological simulations and  
address the kinematics in example discs and giant clumps.  
In \se{stat} (and Appendix \se{clumps}) 
we analyze the clump support in a sample of 77 \insitu clumps drawn 
from the cosmological simulations at $z=2-3$. 
In \se{subs} (and Appendix \se{ramses})  
we study the effect of resolution on the clump support 
using isolated simulations of gas-rich discs 
with resolution that ranges from the resolution 
of the cosmological simulations to a resolution 60 times better. 
In \se{obs} we address the detectability of clump rotation  
in \Halpha\ observations with a given beam width. 
In \se{exsitu} we consider the minor population of \exsitu clumps  
that joined the discs as minor mergers, and highlight the case of retrograde 
rotators.  
In \se{age} we explore the clumps stellar age and gas fraction gradients
 across the disc  
as a testable prediction for the scenario simulated. 
In \se{gmc} we investigate the non-rotational clump support in isolated  
simulations of gas-poor discs that mimic low-$z$ galaxies.  
In \se{conc} we summarize our results, discuss them, and draw conclusions.

\section{Analytic Model} 
\label{sec:theory} 
 
Our analysis is motivated by a simple analytic model,  
that will guide our interpretation of simulation results. 
 
\subsection{Clump Support} 
\label{sec:Jeans} 
 
We approximate the clump as an axisymmetric rotator with an isotropic velocity 
dispersion. The Jeans equation inside the clump then 
reads \citep[e.g.][eq. 4.230]{bt08} 
\be 
\Vphi^2(r,z) = r\frac{\pa\Phi}{\pa r} 
+ \frac{r}{\rhoc} \frac{\pa(\rhoc\,\sigc^2)}{\pa r} \, . 
\label{eq:jeans1} 
\ee 
Here $r$ is the polar-coordinate radius from the clump centre,  
and all other quantities are functions of $r$ and $z$ within the clump: 
$\Vphi$ is the average rotational velocity, 
$\Phi$ is the gravitational potential, 
$\rhoc$ is the mass density, 
and $\sigc$ is the 1D velocity dispersion. 
Thermal pressure gradients are ignored, as the temperature is $\leq 
10^4$K, equivalent to $\sigc \leq 10 \kms$. 
In the equatorial plane, the potential well defines the circular velocity  
of the clump, 
\be 
\Vcc^2 = r \frac{\pa\Phi}{\pa r} \simeq \frac{G M(r)}{r} \, , 
\ee 
so 
\be 
\Vcc^2 \simeq {\Vphi}^2 - \frac{r}{\rhoc} \frac{\pa(\rhoc\,\sigc^2)}{\pa r}\, . 
\ee 
This separates the contributions of rotation and pressure to the 
support of the clump against gravitational collapse. 
If the clump is approximately an isothermal sphere,  
i.e., $\sigc=const.$ and $\rhoc \prop r^{-2}$, we obtain\footnote{If the
clump was a 
self-gravitating exponential disc with scale length $r_{\rm exp}$, the pressure 
term should have beeen multiplied by $r/r_{\rm exp}$ \citep{Burkert10}.} 
\be 
\Vcc^2 =\Vphi^2 + 2\sigc^2 \, . 
\label{eq:jeans2} 
\ee 
The relative contribution of the centrifugal force to the clump support 
can be expressed by the {\it rotation parameter}  
\be 
\Rot \equiv \frac{\Vphi^2}{\Vcc^2} \, . 
\label{eq:R} 
\ee 
According to \equ{jeans2}, $\Rot$ is related to the familiar ratio of 
rotation to dispersion by 
\be 
\left( \frac{\Vphi}{\sigc}\right)^2 = 2\, (\Rot^{-1} -1)^{-1} \, . 
\label{eq:Voversigma} 
\ee  
One can refer to the clump as ``rotationally supported" when 
$\Rot>0.5$, namely $\Vphi/\sigc>\sqrt{2}$.

\subsection{Clumps in Wildly Unstable Discs} 
\label{sec:Sigma} 
 
\def\Vd{V_{\rm d}} 
\def\sigd{\sigma_{\rm d}} 
\def\Sigd{\Sigma_{\rm d}} 
\def\Sigc{\Sigma_{\rm c}} 
\def\Omd{\Omega_{\rm d}} 
\def\Omci{\Omega_{\rm c,i}} 
 
As described for example in DSC09, 
the wildly unstable discs self-regulate themselves in a marginally unstable
state with a Toomre parameter $Q \simeq 1$. 
The mutual gravitational interaction between the perturbations in the disc, 
both the transient features and the bound giant clumps, 
keeps the disc velocity dispersion at a level that maintains $Q \simeq 1$. 
 
Generalizing \S 2 of DSC09 to 
a disc with a rotation curve $\Vd \prop \rdi^\alpha$, we have 
\be 
Q= [2(1+\alpha)]^{1/2} \frac{\sigd}{\Vd} \delta^{-1} \, , 
\ee 
where $\Vd$ and $\sigd$ are the disc circular velocity and one-dimensional  
velocity dispersion respectively, and $\delta = \Md/\Mt$ is the ratio of 
cold disc mass to total mass inside the disc radius,  
involving also a stellar 
spheroid and the inner part of the dark-matter halo. 
A typical value for an unstable disc in steady state at high redshift 
is $\delta \sim 0.33$ (DSC09). 
Note that solid-body disc rotation is $\alpha=1$,  
a self-gravitating uniform disc is $\alpha=0.5$,  
flat rotation curve is $\alpha=0$, 
and Keplerian rotation is $\alpha=-0.5$. 
 
Marginal instability $Q \simeq 1$ corresponds to 
\be 
\delta \simeq [2(1+\alpha)]^{1/2} \frac{\sigd}{\Vd} \, . 
\label{eq:delta} 
\ee 
The characteristic clump initial radius $\Rci$ and mass $\Mc$ relative to 
the disc radius $\Rd$ and mass $\Md$   
as estimated from the ``most unstable" mode of Toomre instability  
are given by 
\be 
\frac{\Rci}{\Rd}  
\simeq \frac{\sqrt{2}\pi}{4(1+\alpha)^{1/2}}\,\frac{\sigd}{\Vd} \, , 
\quad 
\frac{\Mc}{\Md} \simeq \left(\frac{\Rci}{\Rd}\right)^2 \, . 
\label{eq:Rci} 
\ee 
For example, 
with $\Vd/\sigd \sim 5$ 
and $\alpha=0$, 
the typical clump mass is $\sim 5\%$ of the disc mass.
For a disc of $\sim 10^{11}\msun$, we expect clumps of a few times $10^9\msun$.
If the disc radius is $\sim 5\kpc$, the initial clump radius is $\sim 1\kpc$.

Assume that the clump has contracted by a factor $c$, 
from its initial radius to a final radius $\Rc$ where it is supported by 
rotation and pressure, 
\be 
c \equiv \frac{\Rci}{\Rc} = \left( \frac{\Sigc}{\Sigd} \right)^{1/2} \, , 
\label{eq:c} 
\ee 
with $\Sigc/\Sigd$ the 2D overdensity in the clump. 
Note that the contraction factor is not expected to be much larger 
than two or a few, 
because when $Q \simeq 1$, 
the initial unstable perturbation is already not far from being supported 
against gravitational collapse both by pressure and by rotation. 
Indeed, the simulations discussed below show typical overall clump  
overdensities of $\Sigc/\Sigd \sim 10-20$, namely $c \sim 3-5$  
\citep[Figures 4-6 of][]{CDB}. 
 
Based on \equ{delta} and \equ{Rci},  
and using the fact that $\Vd^2=G\Md\delta^{-1}/\Rd$, 
the circular velocity of the bound  
clump (independent of the disc rotation curve slope $\alpha$) is 
\be 
\Vcc^2 \simeq \frac{G\Mc}{\Rc}  
\simeq \frac{\pi}{2} c\,\sigd^2 \, . 
\label{eq:Vcc} 
\ee 
 
\subsection{Rotation of a Protoclump Patch due to Disc Rotation} 
 
We assume that the proto-clump is a small cylindrical patch cut from a  
purely rotating disc.  
Let the disc centre be $\vec{r}_{\rm d}=0$ and the patch centre  
$\vec{r}_{{\rm d}0}$, 
and let the disc rotation curve be $\Vd(\rdi) = \rdi^\alpha$ near $\rd0$, 
with radii and velocities measured in units of $\rd0$ and  
$\Vdi0=\Vd(\rd0)$ respectively. 
 
In the non-rotating inertial frame that is momentarily moving with the 
centre of the patch, consider a ring of radius $\rt$. 
At a longitude $\phi$ along the ring 
(measured from the radius vector $\hat{r}_{\rm d0}$), 
the tangential velocity is 
\be 
v(\phi) = \rdi^{\alpha-1}\, (\cos \phi +\rt) -\cos \phi \, , 
\ee 
where 
\be 
\rdi=(1+2\rt \cos\phi +\rt^2)^{1/2} \, . 
\ee 
In the limit of a small patch, $\rt \ll 1$, simple algebra yields 
\be 
v(\phi) \simeq \rt\, [1+(\alpha-1) \cos^2\phi] \, . 
\label{eq:vphi} 
\ee 
 
\begin{figure} 
\vskip 8.7cm 
\includegraphics{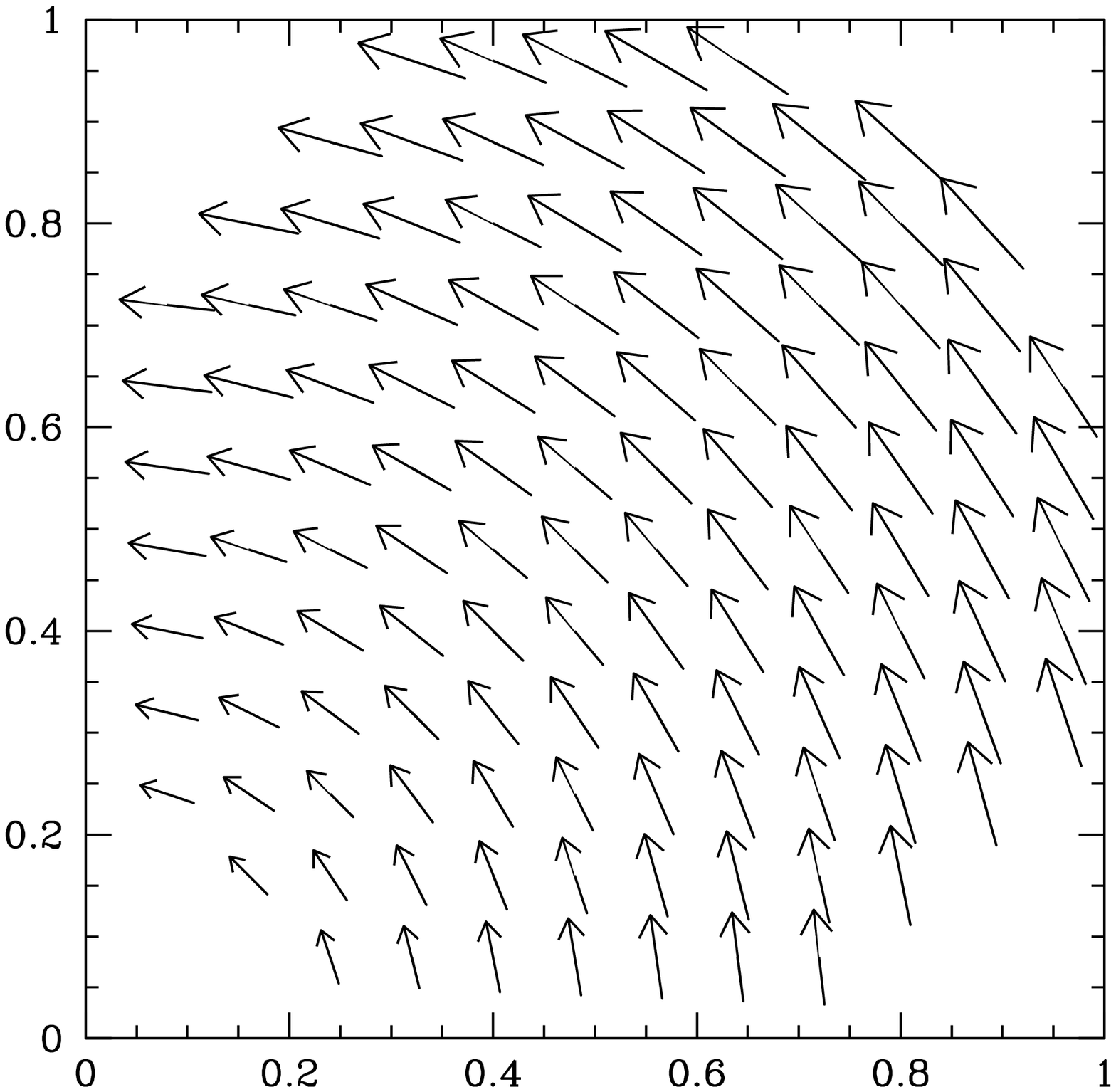} 
\includegraphics{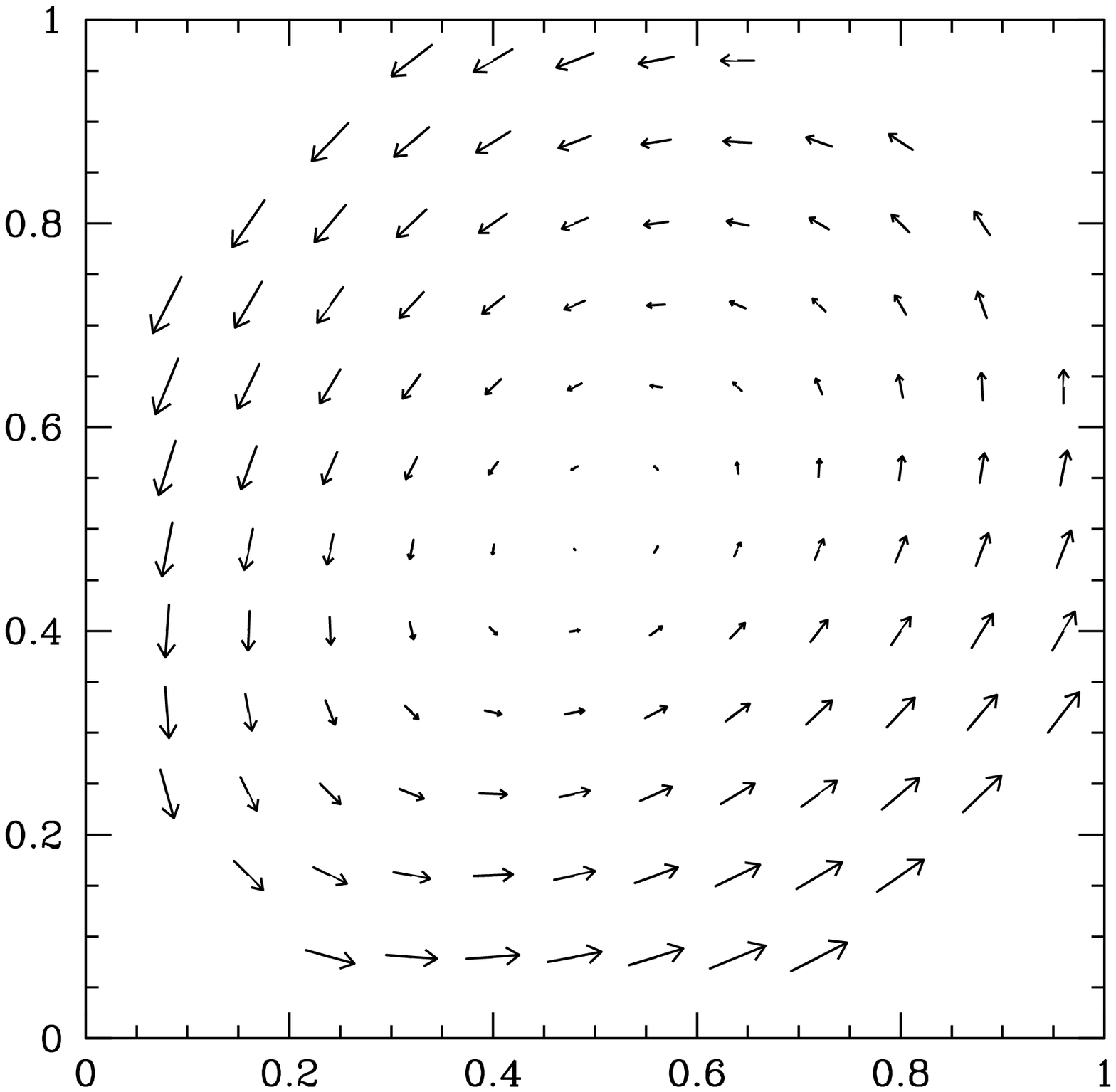} 
\includegraphics{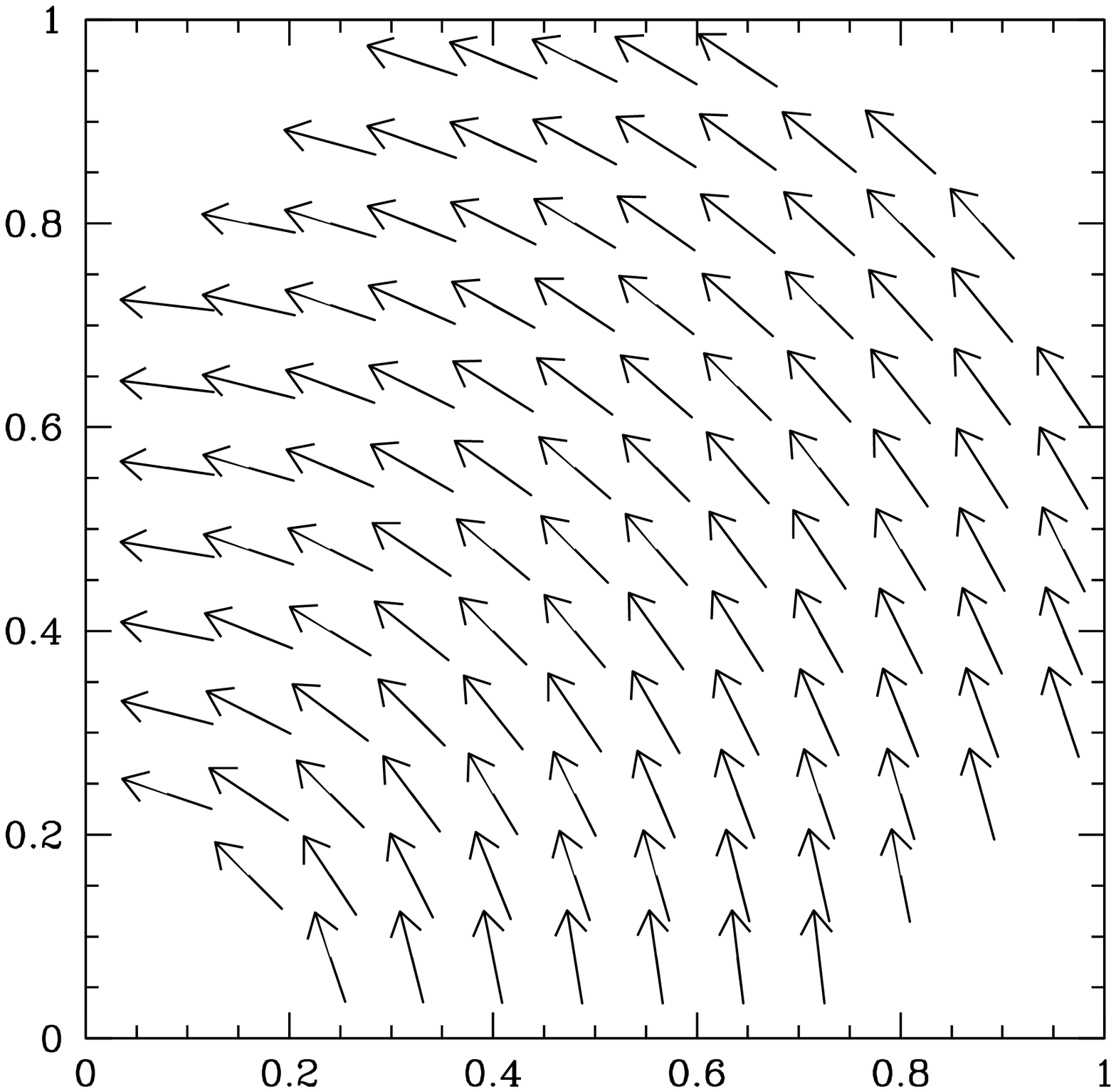} 
\includegraphics{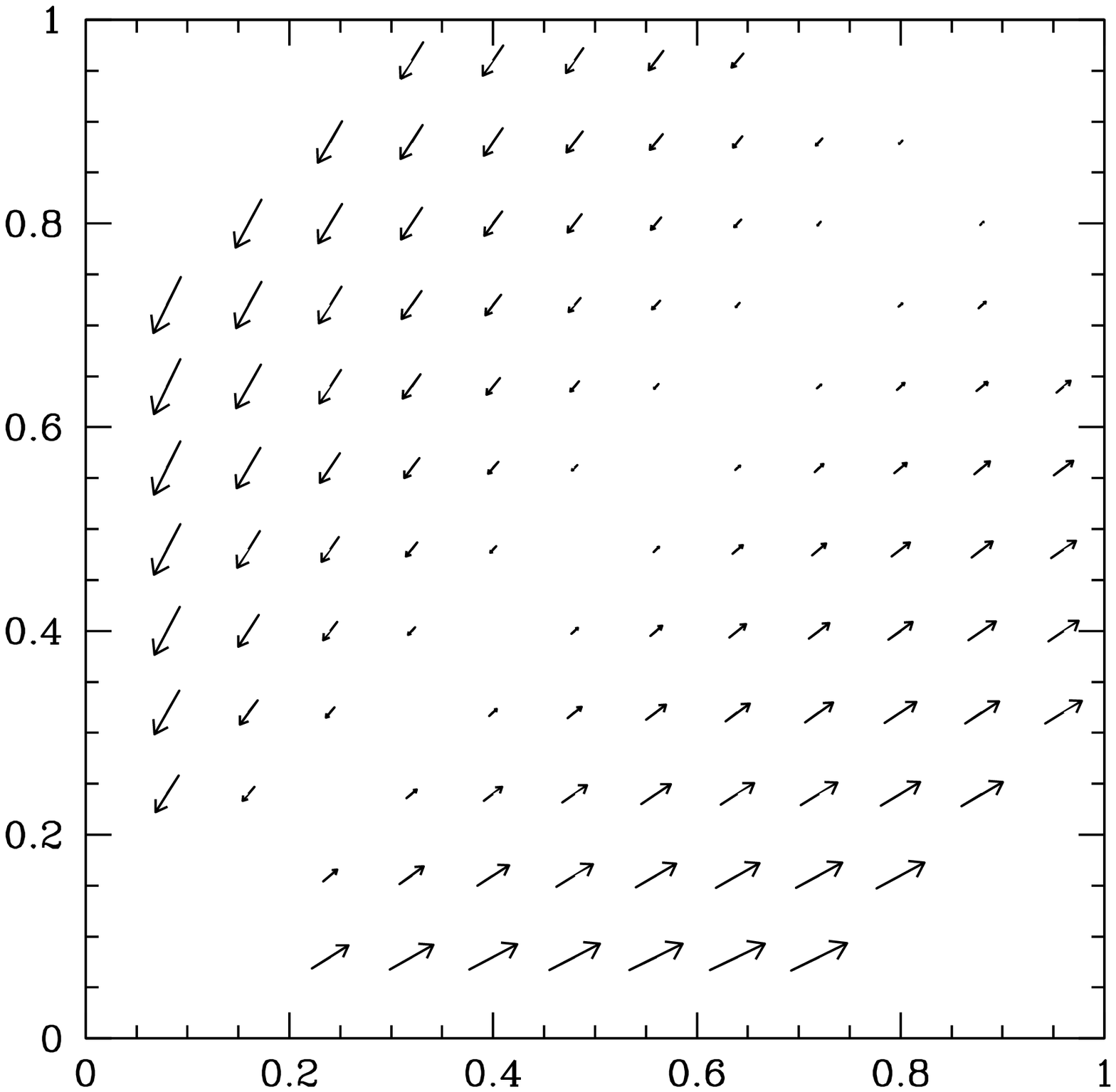} 
\caption{Velocity in a circular patch [centred at (0.5,0.5)], 
part of a rotating disc [centred at (0,0)]. 
Left: in the disc frame. Right: about the patch centre. 
Top: self-gravitating uniform disc, Bottom: flat rotation curve. 
The tangential component in the patch frame is given in \equ{vphi}. 
} 
\label{fig:toy} 
\end{figure} 
 
We see that for a disc in solid-body rotation, 
$\Omd =\Vd/\rdi=const.$, the patch is also in 
prograde solid-body rotation with $\Omci = \Omd$, as expected. 
For any smaller $\alpha$, at $\phi=\pm\pi/2$, namely at $\rdi=\rd0$, 
the patch tangential velocity remains the same, $v=(\rt/\rd0)\Vdi0$, 
independent of $\alpha$. 
On the other hand, at $\phi=0$ and $\pi$, along the disc radius vector, 
the disc differential rotation has a retrograde contribution that brings 
the patch tangential velocity to its $\alpha$-dependent 
minimum, $v=\alpha (\rt/\rd0) \Vdi0$. 
This is illustrated in \fig{toy}. 
 
Assuming that the disc is uniform within the small patch, 
by integrating \equ{vphi} over a circular ring of radius $\rt$ one obtains 
the average tangential velocity 
\be 
\la v (\rt) \ra \simeq 0.5(1+\alpha) \frac{\rt}{\rd0} \Vdi0 \, , 
\label{eq:avv} 
\ee 
with the corresponding average angular frequency independent of $\rt$ 
\be 
\la \Omci \ra \simeq 0.5 (1+\alpha) \Omd(r_0) \, . 
\ee 
For the four types of rotation curves $\alpha=1,1/2,0,-1/2$, 
the patch has $\la \Omci \ra/\Omd \simeq 1,3/4,1/2,1/4$ respectively. 
In all cases the overall patch rotation is prograde. 
Note that by assuming that the proto-clump is a cylindrically symmetric patch 
we are ignoring the distortions by the shear, which are expected to be more  
severe as the disc rotation curve deviates from solid-body rotation. 
 
In a $Q \simeq 1$ disc, using \equ{avv} and \equ{Rci}, 
the rotation velocity in the initial clump at radius $\Rci$ is thus 
\be 
\Vci \simeq 0.5\,(1+\alpha) \frac{\Rci}{\Rd} \Vd 
\simeq  \frac{\sqrt{2}\pi}{8}\,(1+\alpha)^{1/2}\, \sigd\, . 
\ee 
For a flat-rotation-curve disc, this is $\Vci \simeq 0.56\,\sigd$. 
 
The specific angular momentum at r in the protoclump  
is $j(\rt)=\Omci r^2$.  
For a flat rotation curve, the average specific angular momentum 
at the proto-clump edge $\Rci$ is 
\be 
j(\Rci) \simeq \frac{\pi^2}{32} \delta^2\, \Rd \Vd \, .
\label{eq:jedge} 
\ee 
 
If the clump surface-density profile mimics a projection of an isothermal  
sphere, $\Sigma \prop r^{-1}$, the average specific angular momentum
over the whole clump is one-third its value at the edge. 
 
\subsection{Rotation of a Collapsed Clump} 
 
We next assume that the circular patch contracts by a factor $c$ 
to form a clump in equilibrium, and tentatively 
make the assumption that angular momentum is conserved during  
this contraction, to be tested below using the simulations.  
If angular momentum is conserved, the clump is spun up by the contraction 
and its rotation velocity becomes 
\be 
\Vphi \simeq c \Vci 
\simeq \frac{\sqrt{2}\pi}{8} (1+\alpha)^{1/2} c\, \sigd 
\simeq 0.56\, c\, \sigd  \, , 
\ee 
where the last equality is for $\alpha=0$. 
 
A comparison with the circular velocity of the bound clump (\equ{Vcc}) gives 
\be 
\Rot = \frac{\Vphi^2}{\Vcc^2}  
\simeq  \frac{\pi\,(1+\alpha)\,c}{16}  
\simeq 0.2\, c \, .   
\label{eq:Vrot} 
\ee 
Using \equ{Voversigma}, the latter corresponds to 
\be 
\frac{\Vphi}{\sigc} 
\simeq \sqrt{2}\, ( 5.1\,c^{-1} -1 )^{-1/2} \, . 
\ee 
For $\alpha=0$, full rotation support ($\Rot=1$, $\Vc/\sigc \gg 1$)  
is obtained for 
a collapse factor $c \simeq 5$.  
Equal contribution from rotation  
and pressure ($\Rot=0.5$, $\Vc/\sigc=\sqrt{2}$)  
is obtained for $c \simeq 2.5$. 
For a rotation curve mimicking a uniform disc, $\alpha=0.5$,  
full rotation support is obtained for $c \simeq 3.4$. 
Thus, if the clumps conserve most of their angular momentum as they  
collapse to 2D overdensities of $\sim 10$, the rotation 
provides most of the support. 
 
The assumptions of a circular proto-clump patch and conservation of 
angular-momentum during collapse are quite uncertain. 
The actual proto-clump is expected to deviate from circularity because  
of shear, and the disc exerts torques on the clump. 
These torques could involve the transient features in the disc,  
clump-clump encounters, dynamical friction and gas drag.  
Angular-momentum loss could also be associated with mass exchange  
between the clumps and the disc. 
These effects are not easy to estimate analytically, 
and it is not even clear a priori whether the clump rotation as estimated 
above is an overestimate or an underestimate.  
Nevertheless, it may serve as a reference for the results obtained  
from the simulations. 
 
Velocity dispersion provides the rest of the support against gravitational 
collapse.  If a clump has lost all its angular momentum, its internal  
one-dimensional velocity dispersion, from \equ{jeans2}, is   
\be 
\sigc \simeq \frac{1}{\sqrt{2}} \Vcc 
\simeq \frac{\pi^{1/2}}{2} c^{1/2}\,\sigd  
\label{eq:sigc}
\ee 
for any $\alpha$.  With a collapse factor 
$c \simeq 3$ we get a maximum possible velocity dispersion  
of $\sigc \simeq 1.5 \sigd$. 
Realistically, with a significant contribution from rotation,  
the internal velocity dispersion in the clumps is expected to be 
comparable to and somewhat smaller than that of the disc. 
Some variations may be expected from clump to clump,
but in typical clumps, one does not expect large coherent dispersion 
residuals to be associated locally with the clumps.  
In particular, based on \equ{sigc}, any correlation between the clump
internal velocity dispersion and the clump surface-density contrast 
is expected to be rather weak,
$\sigc \prop \Sigc^{1/4}$. 
 
If the resolution allows measuring the velocity dispersion on sub-clump scales, 
where dissipative collapse may correspond to $c \gg 1$, 
one may obtain a somewhat larger velocity dispersion in the clumps, 
but growing only in proportion to 
$\Sigc^{1/4}$. If the dispersion is weighted by \Halpha\ density, 
and if it traces regions of surface-density contrast $\sim\!100$, say, 
one may expect a positive dispersion residual but still comparable to 
$\sigd$, $\sigc\!\sim\!2 \sigd$.

\section{Giant Clumps in Cosmological Simulations} 
\label{sec:sim} 
 
\subsection{The cosmological simulations}

We use zoom-in hydro cosmological simulations of five moderately massive 
galaxies with an AMR maximum resolution of $70\pc$ or better, evolved till 
after $z \sim 2$. They utilize the ART code
\citep{Kravtsov97,Kravtsov03}, which accurately follows the evolution of a
gravitating N-body system and the Eulerian gas dynamics using an adaptive mesh.
Beyond gravity and hydrodynamics, the code incorporates at the subgrid level
many of the physical processes relevant for galaxy formation.
They include gas cooling by atomic hydrogen and helium, metal and molecular
hydrogen cooling, photoionization heating by a UV background with partial
self-shielding, star formation, stellar mass loss, metal enrichment of the ISM,
and feedback from stellar winds and supernovae, implemented as local injection
of thermal energy.
More details concerning the simulation method are provided in an appendix,
\se{art}, as well as in \citet{Ceverino09} and CDB10.

The five dark-matter haloes were drawn from N-body simulations
of the $\Lambda$CDM cosmology with the WMAP5 parameters (\se{art}),
in a comoving cosmological box. 
The haloes were selected to have a virial mass in a desired mass   
range at $z=1$. 
The only other selection criterion was that they show no ongoing major merger
at that time. This eliminates less than $10\%$ of the haloes,
and has no noticeable selection effect at $z \geq 2$, where our main analysis
is performed.
Galaxies A, B and C, which have been studied in some detail in CDB10,
were selected to have a virial mass $\Mv \sim 10^{12}\msun$ at $z=1$
(intended to end up as $(3-4)\times 10^{12}\msun$ today, somewhat more
massive than the Milky Way).  Two new galaxies, D and E, were selected to have
$\Mv \sim 4 \times 10^{12} \msun$ at $z=1$, four times more massive than the
other galaxies.
The virial properties of the five dark-matter haloes in the snapshots analyzed
in the redshift range $z\sim 1.9-3$ are listed in \tab{0a}.
As expected, the halos of galaxy D and E are more massive than A-C at all
times. For example, at z=2.3,  
galaxies A-C have a virial mass of $(0.40, 0.35, 0.61) \times 10^{12} \msun$,
whereas galaxy D and E have $\Mv = (0.94, 1.54) \times 10^{12} \msun$, 
roughly two and four times more massive than galaxy A.

\begin{figure} 
\includegraphics[trim = 0.9cm 0.9cm 0.3cm 1.5cm, clip, width=0.48\textwidth]
               {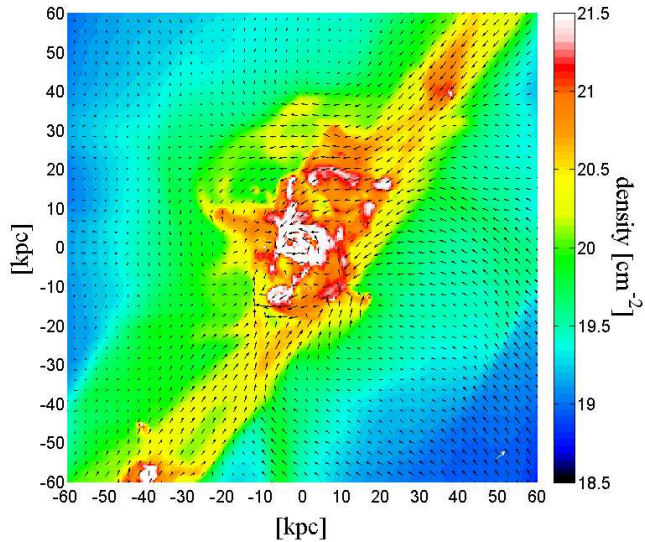}
\caption{
Gas surface density and projected velocity field within the halo of galaxy A
at $z=2.3$.
The box roughly encompasses the virial sphere of $\Rv=70\kpc$, and
the projection depth along the line of sight is $80\kpc$.
The arrows represent the velocity field, mass-weighted average along the
line of sight, with the white arrow denoting $200\kms$.
Narrow, long streams of cold gas and merging galaxies feed a central rotating
disc through a messy interphase region. The disc is seen nearly face on.
}
\label{fig:streams}
\end{figure}

The initial conditions corresponding to each of the selected haloes
were filled with gas and refined to a much higher resolution on an adaptive
mesh within a zoom-in Lagrangian volume that encompasses the mass within
twice the virial radius at $z =1$, roughly a sphere of comoving radius $1\Mpc$.
This was embedded in a comoving cosmological box of side $20$ and $40\hmpc$
for galaxies A-C and D-E respectively.
Each galaxy has been evolved with the full hydro ART and subgrid physics on an
adaptive comoving mesh refined in the dense regions
to cells of minimum size between 35-70 pc in physical units.
This maximum resolution is valid in particular throughout the cold discs
and dense clumps, allowing cooling to $\sim 300$K and gas densities of
$\sim 10^3\cmc$.
The dark-matter particle mass is $6.6\times 10^5\msun$, and the particles
representing stars have a minimum mass of $10^4\msun$.

As listed in \tab{0c}, 
the sample spans roughly an order of magnitude in stellar mass inside
the disc radius, ranging from 10$^{10}$ to 10$^{11} \msun$, 
in an order that reflects the halo mass.
Galaxy B is the smallest, with $\Ms = 1.3 \times 10^{10} \msun$ 
inside $\Rd = 3\kpc$ at z=2.3, and galaxy E is the biggest, with 
$\Ms = 1.4 \times 10^{11} \msun$ inside $\Rd = 8.5\kpc$ at $z=2.3$.
Although this is a small set of galaxies, 
they span the mass range of typical observed massive star-forming galaxies at 
$z \sim 2$ \citep{Forster09}, and otherwise no selection criteria was 
imposed on their properties at $z \sim 2-3$.
As shown in CDB10, galaxies A-C are consistent with the observed 
scaling relations of $z \sim 2$ galaxies, including the relation between
SFR and stellar mass and the Tully-Fisher relation \citep{Forster09, Cresci09}. 
We can therefore assume that this is a fair sample of galaxies in the relevant
mass and redshift range.

\subsection{Cold streams and clumpy discs} 

In order to get familiar with the large-scale context of the disc giant clumps,
\fig{streams} shows the gas in the halo of galaxy A at redshift
$z \sim 2.3$ -- a prototypical case of a high-$z$ clumpy disc (see CDB10).
The central disc, of radius $\sim 6\kpc$,
is continuously fed by a few, co-planar
narrow streams that extend to hundreds of kpc
as they ride the dark-matter filaments of the cosmic web
\citep[][Danovich et al. in prep.]{dekel09}.
The streams consist of gas at $\sim (1-5)\times 10^4$K as well as clumps of
all sizes, the biggest of which are actual galaxies with gas, stars and
dark-matter haloes, to be merged with the central galaxy.
These supersonic streams penetrate to the central regions of the dark-matter
halo where they interact with other streams and the disc and blend into
a turbulent interphase region that encompasses about 20\% of the virial radius
before the gas and stars settle in the central disc and bulge.
The complex structure and kinematics in this interphase region, where
energy, momentum and angular momentum are being transferred among the
different components of gas, stars, dark matter and radiation,
is yet to be investigated.
A significant fraction of the gravitational energy gained by the infall
into the halo potential well is released as Lyman-alpha radiation
\citep[][Kasen et al. in prep.]{Goerdt10},
while the cold streams can also be detected in absorption
mostly as Lyman-limit systems \citep{Fumagalli11}.
The continuous intense input of cold gas drives the disc into
violent disc instability, and helps maintain this configuration in a
self-regulated steady state for cosmological times (DSC09, CDB10). 
Although the disc is perturbed by the continuous instreaming,  
it does maintain a global disc shape and organized rotation pattern
for long periods, and it behaves according to 
the expectations from Toomre instability (DSC09,CDB10).

\begin{figure*} 
\subfigure{\includegraphics[trim = 1.2cm 0.9cm 0.3cm 1.0cm, clip,
           width=0.495\textwidth]{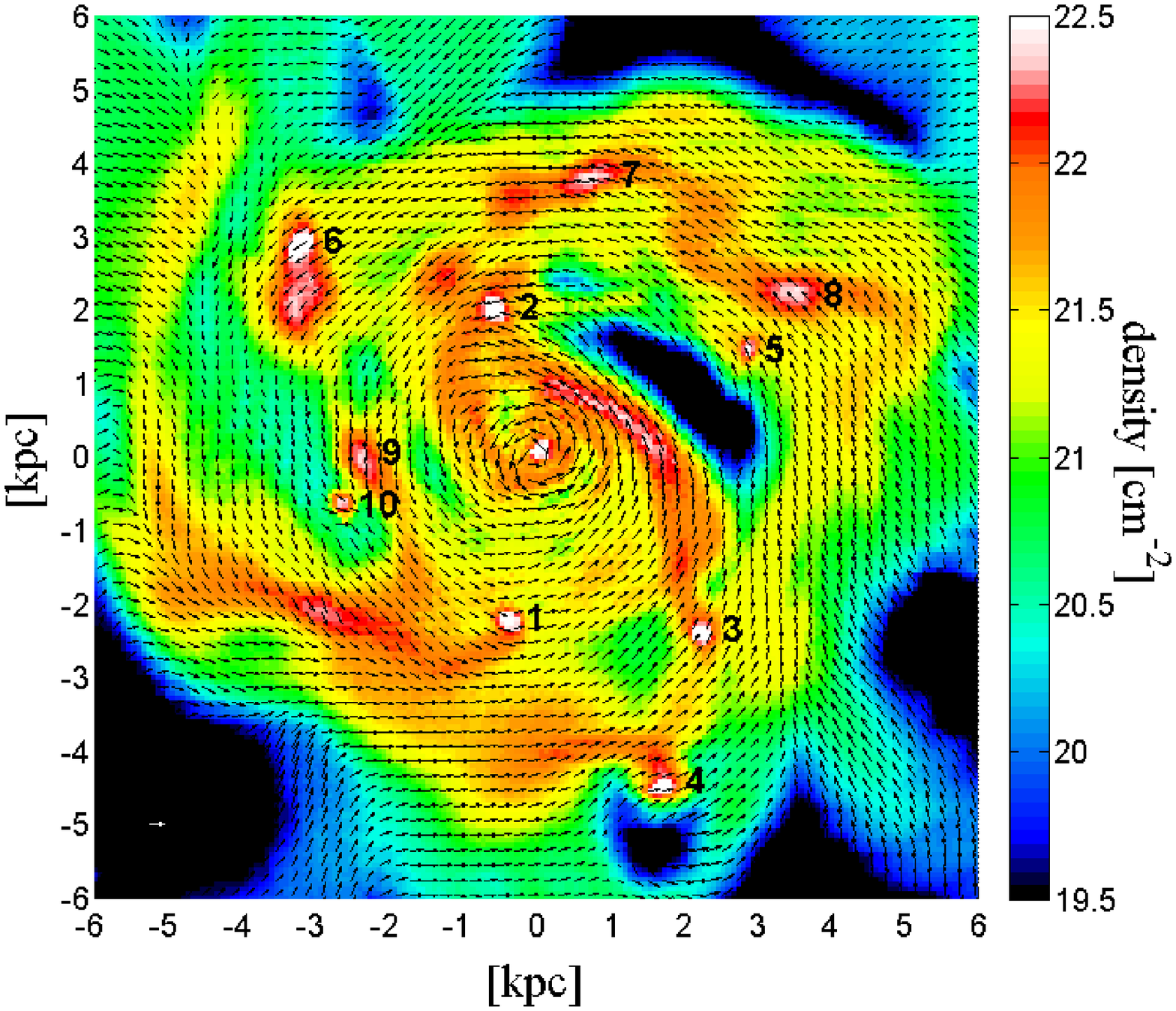}}
\subfigure{\includegraphics[trim = 1.2cm 0.9cm 0.3cm 1.0cm, clip,
           width=0.495\textwidth]{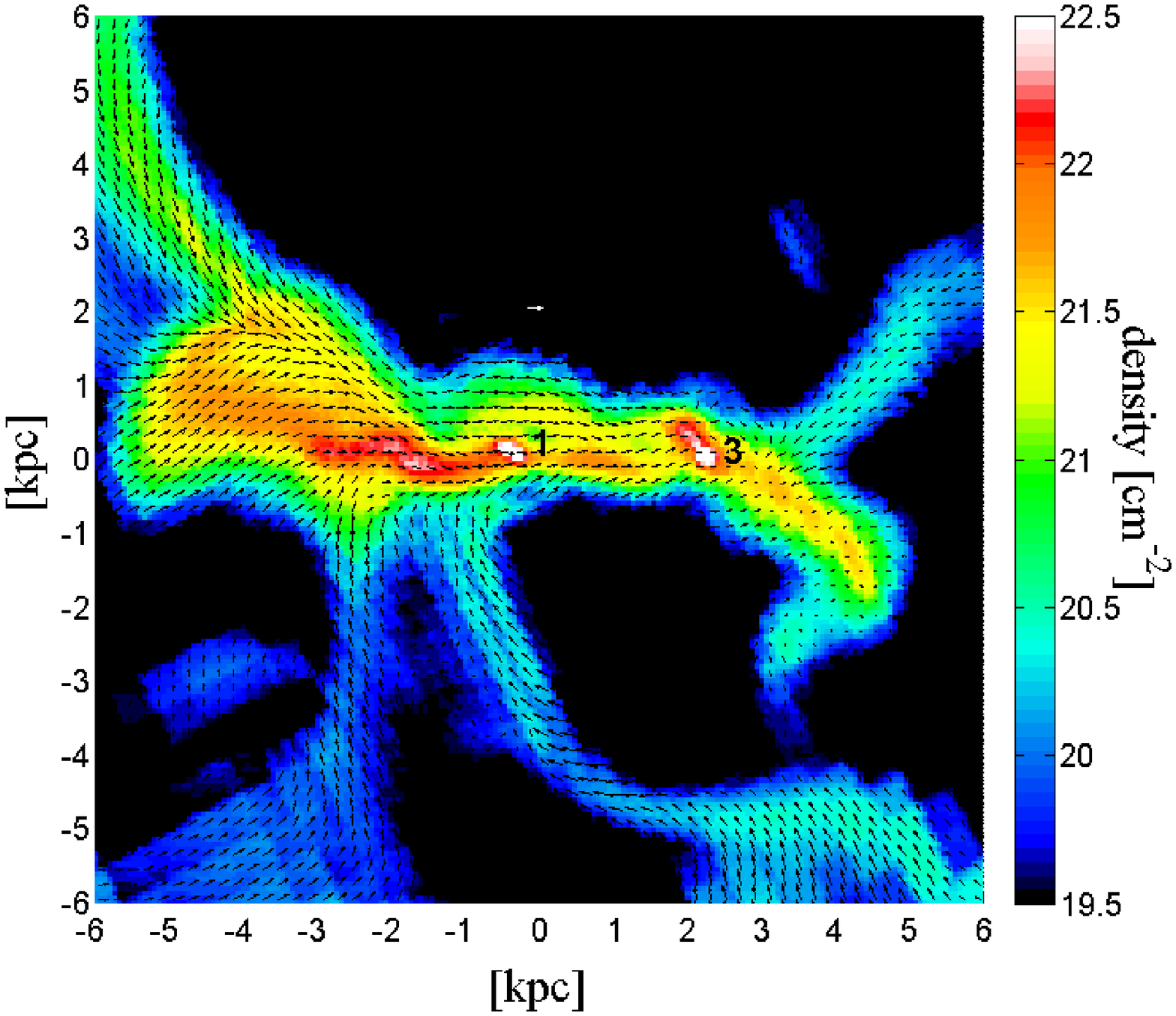}}
\caption{
Gas surface density and projected velocity field in the disc of galaxy A
at $z=2.3$, face on and edge on (details similar to \fig{streams}). 
The box encompasses the main body of the gas disc, a radius of $\sim 6\kpc$.
The face-on projection depth is $2\kpc$, 
and the edge-on depth is $1.2\kpc$, about clumps 1 and 3.
The reference velocity (white arrow) is $200\kms$. 
The rotating disc shows a highly-perturbed morphology, with large
elongated transient features and ten compact giant clumps marked by
numbers (two of which are \exsitu clumps, \se{exsitu}),
see \tab{clumps}.
The mass of a typical clump is a few percent of the disc mass, and the
total mass in clumps is $\sim 0.2$ of the disc mass. 
The disc is highly warped locally, sometimes tilting the clumps relative
to the global disc plane.}
\label{fig:disc_mw3} 
\end{figure*}

\begin{figure*} 
\subfigure{\includegraphics[trim = 1.2cm 0.9cm 0.3cm 1.0cm, clip,  
       width=0.495\textwidth]{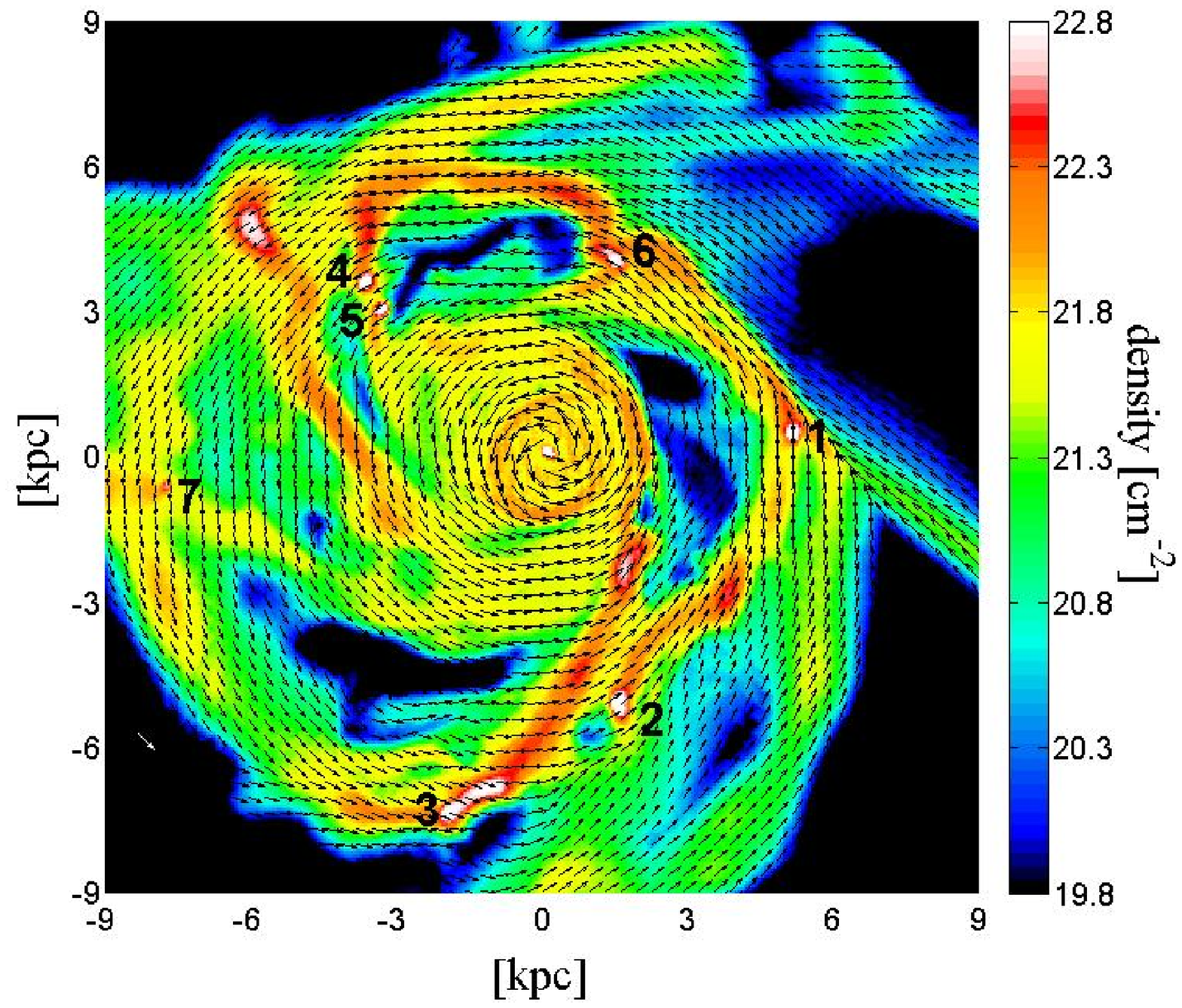}} 
\subfigure{\includegraphics[trim = 1.2cm 0.9cm 0.3cm 1.0cm, clip,  
       width=0.495\textwidth]{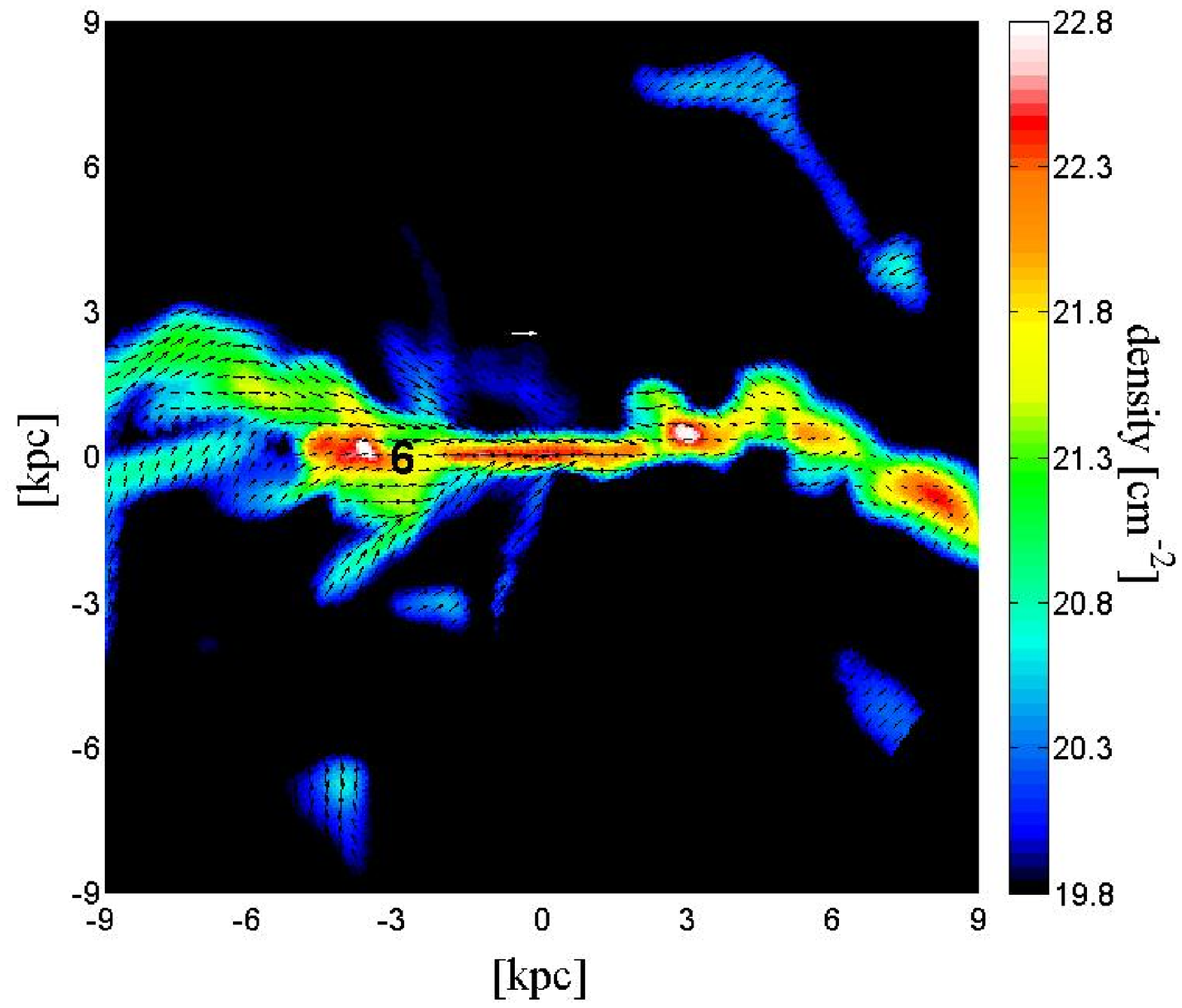}} 
\caption{
Gas surface density and projected velocity field in the disc of galaxy E 
at $z=2.4$, face on and edge on (details as in \fig{disc_mw3} and 
\fig{streams}). 
The box encompasses the main body of the gas disc, 
which has a radius of $\sim 9\kpc$,
but is asymmetric about the kinematic center. 
The projection depths are $2 \kpc$, with the edge-on slice including   
clump 6 and another perturbation.
The white arrow marks a velocity of $500\kms$.  
The rotating disc shows a highly-perturbed morphology, with large 
elongated transient features and seven compact giant clumps marked by numbers 
(see \tab{clumps}). 
The mass of a typical clump is a few percent of the disk mass, and the 
total mass in clumps is $\sim 0.2$ of the disk mass. 
The disc is highly warped locally, sometimes tilting the clumps relative 
to the global disc plane.
}
\label{fig:disc_sfg1} 
\end{figure*} 

Although a detailed analysis of the galaxy properties in these simulations 
is beyond the scope of the present paper, \tab{0b} and \tab{0c} list 
relevant global properties for the five clumpy discs
at the different snapshots used for the analysis of giant clumps.
The disc of stars and gas and the stellar bulge are typically comparable
in mass.  The disc mass ranges from $\Md=7 \times 10^{9}\msun$ in galaxy B 
to $\Md=7 \times 10^{10}\msun$ in galaxy E at $z\sim 2$. 
Despite their perturbed morphology, the discs rotate with a rotation speed 
that ranges between $\Vd \sim 200 \kms$ 
in galaxy A and B to $\Vd \sim 400 \kms$ in galaxy E. 
The gas velocity dispersion is high, $\sigd = 20-60 \kms$, 
which should lead to
Toomre giant clumps with masses of a few percent 
of the disc mass, $\Mc=10^8-10^9 \msun$. 
Many of the simulated galaxy properties resemble the properties of observed
clumpy discs at high redshift 
\citep{Genzel06,Elmegreen06,Genzel08,Genzel11,Forster11a}, 
although the simulated gas fractions of 0.1-0.35 are somewhat low
compared to the values 0.3-0.6 observed at $z \sim 2$ 
\citep{Daddi09,Tacconi10}, probably due to an overproduction of 
stars in the simulation at earlier times by a factor of $\sim 2$ (CDB10). 
The underestimated gas fractions suggest that our simulations  
conservatively underestimate the actual effects of gravitational instability
in real galaxies at $z \sim 2$.
In particular, the modest gas fractions are likely to cause an underestimate 
of the degree of dissipation and therefore the clump contraction factor, 
thus lowering the degree of clump rotational support (\se{theory}). 

\Fig{disc_mw3} shows the gas in the central disc of galaxy A at redshift 
$z=2.3$, face on and edge on. 
The total mass of baryons in the disc is $1.1\times 10^{10}\msun$,  
and the gas fraction is 0.35.  A comparable baryonic mass is in a stellar 
bulge. 
The disc shows a systematic rotation of $\sim 180 \kms$ extending to a radius  
of $\sim 6 \kpc$, and certain local perturbations. 
The density is highly perturbed, with elongated transient features and eight  
giant clumps, of baryonic masses $(1.8-4.0)\times 10^8\msun$ (\tab{clumps}). 
The clumps labelled 1-8 were all formed \insitu in the disc.  
They contain stars and gas but show no trace of local dark-matter haloes. 
There are also two clumps with dark-matter components 
(labelled 9 and 10), which were formed externally as small
galaxies and merged into the disc 
(\se{exsitu}). 

\Fig{disc_mw3} also shows the gas in an edge-on slice of galaxy A at $z=2.3$.
The $x$ axis is the same as in the face-on view, and the depth of the slice
is from $y=-2.8$ to $y=-1.6\kpc$, chosen to show clumps
1 and 3. While there is clearly a global disc configuration, it is highly
perturbed, twisted and warped, with indications of incoming streams.
The clumps seem to be oblate, 
with the minor axis along the normal to the local disc plane.

\Fig{disc_sfg1} shows the gas in the central disc of galaxy E at redshift 
$z=2.4$, face on and edge on.
This galaxy is more massive, with a baryonic disc mass  
$5.8\times 10^{10}\msun$, a bulge of $8.2\times 10^{10}\msun$, and
gas fraction $0.14$. The rotation velocity is $400\kms$.
The rotating disc extends to $\sim 9\kpc$,
and on top of being locally inhomogeneous and warped, it
shows a global asymmetry, indicating intense fresh gas supply.
The seven marked clumps have masses in the range $(0.4-1)\times 10^9\msun$ 
(\tab{clumps}).
The edge-on slice is $2\kpc$ thick, centered on clump 6 and 
chosen to contain both the global angular momentum vector of the disc 
and the angular momentum vector of clump 6. 
The y axis in this projection coincides with the global 
angular momentum vector of the disc and the x axis extends from 
the top right to the center left in the face on projection. 
Only clump 6 is visible in the image, but it also shows a cut through 
the perturbation near the top left of the face on image, 
(between clump 7 and clumps 4 and 5 in that image).

\begin{figure*} 
\includegraphics[width =1.0 \textwidth]{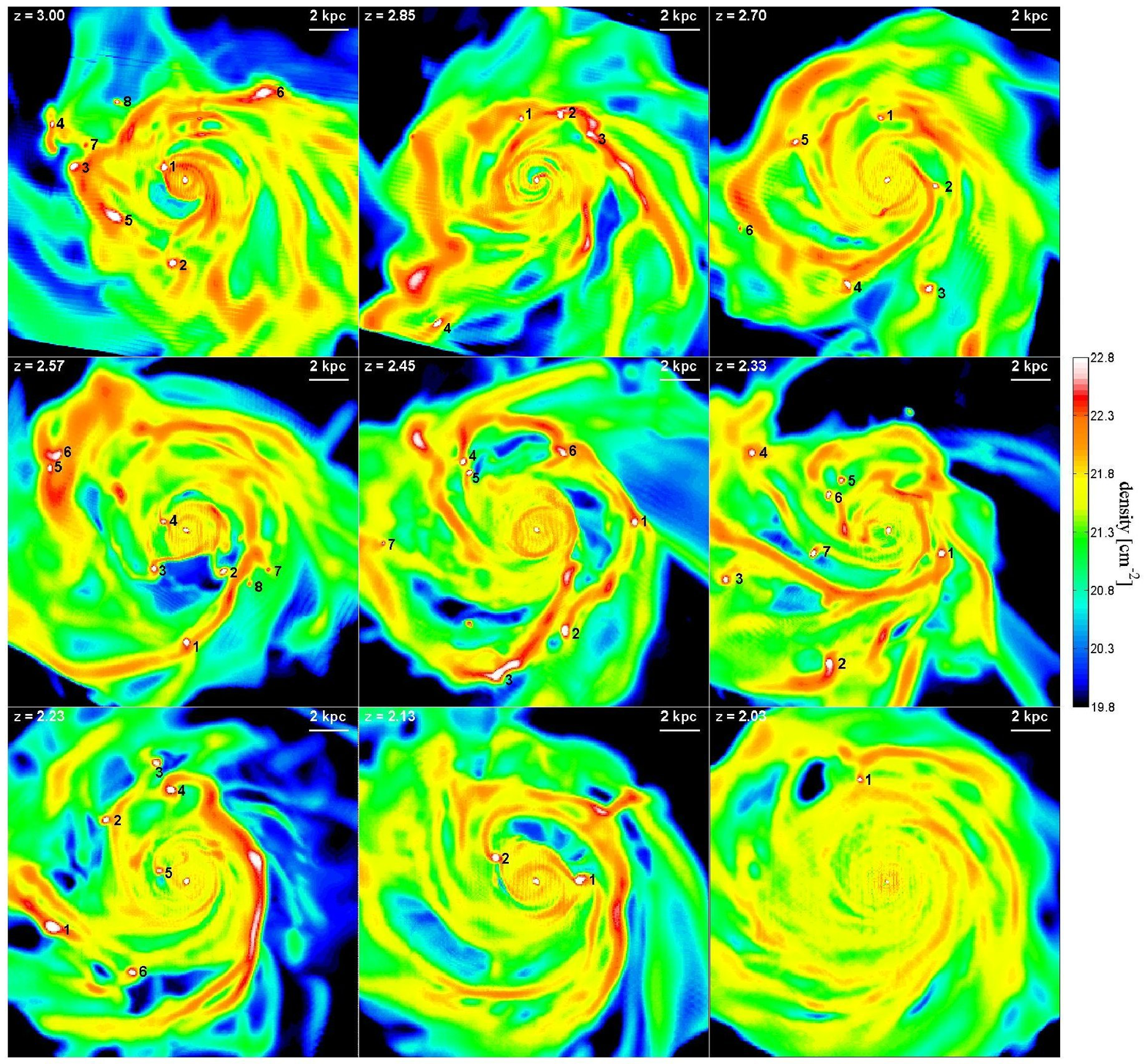}
\caption{
Evolution of gas surface density in the disc of galaxy E from $z=3$
to $z=2$. 
The orientation is face on and the depth is $5\kpc$.
The box encompasses the main body of the disc, face on in a slice of
$18\kpc$ on the side and $5\kpc$ depth.
The snapshots are equaly spaced in expansion factor,
$\Delta a= 0.1$, corresponding to $\Delta z \simeq 0.123$.
All identified clumps are marked in each panel with a (random) serial number
corresponding to \tab{clumps}.
The disc is large and highly perturbed with many clumps until $z=1.4$,
except near $z=2$ whne it is temporarily featureless.}
\label{fig:sfg1}
\end{figure*}

\Fig{sfg1} shows the evolution of the face-on gas disc in galaxy E
in parts of its continuous clumpy phase, from $z=3$ to $z=2$.  
At $z=2$, the disc seems to tentatively stabilize in a less perturbed 
configuration,
but it resumes its clumpy phase soon thereafter, to fade away only 
at $z \sim 1.4$, after which the gas disc shrinks and becomes less perturbed.
Right before $z=3$, the galaxy suffered a 1:4 merger, which makes
the bulge-to-disc mass ratio as high as $\sim 2$ at $z=3$. 
Between $z=3$ and $z=2$, the disc is growing by smoother gas inflow. 
This makes the disc double its mass (and increase its size by 50\%),
while the bulge mass grows only by 30\%, reducing the bulge-to-disc mass ratio
to $\sim 1$ at $z=2$.
During this period, the disc-to-total mass ratio within the disc radius 
remains near 0.2. 
This configuration agrees with the cosmological steady state of a clumpy
disc, as predicted by DSC09.
The onset and termination of the violent instability phase is determined
by several factors including the evolution of the cosmological accretion rate
and the accretion and merger history of the specific galaxy,
the star-formation history and evolution of stellar fraction in the disc,
and the growth of the stabilizing bulge
\citep[DSC09,][]{Agertz09b,Martig09,Cacciato11}.

\subsection{Zoom-in on a few typical clumps} 
\label{sec:zoom} 

\begin{figure*}  
\subfigure{\includegraphics[width=0.49\textwidth]{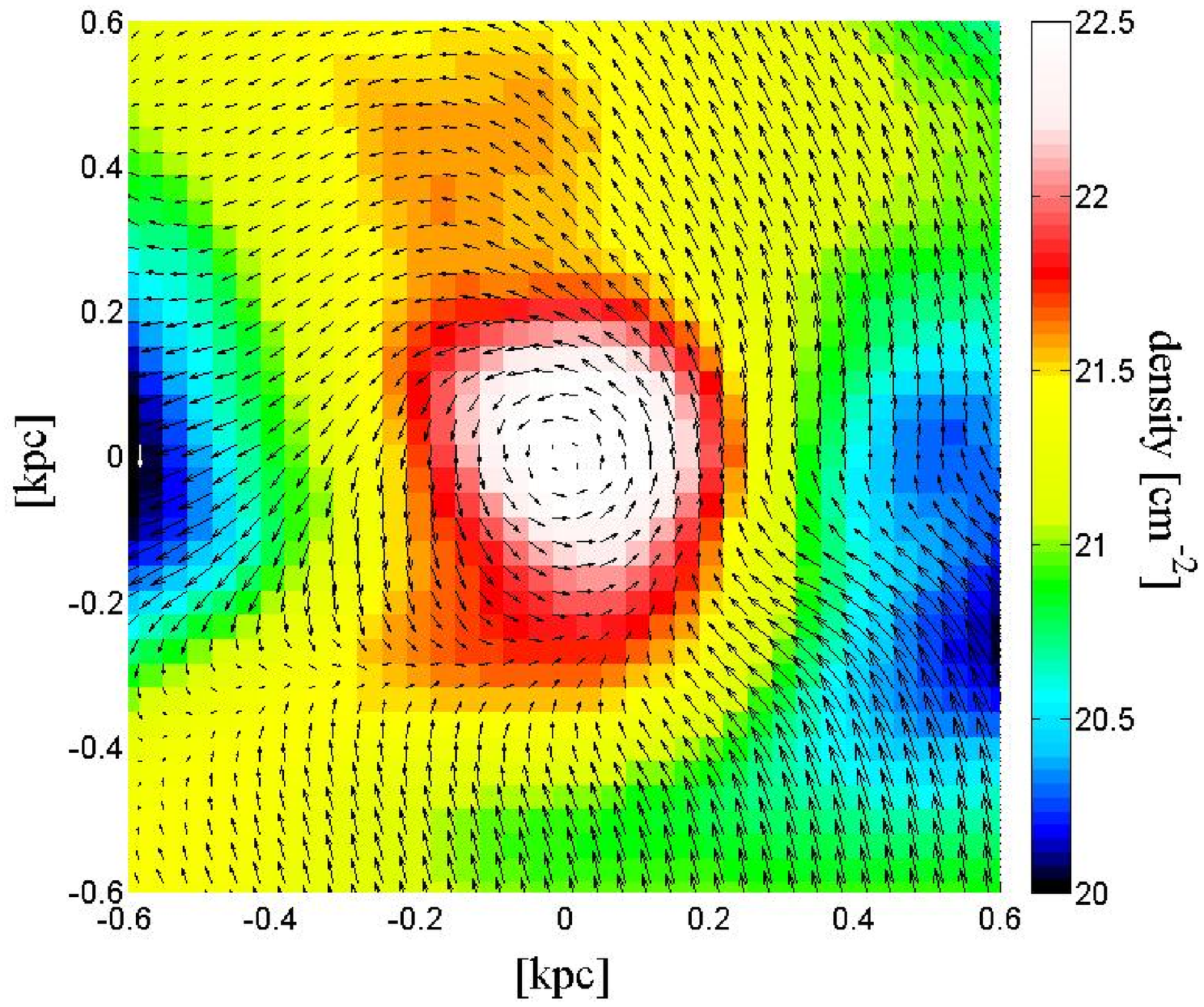}} 
\subfigure{\includegraphics[width=0.49\textwidth]{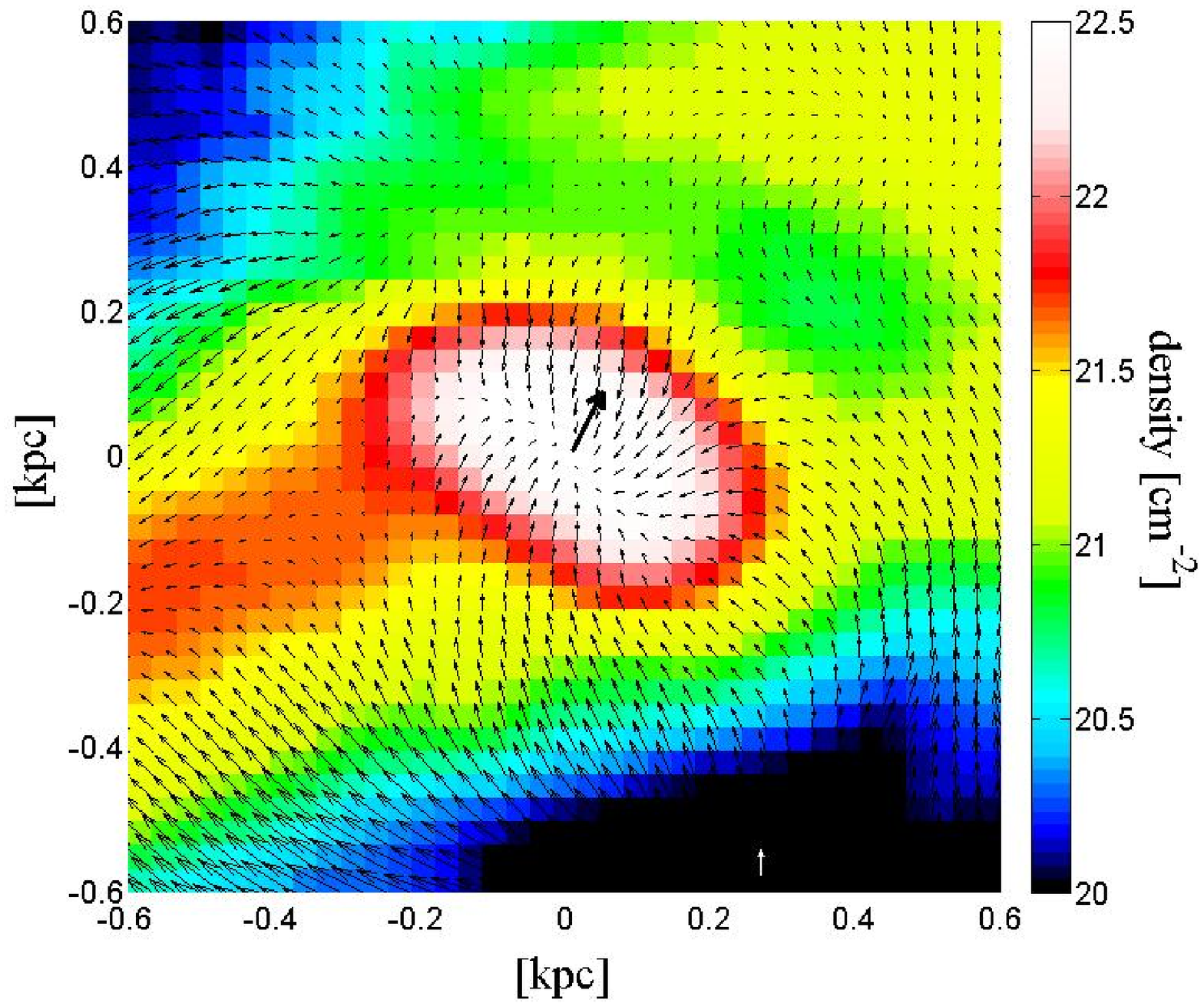}} 
\caption{
Zoom-in on the gas in clump 1 of galaxy A at $z=2.3$, face on and edge on 
in the clump frame. 
The depth of the slice is $0.6\kpc$.  The velocity field is as in
\fig{streams}, with the white arrow denoting $70\kms$.
The clump, of radius $\sim 450\pc$,
shows systematic rotation of $\sim 63\kms$ 
and is highly rotation supported with $\Rot=0.86$. 
The $y$ axis in the edge-on orientation is alligned with the global disc
rotation axis, while the thick black arrow marks the clump spin axis.
A small tilt of $\sim 20^\circ$ ($\cos({\rm tilt})=0.94$) between the clump 
spin and the global disc angular momentum is evident.
The apparent inflow along the minor axis involves only $\sim 10\%$ of the clump
mass and is largely an artifact of the marginal resolution along this axis.}
\label{fig:clump1_mw3}
\end{figure*}

\begin{figure*}  
\subfigure{\includegraphics[width=0.49\textwidth]{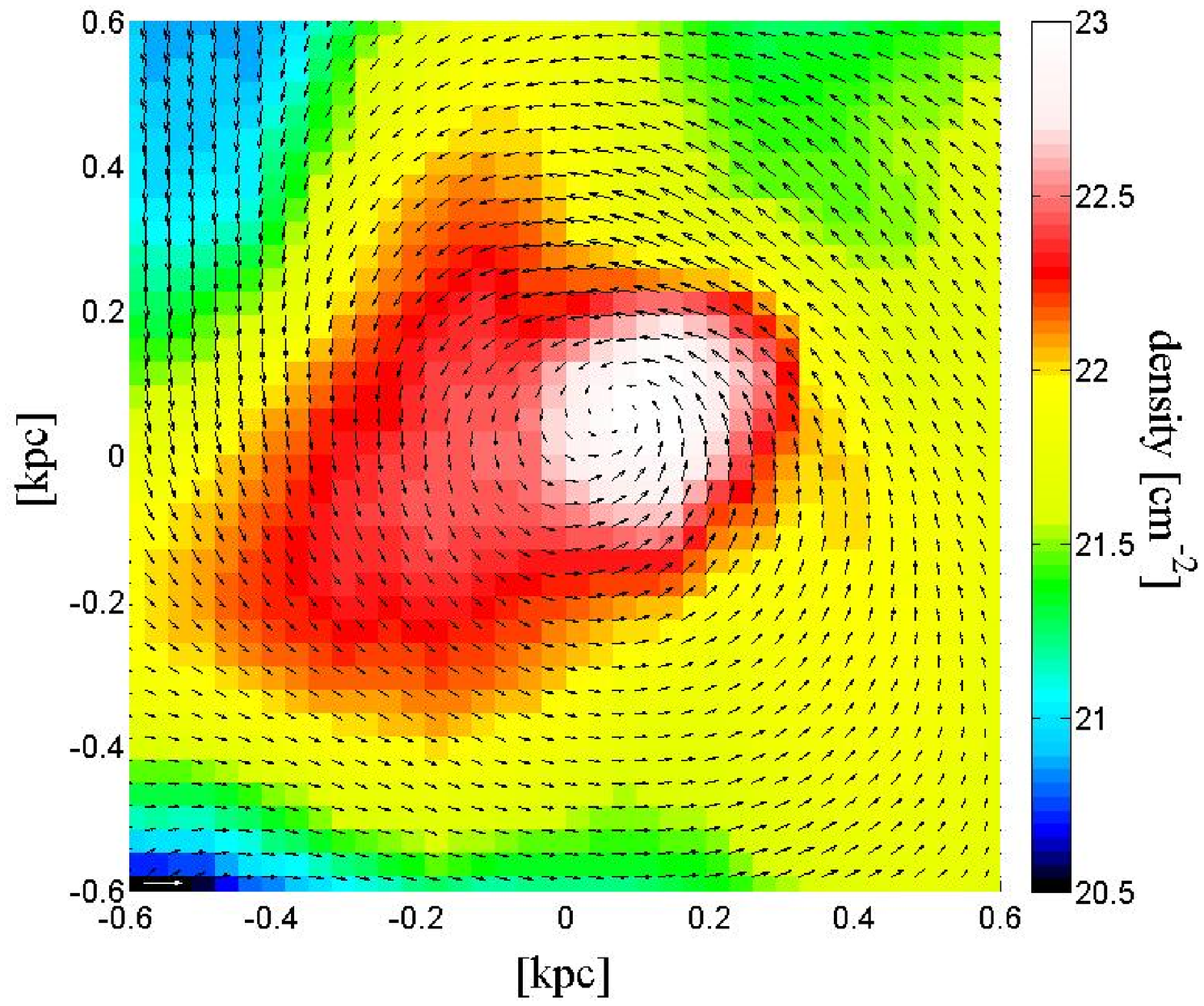}} 
\subfigure{\includegraphics[width=0.49\textwidth]{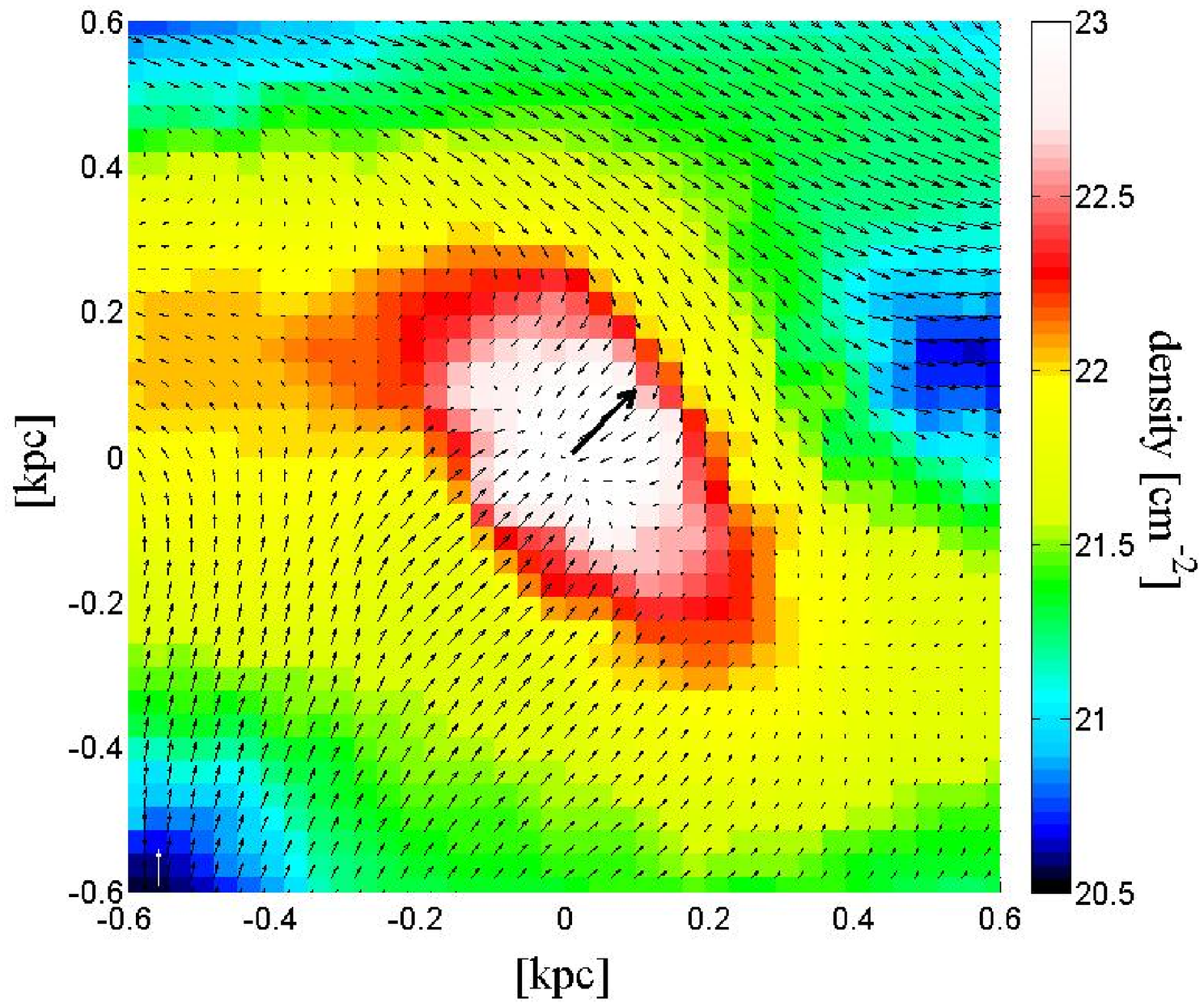}} 
\caption{
Zoom-in on clump 6 of galaxy E at $z=2.4$ (more massive than
the clump shown in \fig{clump1_mw3}), 
face on and edge on in the clump frame.
The depth of the slice is $0.6\kpc$.  The velocity field is as in
\fig{streams}, with the white arrow denoting $200\kms$.
This clump is rotating with $\Vc=108\kms$ and is highly rotation supported
with $\Rot \simeq 1$. 
The $y$ axis in the edge-on orientation is alligned with the global disc
rotation axis, while the thick black arrow marks the clump spin axis.
The clump spin is tilted compared to the global disc angular 
momentum, by $\sim 47^\circ$ ($\cos({\rm tilt})=0.68$),
but it appears to be alligned with the local disc plane.
}
\label{fig:clump7_sfg1}
\end{figure*}

\begin{figure*} 
\includegraphics[width =0.9 \textwidth]{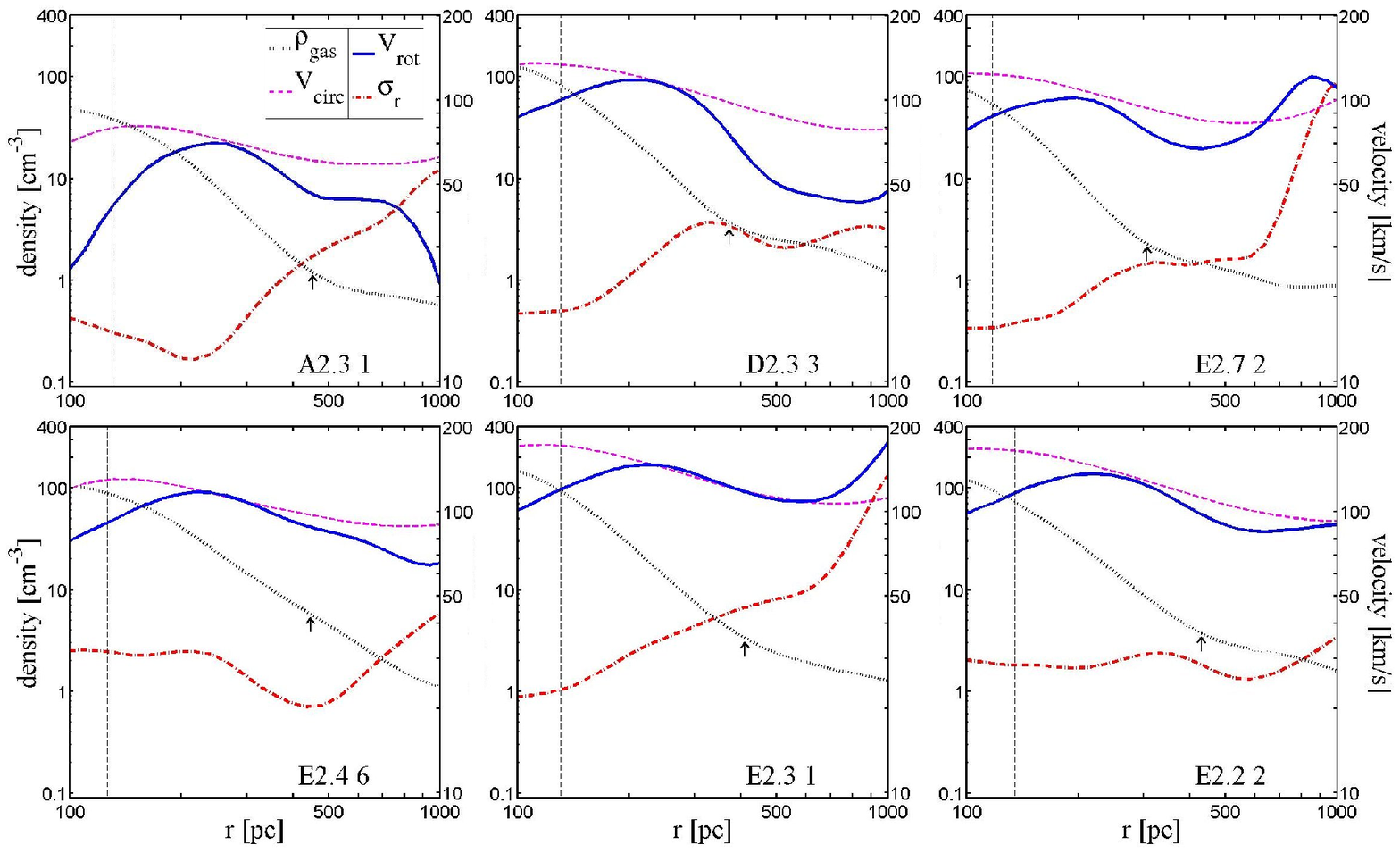}
\caption{
Profiles of the gas in the equatorial plane for a sample of
six high-$z$ clumps from the cosmological simulations.
Each panel shows the profiles of density (left axis), circular velocity,
rotation velocity and radial velocity dispersion (right axis).
The clump masses are $(3.9, 8.3, 6.2, 9.6, 13, 12)\times 10^8\msun$.
within the clump radii marked by an arrow.
The force resolution is marked by a vertical line.
The rotation curves rise to a maximum and then decline toward the
clump edge, beyond which they are dominated by the disc rotation.
All the clumps shown are rotation supported with $\Rot > 0.8$.
}
\label{fig:prof_cos}
\end{figure*}

We describe in \se{clumps} how we identify the clumps in the simulated discs
and measure their properties. This includes identifying the clump center
and its major plane, measuring the density and velocity profiles within the
clump, obtaining a clump radius and mass in the different components,
as well as the associated circular velocity,
evaluating the effective rotation velocity, velocity dispersion and
rotation parameter, and estimating the clump density contrast compared to the
disc and the associated contraction factor.

We start with a zoom-in on the dynamical properties of a few typical clumps 
formed \insitu in our simulated discs. 
\Fig{clump1_mw3} shows face-on and edge-on views of the gas density and 
velocity field in clump 1 of galaxy A at $z \simeq 2.3$.
This clump is located roughly half way between the galaxy centre and the disc 
edge.  It is the most massive clump in that snapshot, with baryon mass 
$\Mc =3.9\times 10^8\msun$ inside the clump radius $\Rc = 455\pc$. 
Its potential well is characterized by $\Vcc = 68\kms$ at $r=310\pc$, 
the center of the shell in which $\Vc$ is computed.  
The face-on and edge-on views are selected by the total angular momentum
vector of the cold gas ($T \leq 10^4$K) in a spherical shell between 
$0.5\Rc$ and $\Rc$.
The surface density and velocity field shown are generated from a cube of 
side $1.2\kpc$ centred on the clump centre of mass.
The clump is centrally condensed, embedded in an elongated large-scale 
perturbation (see also \fig{disc_mw3}) with a relative gas overdensity of 
$\sim 30$ compared to the background disc.
It has an oblate morphology with an apparent axial ratio of about 2:3.
It is important to realize that this clump is only marginally resolved, 
with $\sim 14$ cells across its major axis and $\sim 9$ cells across its 
minor axis.
It therefore shows no substructure (see \se{subs}).
The resolution effects are more severe along the minor axis, where
they may affect the flattening of the clump. 
We therefore limit our analysis of the density profile to the clump 
equatorial plane.

The velocity field shown in \fig{clump1_mw3} is in the rest frame of the clump,
and is computed as the mass-weighted average along the line of sight
across $0.6\kpc$.
The face-on view shows systematic rotation from the centre of the clump to 
about $300 \pc$, with a velocity of $\sim 60 \kms$. The clump is thus
highly rotation supported with $\Rot=0.86$, and with a tilt of 
$\cos(\rm tilt)=0.94$
between the clump spin and the global disc angular momentum.
Outside the clump radius we notice bulk motions in different directions,
towards the clump and away from it, reflecting the disc rotation as well as 
the perturbations about it.
The apparent infall pattern along the minor axis in the edge-on view involves 
only $\sim 10\%$ of the clump mass, and it is largely an artifact of the 
limited resolution along this axis. In fact, most of the clump mass has a 
rather isotropic velocity dispersion, 
computed as the standard deviation of each cylindrical component within 
a given shell.

\Fig{clump7_sfg1} shows zoom-in views of clump 6 of galaxy E at $z=2.4$,
similar to \fig{clump1_mw3}, but more massive,
$\Mc=9.6 \times 10^8\msun$ within $\Rc=432\pc$.
This clump is rotating with $\Vc=108\kms$ and it is highly rotation supported
with $\Rot = 0.94$. However, the clump spin is quite tilted 
with respect to the global disc angular momentum, with a tilt of
$\sim 47^\circ$ ($\cos({\rm tilt})=0.68$).
Despite this, the clump does appear to be aligned with the local disc plane.
This example shows that despite the tendency for prograde rotation
and rotation support, the perturbed and warped discs 
can give rise to highly tilted clumps compared to the global disc.

\Fig{prof_cos} shows profiles of gas velocity and three- dimensional gas density
for six representative high-$z$ clumps from the cosmological simulations, 
including the two clumps discussed above. 
The profiles are averages over rings in the clump equatorial plane.
The density profiles crudely resemble the familiar NFW function, 
namely a local
logarithmic slope that steepens with radius from a core of slope flatter than
-1 below the resolution scale, through -2 to -3 in the resolved scales,
before it flattens again near the clump edge toward a uniform background.
The mass-weighted average slope in the clump outside the force resolution 
scale is close to -2, implying that for certain purposes the clump profile 
can be very crudely approximated by the profile commonly associated with
an isothermal sphere.
As described in \se{clumps}, 
we define the clump radius $\Rc$ where the local slope 
becomes flatter than -2 in its approach to a constant at larger radii, 
or where there appears to be a "shoulder" in the density profile 
(as in clump 6 from galaxy E at $z=2.4$).
This definition is very similar to the standard definition of subhalo radius as an upturn or inflexion point, generally used in dark-matter halo-finders \citep[see review in][]{Knebe11}.
The background density outside the clump varies from clump to clump and from
host disc to host disc, typically in the range $n \sim (2-5) \cmc$,
depending on the large-scale perturbation that the clump is embedded in.

The six clumps have maximum rotation velocities of 
$(70, 118, 117, 146, 102, 136)\kms$ at $r \sim 200-300\pc$, 
and they then decline slowly toward the clump edge and sometimes beyond.
The $\sigmar$ profiles tend to be rather flat within the clump,
with values of $(24, 32, 24, 25, 41, 31)\kms$ near the clump edge.
The profiles of the other components of the velocity dispersion 
are similar to the $\sigmar$ profiles, 
implying that the velocity dispersion is rather isotropic for most of the 
clump material.
The rotation support parameters are 
$\Rot = (0.86, 0.90, 0.83, 0.94, 1.01, 0.88)$, and 
$\Vc/\sigmar$ is given by \equ{Voversigma}.
These clumps, like most of the \insitu giant clumps in the high-redshift
discs studied here (\se{stat} below), can indeed be approximated
as isotropic rotators mostly supported by rotation, 
with the pressure that is built up inside the clumps 
providing the rest of the support against gravitational collapse.

\section{Statistical analysis of clump support}
\label{sec:stat}

\subsection{A sample of clumps} 

\tab{clumps} presents a sample of 86 disc clumps selected
from our simulations for a crude statistical analysis of their properties. 
The clumps, above a circular-velocity threshold $\Vcc > 30 \kms$, were 
extracted from the five simulated discs, in the redshift range 
$z \sim 1.9-3.0$, from snapshots where they show gravitationally unstable 
discs with several giant clumps. A separation of $\Delta z \sim 0.2$ between 
snapshots corresponds at $z \sim 2.5$ to $\sim 220 \Myr$, which is on the 
order of the disc orbital time. This is comparable to the lifetime of a 
clump between formation and end of migration into the central bulge 
\citep[DSC09, CDB10,][]{Genzel11},
so each clump is sampled once or maybe twice during its lifetime. 
In galaxy E, we sample the clumps in snapshots equaly spaced in 
expansion factor $\Delta a = 0.1$, corresponding to  $\Delta z \simeq 0.123$, 
so each clump is sampled more than once during its lifetime. 
However, the $\sim 130\Myr$ between snapshots is several times 
the internal dynamical time in the clump, so it is allowed to evolve 
considerably between snapshots.   
While this sample is not strictly a proper statistical sample,
we consider it to be a crude approximation for a fair sample spanning
the clump properties in clumpy discs of baryonic mass 
$10^{10}-2\times 10^{11}\msun$, or halo mass $10^{11} - 2\times 10^{12}\msun$, 
in the redshift range 2-3. 

In each snapshot the clumps are marked by a random serial number.
\tab{clumps} lists for each clump the clump radius $\Rc$, 
and the baryonic mass $\Mc$ within that radius. 
The gas circular velocity $\Vcc$, rotation velocity $\Vc$, 
radial velocity dispersion $\sigmar$, and 
the rotation parameter $\Rot = \Vc^2/\Vcc^2$ (\equ{R})
are obtained at the outer half of the clump, as explained in \se{clumps}.
The table then quotes the alignment parameter, defined as the cosine of the 
tilt angle between the clump spin and the global disc angular momentum.
Following is the baryonic surface density in the clump $\Sigc$, 
and the contraction factor $c = (\Sigc/\Sigd)^{1/2}$ (\equ{c}).  
The relevant disc surface density $\Sigd$ for quantifying a contraction
factor is quite uncertain, so the values of $c$ should be taken with a
grain of salt.
The table also lists the mean stellar age in the clump,
the dark-matter fraction and gas fraction within $\Rc$,
the position of the clump center in the disc in polar coordinates $r$ and $z$,
and its center-of-mass velocity components $V_r$, $V_z$, and $V_\phi/\Vd$,
where $V_d$ is the average disc rotation velocity at $r$.
Comments indicate whether the clump is closely interacting with other clumps,
whether it has a surface density contrast lower than 3,
and whether it is an \exsitu clump (\se{exsitu}). 
We also list the progenitor clumps in the preceding snapshot in cases where 
such progenitors are identified. 

\begin{figure}  
\includegraphics[trim = 0.4cm 0.5cm 1.6cm 0.4cm, clip, 
     width =0.48\textwidth]{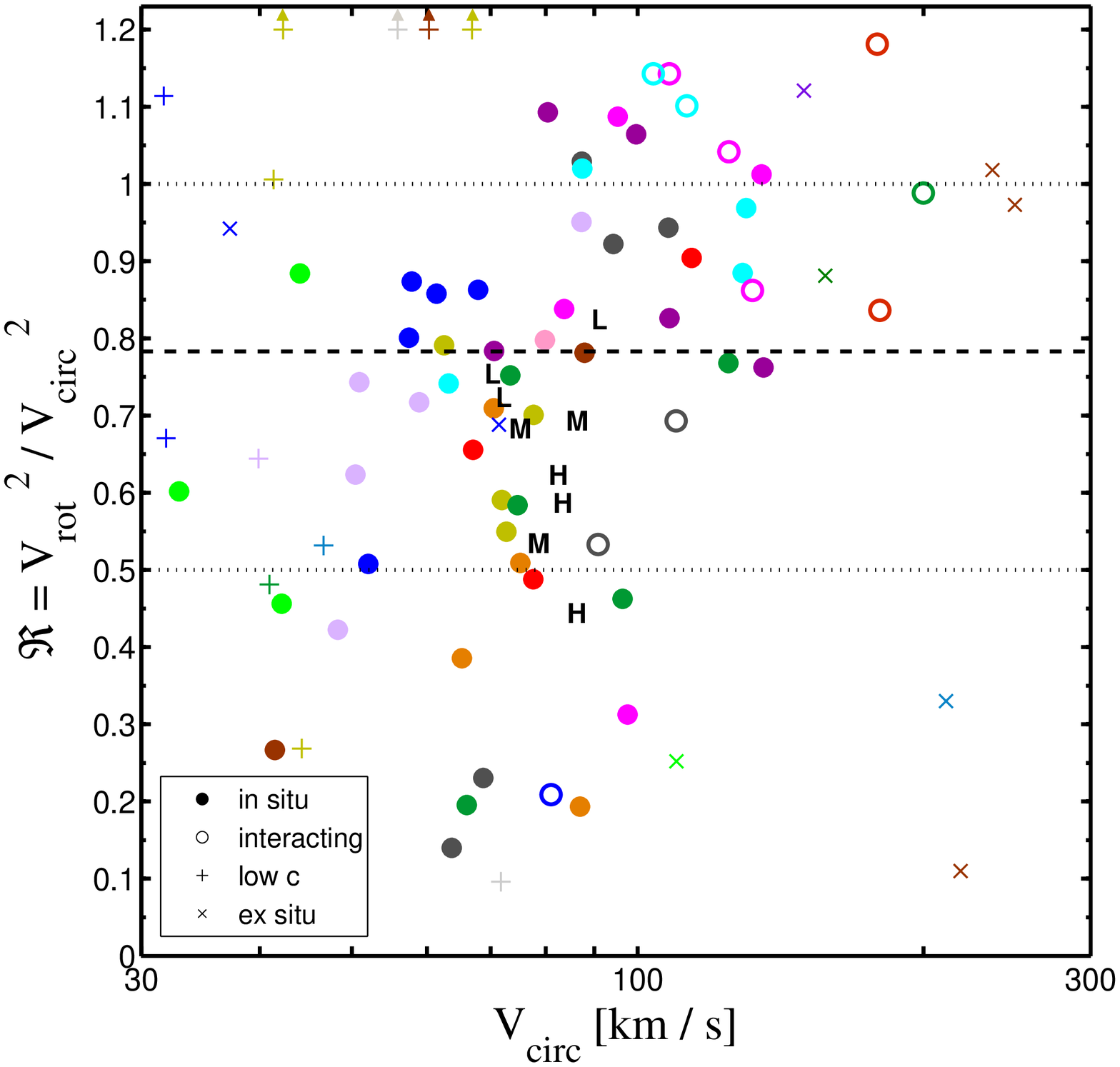} 
\caption{Rotation parameter, $\Rot = \Vc^2/\Vcc^2$, as a function of 
clump circular velocity, $\Vcc$, for the sample of disc clumps 
(\tab{clumps}). The clumps are extracted from the cosmological 
simulations of clumpy disc galaxies of baryonic mass 
$10^{10}-2\times 10^{11}\msun$ in the redshift range $1.9-3.0$. 
The colors of the symbols mark the relevant galaxy and snapshot as labeled in 
\fig{jeans}. 
Normal \insitu clumps are marked by filled circles. 
In-situ clumps that are closely interacting or are remnants of a recent merger 
are marked by open symbols.  
Clumps with a surface-density contrast lower than 3 are marked by a ``$+$". 
Ex-situ clumps that contain dark-matter are marked by a ``$\times$", 
but are not included in the current analysis. 
The median of $\Rot=0.78$ is denoted by a dashed horizontal line. 
We see that the majority of the clumps are supported mainly by rotation. 
Shown in comparison (marked ``L", ``M" and ``H") are the clumps extracted  
from isolated galaxy simulations (\se{subs}) at different resolutions 
of 70 pc (LR), 12 pc (MR), and 2 pc (HR). At the highest resolution, $\Rot$ 
is reduced by $\sim 30\%$, but the rotation still provides a significant 
fraction of the clump support. 
}
\label{fig:R} 
\end{figure}

\begin{figure}  
\includegraphics[trim = 0.4cm 0.5cm 1.6cm 0.4cm, clip, width =0.48\textwidth] 
                 {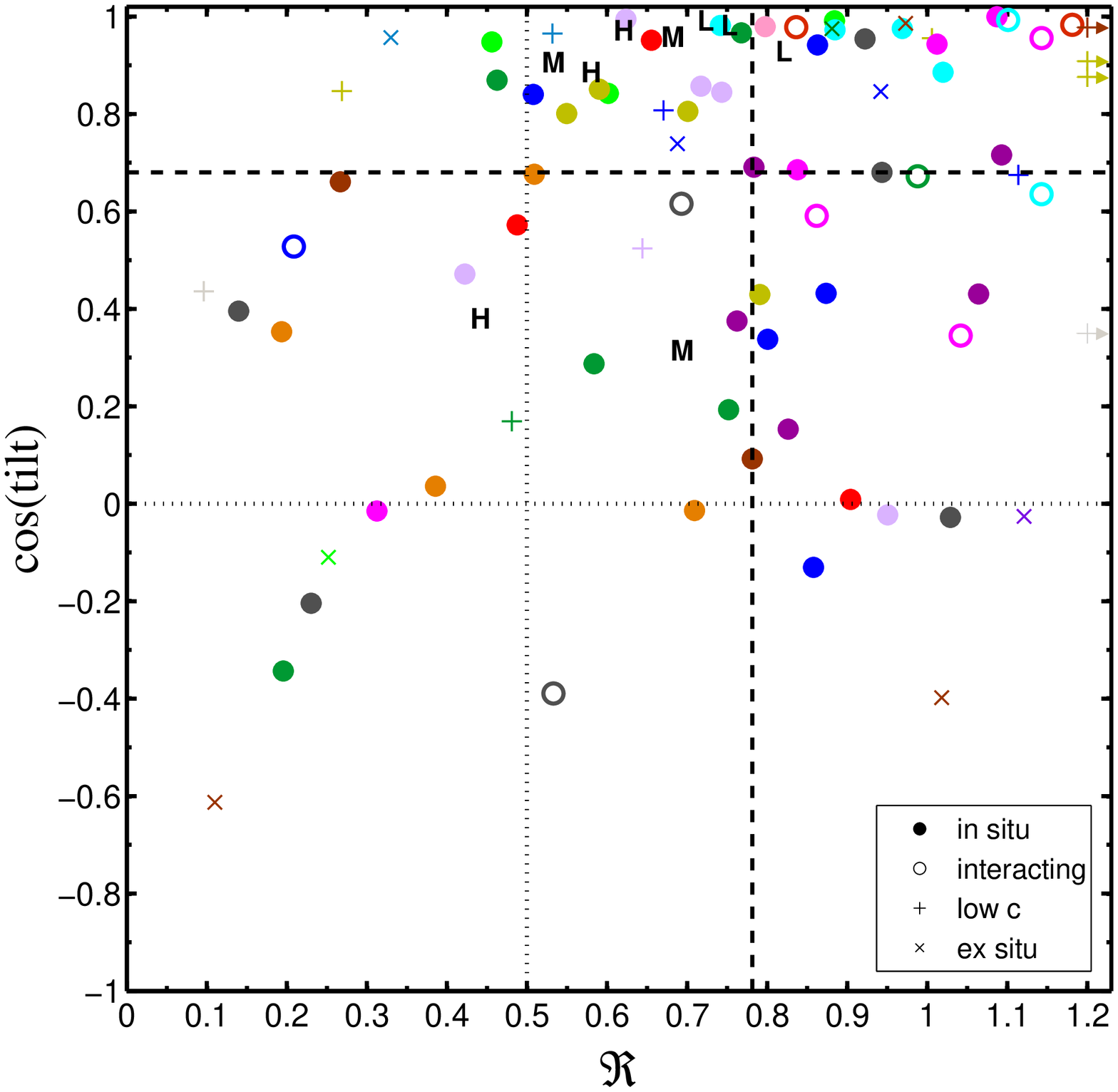} 
\caption{ 
The alignment of clump and disc angular momentum, $\cos({\rm tilt})$, 
versus rotation parameter $\Rot$. 
The symbols are as in \fig{R}. 
The medians of $\cos({\rm tilt})=0.68$ and 
$\Rot=0.78$ are marked by dashed lines. 
The average in the case of a random tilt is $\cos({\rm tilt})=0$. 
There is a general tendency for prograde clump rotation, but large tilts 
occur rather often, and 10\% of the \insitu clumps actually rotate 
in a retrograde sense, with $\cos({\rm tilt})<0$.  
There is no significant correlation between the tilt 
and $\Rot$. There is a marginal tendency for the more massive clumps 
to be more aligned with their host discs (not shown). 
The clumps of the isolated galaxy simulations tend to be more aligned 
with their discs, but there are also tilted cases.} 
\label{fig:tilt} 
\end{figure}

Out of the 86 clumps in the sample, 9 (10\%) are \exsitu clumps 
coinciding with local peaks in the dark matter density  
and containing an older stellar population, which joined
the disc as minor-merging galaxies. 
We include them in \fig{R} to \fig{c},
marked by a ``$\times$" symbol, but we exclude them from the statistical
analysis of the current section, to be addressed separately
in \se{exsitu}.
The other 77 clumps (90\%) were formed \insitu in the disc
and have dark matter fractions comparable to the local background. 
Out of the 77 \insitu clumps, 11 clumps (14\%)
are tagged as closely interacting with other clumps or as remnants of 
recent binary mergers of clumps.  We include these in our analysis, 
but distinguish them in \fig{R} to \fig{c} by open circles. 
Out of the 77 \insitu clumps, 12 (16\%) seem to have a surface-density 
contrast below 3, and are marked low-contrast clumps. We include these in 
our analysis, but distinguish them in \fig{R} to \fig{c} by ``$+$" symbols.
Some of their properties are less certain.
All other \insitu clumps are marked by filled circles.

\subsection{Rotation support} 

\Fig{R} shows the distribution of the rotation parameter $\Rot$ versus $\Vcc$
for the cosmological disc clumps.
The median value for the \insitu clumps is found to be $\Rot \simeq 0.78$, 
saying that the support in most of the clumps is heavily dominated by rotation.
In 82\% of the clumps $\Rot > 0.5$, i.e. the rotation provides most of the
support.
The massive \insitu clumps, 17 clumps with $\Vcc \geq 100\kms$, have 
$\Rot > 0.75$ (except of one clump of $\Rot = 0.7$ associated with a merger), 
namely less than 6\% have $\Rot < 0.7$.
The less massive clumps show a tail of intermediate and low
rotation support down to $\Rot \sim 0.1$.
The clumps that are associated with clump interactions or clump mergers 
tend to have the highest $\Vcc$ and be
highly supported by rotation.

\Fig{R} also shows the clumps extracted from simulations of isolated discs with
different resolutions, described in \se{subs} below.
The three clumps simulated with a resolution similar to the resolution in the
cosmological runs (LR) show rotation parameter values similar to the
cosmological simulations, $\Rot=0.72-0.82$.
The three clumps simulated with a much higher resolution of 2 pc (HR),
exhibit lower rotation parameters, $\Rot=0.44-0.62$.
While the higher resolution results in a reduction of $\sim 30\%$ in $\Rot$,
the rotation still tends to be a major player in the support of the
highly resolved clumps.

One should note that the measured rotation parameter depends on the actual 
radius within the clump where it is obtained. We have tried modifications
of the algorithm described in \se{clumps}, including measuring $\Rot$ at the 
maximum of the rotation curve or at the clump edge, and found no qualitative
change in the result. The typical variation in $\Rot$ is limited to
$\sim 15\%$, reaching $\sim 30\%$ in exceptional cases where the measurement 
is contaminated by nearby perturbations in the disc.
The choice of obtaining $\Rot$ as the average in the
outer half of the clump turned out to be a sensible compromise.

In 13 of the \insitu clumps, the measured rotation parameter is apparently
slightly above unity, $1.0<\Rot<1.2$, namely an excess of 10\% or less for
$\Vc$ over $\Vcc$.  This could reflect small deviations from pure rotation 
or from spherical symmetry at the radius where these quantities are measured, 
and they are within the expected scatter of the mean. 
We therefore do not interpret these deviations as indicating departures from
Jeans equilibrium, and adopt $\Rot=1$ for these clumps.
Four of the low contrast, low mass, clumps have very high values of 
$1.5<\Rot<2.5$, marked in \Fig{R} as ``$+$" symbols with upward pointing 
arrows. These clumps appear to be unbound, at least at the large radii adopted
for these clumps. We also note that these small clumps are rather poorly 
resolved and the statistical ``measurement" error for them is large.


\subsection{Tilt relative to the disc} 
 
\begin{figure}  
\includegraphics[trim = 0.4cm 0.5cm 1.6cm 0.4cm, clip, width =0.48\textwidth] 
                {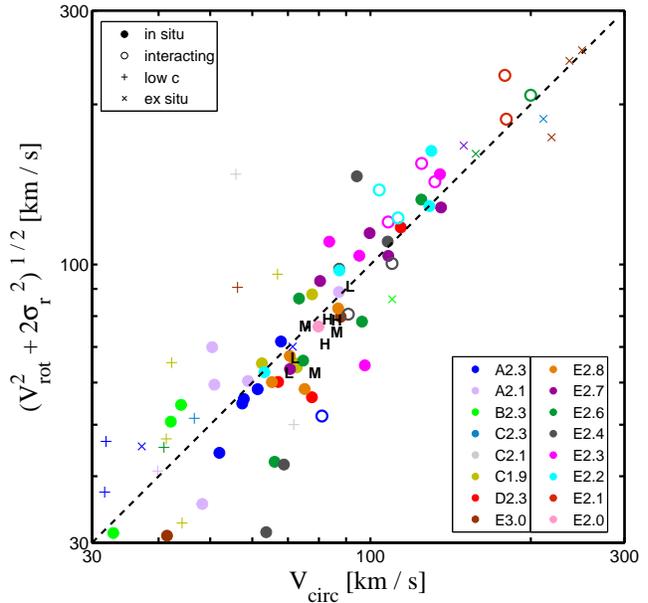} 
\caption{ 
Validity of Jeans equilibrium. 
The square root of the two sides of \equ{jeans2}, 
$(\Vc^2+2\sigmar^2)^{1/2}$ versus $\Vcc$, representing support versus gravity. 
The symbols are as in \fig{R}. 
The clumps roughly obey the Jeans equation for an isotropic, isothermal 
rotator (dashed line).  
The outliers downwards tend to have a low mass and low $\Rot$ (not shown). 
These clumps come closer to the line  
when the factor 2 multiplying $\sigma^2$ in the 
support term is replaced by a larger factor. 
The low-contrast clumps tend to lie above the line due to their high velocity 
dispersions. 
} 
\label{fig:jeans} 
\end{figure} 
 
In the analytic model outlined in \se{theory},  
the disc is assumed to be rotating 
uniformly and the clumps are predicted to end up in prograde rotation 
with their spins and minor axes aligned with the disc axis. 
However, the high-redshift discs in our cosmological simulations are highly 
perturbed, with the local ``plane" sometimes deviating from the global disc 
plane (as defined either by inertia or by angular momentum), 
as can be seen in the edge-on slices in \fig{disc_mw3} and \fig{disc_sfg1}. 
One may therefore expect some of the clump minor axes to be tilted 
relative to the global disc minor axis, possibly following their local disc 
neighborhood (\Fig{clump7_sfg1}). 
 
\Fig{tilt} shows the distribution of the alignment cosine $\cos({\rm tilt})$ 
versus $\Rot$ for our sample of clumps.  
The median is at $\cos({\rm tilt})=0.68$, corresponding to a tilt of  
$\sim 47^\circ$. 
There is a tail of large tilts extending to $\cos({\rm tilt}) \sim 0$, 
and 10\% of the clumps actually rotate in a retrograde sense,  
$\cos({\rm tilt}) < 0$. 
We conclude that the expected tendency for prograde rotation is clearly there,  
but the occurrence of significant tilts is not infrequent. 
There is no significant correlation between the tilt and $\Rot$. 
There is a marginal tendency for the more massive clumps, especially the 
ones associated with merging clumps, to be more aligned with the global disc 
plane (not shown). For example, there are no retrograde clumps among  
the \insitu clumps with $\Vcc > 100\kms$. 
It is yet to be investigated whether the tilts were generated by the 
perturbed velocity field at the time of clump formation or by torques from 
the perturbed environment during the later stages of clump evolution. 
 
The clumps of the isolated galaxy simulations tend to be more aligned 
with their host discs than the cosmological clumps, as expected,
with a median at $\cos({\rm tilt}) = 0.93$, and no retrograde clumps. 
However, two of these clumps have significant tilts, 
$\cos({\rm tilt}) \sim 0.3-0.4$. 
There is no obvious dependence on resolution.

\subsection{Jeans equilibrium} 
 
We wish to find out to what extent the simulated clumps obey the 
Jeans equation for an isotropic rotator, \equ{jeans2}.  
We find the internal velocity dispersion to be close to isotropic, 
and the density profile to be not far from that of
an isothermal sphere (\se{zoom}),  
so we expect \equ{jeans2} to be approximately valid for a clump in equilibrium. 
For this purpose, \fig{jeans} compares the square-roots of the two sides of  
\equ{jeans2}, $(\Vc^2+2\sigmar^2)^{1/2}$ versus $\Vcc$. 
We find that most clumps lie close to the line that marks Jeans equilibrium, 
with a few outliers that correlate with low mass and low $\Rot$ (not shown). 
The fit would improve for the clumps with a low $\Rot$ once the factor 2 in 
\equ{jeans2} is replaced by 3 or even a higher factor, to take into account  
the deviation of the clump from an isothermal sphere profile, 
with a somewhat steeper density profile, as seen in \fig{prof_cos}. 
The low-contrast clumps tend to lie above the line due to their high internal 
velocity dispersions, indicating that some of them may be transients not in  
Jeans equilibrium. 
The clumps of the isolated galaxy simulations roughly obey the Jeans equation, 
independent of resolution. 
We see a tendency of the other less massive clumps to show larger
deviations from Jeans equilibrium, be less rotationally supported, and contain
a younger stellar population (\se{age} below), 
indicating that they may be still collapsing toward equilibrium.

\subsection{Rotation versus contraction factor} 
 
\begin{figure}  
\includegraphics[trim = 0.4cm 0.5cm 1.6cm 0.4cm, clip, width =0.48\textwidth] 
                {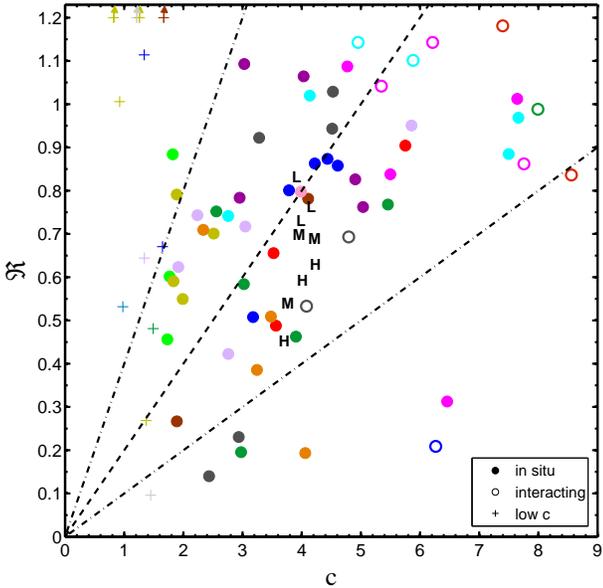} 
\caption{ 
Rotation parameter $\Rot$ versus contraction factor $c$. 
The toy-model prediction assuming conservation of angular momentum, \equ{Vrot}, 
is marked by a dashed line.  
The symbols are as in \fig{R}. 
Except for 10\% outliers, the clumps lie within a factor two from the model 
prediction  (dash-dotted lines).
} 
\label{fig:c} 
\end{figure}

A comparison of the rotation support parameter $\Rot$ and the contraction  
factor $c$ could allow us in principle to
test the validity of the simple model prediction 
spelled out in \equ{Vrot}, $\Rot \sim 0.2\,c$. 
To the extent that the toy model of \se{theory} is a viable approximation, 
this will be a measure of the degree of conservation of internal angular  
momentum during the clump formation and evolution. 
\Fig{c} shows $\Rot$ versus $c$.  
Except for 10\% outliers with low $\Rot$, the bound clumps lie 
within a factor of two from the toy-model prediction, 
and they are spread both above and below the predicted line.
This indicates that, on average, the toy model is a viable crude
approximation and, on average, angular momentum is roughly conserved.
The big scatter, which partly reflects uncertainties in measuring $\Rot$ and 
especially $c$, may imply that some clumps do reach high 
values of contraction factor that may indicate certain angular-momentum 
loss. 
This is especially true for the clumps associated with interactions or mergers, 
and in general for the most massive clumps with $\Vc>100 \kms$. 
There is a correlation between angular-momentum loss and low $\Rot$. 
Most of the low-$c$ clumps are outliers toward a low contraction 
factor and high rotation, either because they gained angular momentum, or  
because they are unbound transients (\fig{jeans}). 
The clumps from the isolated-galaxy simulations are also in general agreement 
with \equ{Vrot}, with a contraction factor $\sim 4$. 
At the highest resolution, they indicate an angular-momentum loss of about  
30\%.  
 
\section{Isolated galaxies with higher resolution} 
\label{sec:subs} 

\begin{table}  
\centering 
\caption{Resolution and physical parameters for the high-resolution idealized 
simulations}\label{table-FB} 
\begin{tabular}{lcccc} 
\hline 
\hline 
Resolution & $\epsilon_\mathrm{AMR}$ (pc) & $m_\mathrm{res}$ ($\msun$) & 
$n_\mathrm{SF}$ (cm$^{-3}$) & $\eta_{\mathrm SN}$\\ 
\hline 
HR & 1.7~pc   &   $2 \times 10^3$  & $5 \times 10^4$  & 50\% \\ 
MR & 10.2~pc &  $2 \times 10^3$  &  $4 \times 10^3$  & 50\% \\ 
LR & 68~pc     &  $2 \times 10^4$  &  $8 \times 10^2$  & 50\% \\ 
\hline 
\end{tabular} 
\end{table} 
 
The simulations start with a pre-formed exponential disc of  
half gas and half stars, and a stellar mass of $\sim 1.5 \times 10^{10}\msun$, 
embedded in a dark-matter halo. 
More details are provided in \se{ramses}. 
These simulations rapidly evolve into turbulent discs with giant clumps  
of gas and young stars.  The clumpy morphology, the clump 
formation rate, their mass, number, and migration to the disc centre, are 
similar to the behavior in the idealized simulations of BEE07 and in the 
cosmological simulations analyzed above (see CDB10).  
 
\begin{figure*} 
\includegraphics[width =1. \textwidth]{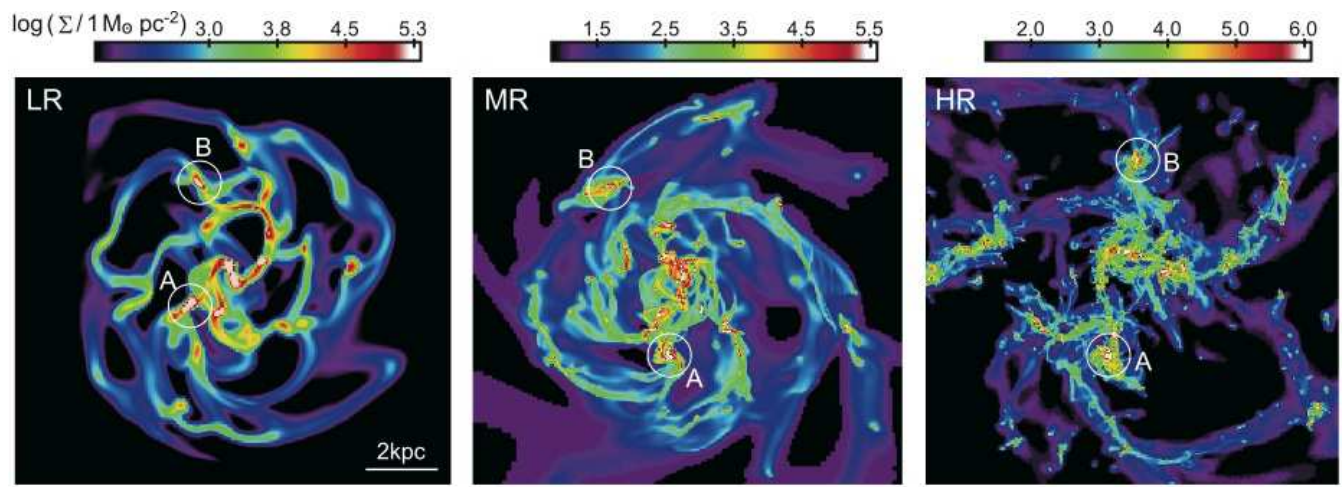} 
\caption{Face-on views of the gas surface density in the simulations 
of an idealized isolated galaxy with different maximum resolutions of  
68 (LR), 10 (MR), and 1.7 pc (HR). 
This snapshot is when the outer disc has completed two rotations. 
There is no one-to-one correspondence between the giant clumps at various  
resolutions, because the initial small-scale fluctuations and noise vary 
with resolution.  
Clumps A and B were chosen on the basis of their similar masses  
and distances from the disc centre. Clump C was picked from a later output.} 
\label{fig:FB1} 
\end{figure*} 
 
\begin{figure*} 
\includegraphics[width =0.9 \textwidth]{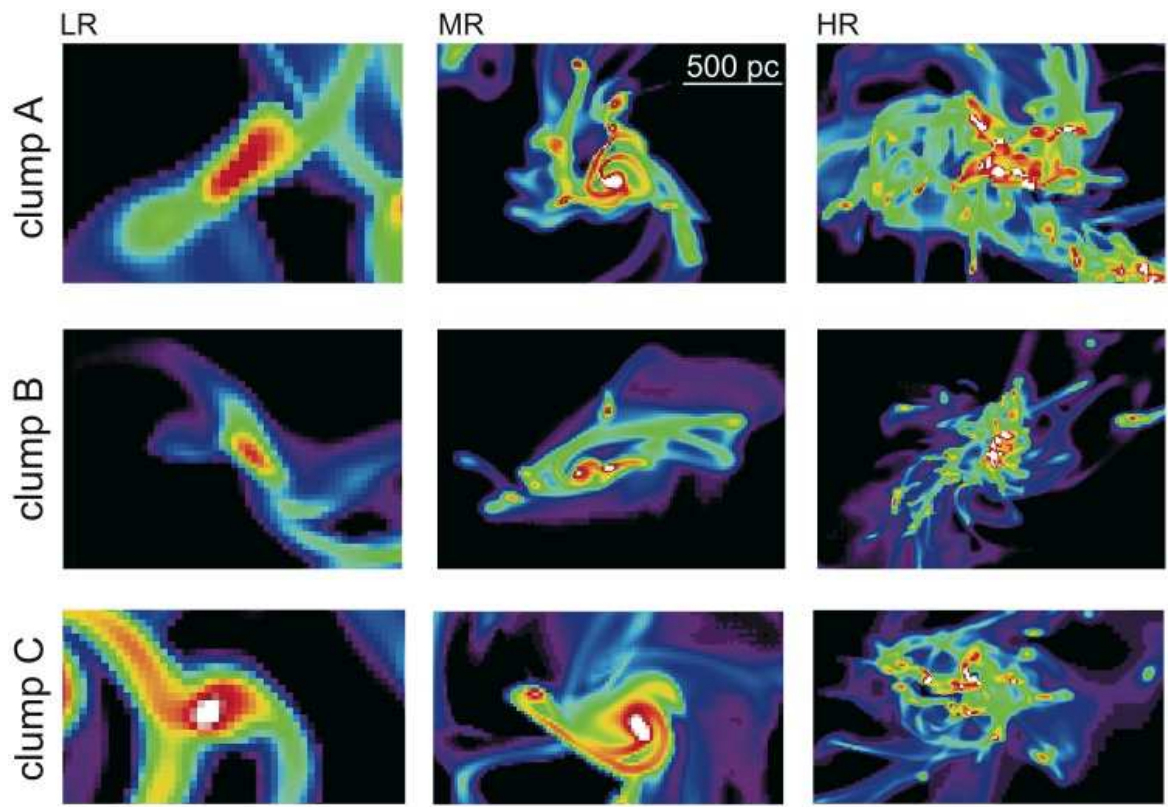} 
\caption{Zoom-in views of clumps A, B and C from the isolated-galaxy 
simulations at the different resolution levels, LR, MR, and HR.  
The orientation is face on with respect to the global disc.  
The morphological appearance of the clumps varies with resolution.  
The LR clumps resemble inflection knots along broad density waves,  
with a smooth body an a weak spiral pattern in their outer regions.  
The MR clumps are mini spiral discs, suggesting local rotation. 
The HR clumps have substructure with dense subclumps. Internal shocks  
suggest supersonic turbulent motions, but the global spin of the giant clumps  
in HR is only $\sim 30\%$ lower than in LR. 
} 
\label{fig:FB2} 
\end{figure*} 
 
\begin{figure*} 
\includegraphics[width =0.9\textwidth]{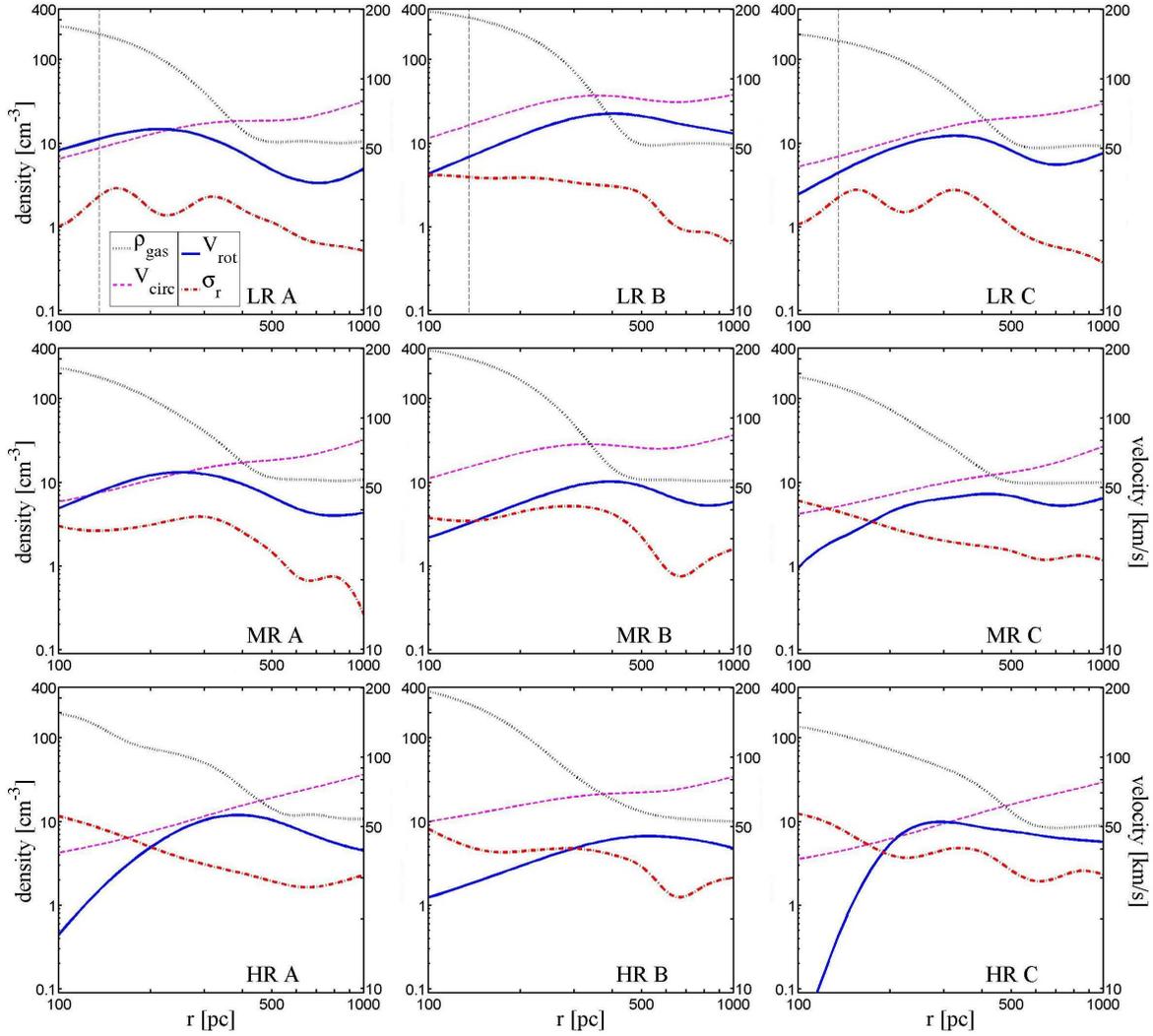} 
\caption{
Profiles of the gas in the equatorial plane for three clumps form each of the
high-$z$ isolated-galaxy simulations LR, MR, and HR (from top to bottom),
with maximum resolution of 68, 10, and $1.7\pc$.
Each panel shows the profiles of density (left axis), rotation velocity and
radial velocity dispersion (right axis).
The profiles are smoothed as described in Appendix \se{clumps}.
The LR clump profiles are similar to those of the cosmological clumps 
when scaled by mass (\fig{prof_cos}).
While $V/\sigma$ is decreasing with improving resolution, the 
clumps remain rotation supported in their outer parts even in the HR case. 
}
\label{fig:FB3} 
\end{figure*} 

\begin{figure*} 
\includegraphics[width =0.9\textwidth]{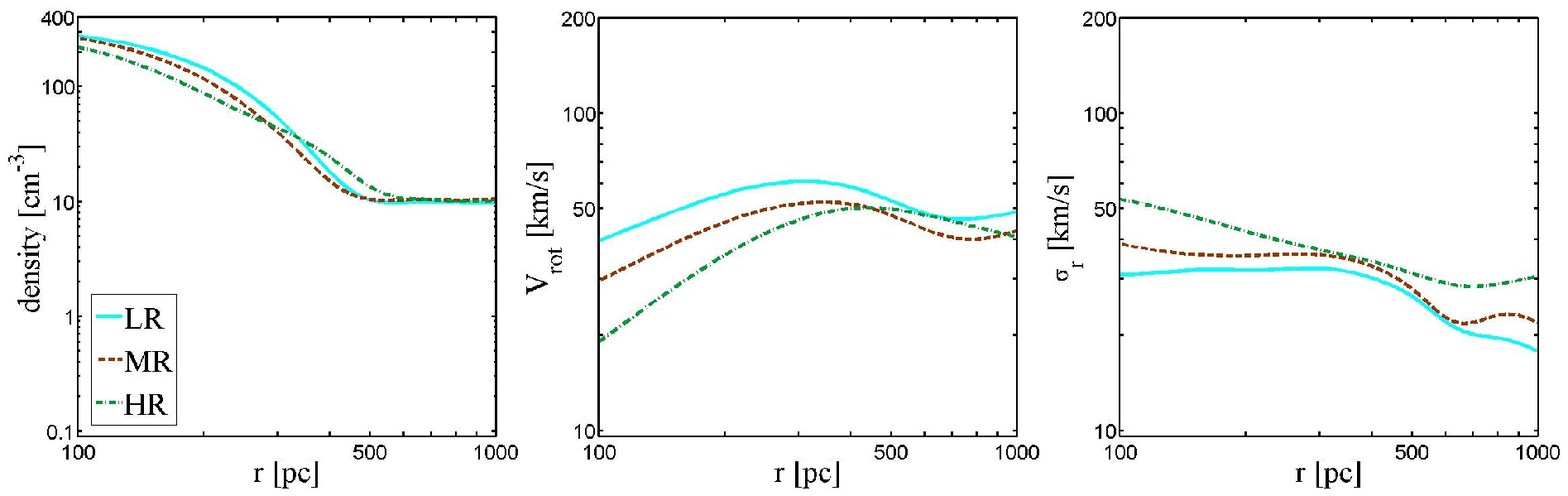} 
\caption{
Average profiles over the three clumps shown in \fig{FB3} for the three
resolution levels, LR, MR, HR.
The rotation velocity in HR is on average $\sim 20\%$ lower than in LR, 
and the velocity dispersion is correspondingly $\sim 20\%$ higher. 
}
\label{fig:FB3a}
\end{figure*}

In the cosmological simulations analyzed in the previous sections,
where the maximum grid resolution is between 35 and 70 pc,
the high-redshift giant clumps are only marginally resolved,
with the collapsed clump diameter ranging from 6 to 30 grid cells.
We should address the possible effects of this marginal resolution
on our conclusions concerning the rotational support of the clumps.
When the resolution is better, the clumps are expected to develop a rich
turbulent substructure including fragmentation to subclumps
(see \fig{FB2} below).
By not including this substructure, the limited resolution may introduce
a bias in the rotation support of the clump in several different ways.
First, if massive fragmentation happens early in the clump collapse,
the collapse might involve less dissipation,
because of the potential small cross-section of the sub-clumps.
If the clumps are less dissipative, the contraction factor may become smaller,
and the clumps could end up with less rotation support 
(\equ{Vrot}). The kinetic energy of the subclump motions within the clump
may add to the pressure support of the giant clump.
Second, the limited resolution could introduce a bias of an opposite sign,
toward less rotation support, by preventing the (small) clumps from contracting
to below a size of a few resolution elements.
Third, the limited resolution may cause an error in the torques
acting on the clump and thus in
the angular-momentum exchange between clumps and
disc, which could be of either sign.
For example, numerical viscosity may lead to reduced torques and 
angular-momentum losses.
The potential artifacts
of numerical viscosity might be non-negligible and hard to estimate,
but they should be smaller at higher resolution.
Fourth, the level of rotation versus dispersion support may depend on
the turbulence within the clumps, which becomes fully developed only when
the resolution is sufficiently high.

To investigate the effects of numerical resolution and small-scale
substructure, we use simulations of idealized, isolated, gas-rich disc galaxies
representative of $z \sim 2$ star-forming galaxies.
The absence of continuous cosmological
gas supply limits the duration of disc evolution that can be followed,
but it permits a much higher resolution than the cosmological simulations.
Overall, the initial conditions and evolution into high-z, clumpy discs
are as described in \citet[][BEE07]{Bournaud07}, but the current simulations
were performed with much better resolution using the AMR code RAMSES
\citep{Teyssier02}, comparable in many ways to the ART code used in the
cosmological simulations utilized in the previous sections.
The sub-grid physical recipes include cooling using a barotropic equation
of state down to $\sim 100$K, star formation and supernova feedback.
More details of the RAMSES simulation technique and the sub-grid physical
recipes are provided in an appendix, \se{ramses},
as well as in \citet{Bournaud10,Teyssier10}.

We performed three simulations of a system that represents
a massive disc at $z \sim 1-3$, at a high resolution (HR), medium resolution
(MR) and low resolution (LR), as listed in Table~\ref{table-FB}.
The resolution in the LR simulation is similar to the resolution
in the cosmological simulations, with slightly lower spatial resolution
but somewhat higher mass resolution.
The HR simulation resolves gas densities up to $10^{7}\cmc$ and sizes of a
few parsecs, comparable to today's molecular clouds. The resulting
density power spectrum for the HR model (Bournaud et al. 2010, section 3.4 and
Figure~15) is characteristic of a fully developed three-dimensional turbulence
cascade, implying that the HR resolution is sufficient for
convergence on the internal properties of kpc-size clumps.

\Fig{FB1} shows the global disc morphology in gas for the three resolution  
levels.  
The discs are shown after two rotations of the outer disc, when 
a clumpy turbulent steady state has been reached, with a constant 
turbulent speed.  
We compare three clumps simulated at the three resolution levels. 
Unfortunately, there is no strict one-to-one correspondence between the  
clumps in the three resolution levels, because the small scale initial 
fluctuations that seeded the instabilities were naturally different.  
We therefore picked from the most massive clumps three that are located 
at similar distances from the galaxy centre, all with masses 
of a few $10^8\msun$, as estimated in a circular aperture of 
radius $\sim 500\pc$. 
Two of the clumps, labeled A and B, are shown in \fig{FB1}, and a third clump, 
labeled C, is selected on the same basis from a later output of the three  
simulations. 
Zoom-in views of the three clumps at the three resolution levels 
are shown in \fig{FB2}. 
 
The morphology of the giant clumps shows substantial variations with  
resolution.  
The LR clumps have a smooth ellipsoidal main body, with elongated extensions  
in the outer parts, somewhat resembling inflection knots along spiral density  
waves where the vorticity is at a maximum.  
The MR clumps show a mini-spiral morphology, suggesting internal  
clump rotation. The spin axis in clumps A and C is aligned with the disc axis,  
while clump B is significantly tilted. 
The HR clumps show a rich substructure with many dense sub-clumps. 
At this resolution, the internal supersonic turbulence  
(velocity dispersion $\sim 50\kms$, gas sound speed $\lsim 5\kms$) 
is properly resolved, as indicated by the power spectrum analysis. 
The turbulent flows generate dense filaments that give rise to even 
denser sub-clumps.   
The outer spiral armlets suggest significant clump rotation. 
 
We analyze the kinematic properties of the clumps in each  
resolution level following a procedure similar to the analysis of the  
cosmological simulations. The density, velocity and velocity dispersion  
profiles of the three clumps at the different resolution levels are shown  
in \fig{FB3}, and the average profiles over the three clumps are shown in
\fig{FB3a}.
Rather surprisingly,  
the density profiles show only a weak systematic variation with the resolution  
level --- no significant change at $r < 100 \pc$ and at $r> 250\pc$, 
and an apparent increase of less than 50\% in the average density at  
$r \sim 100-200\pc$ between HR and LR, which may be insignificant given the 
small-number statistics. 
The global kinematics inside the clumps shows only a marginal variation 
between the HR and MR levels, despite the order-of-magnitude change in  
resolution. However, 
comparing HR with LR we see a systematic trend. For example, the average clump  
rotation curve is $\sim 20\%$ higher in the LR case, and the associated 
velocity dispersion level is $\sim 20\%$ lower in LR compared to HR. 
Thus, the $70\pc$ resolution is responsible for a $\sim 20\%$ overestimate of 
the rotation velocity and a corresponding underestimate of the velocity  
dispersion. Combined with the small variation in the density profile,  
this translates to a $\sim 30\%$ overestimate in the average 
rotation parameter $\Rot$, from $\Rot = 0.57$ to 0.76.  
It seems that the main drivers of lower rotation and higher dispersion 
in the HR case are the reduced dissipation due to substructure, 
and the enhanced pressure due to resolved internal turbulence. 
In the HR clumps, some rotational energy is transferred to random motions,  
as a result of tidal stirring of the subclumps.  
 
The contraction factor is rather insensitive to the resolution, at the level of 
$c \sim 4$ for all clumps.  According to the toy model, \equ{Vrot},  
this level of contraction roughly corresponds to a rotation parameter of  
$\Rot \sim 0.8$ if there is no angular-momentum exchange between clump and disc. 
This is indeed the case for the median clumps in the cosmological simulations 
and the clumps in the isolated discs simulated with a similar resolution, LR. 
The clumps simulated with higher resolution, HR, with an average $\Rot \sim 
0.57$, have lost on average $\sim 30\%$ of their angular momentum. 
This may be due to torques associated with the clump substructure. 
We find no variation in the clump size or angular momentum in the direction expected for numerical viscosity. 
If anything we see small variations in the opposite sense, indicating that the behavior may be driven by physical torquing and dissipation in shocks rather than by numerical viscosity, even at LR and equivalently in the cosmological  simulations.
 
We report in \se{gmc} that simulations similar to those described 
here but with a lower gas fraction that resembles low-$z$ galaxies 
produce smaller clumps that are not supported by rotation.

 
\section{Observable signature of rotating clumps} 
\label{sec:obs} 
 
\begin{figure}  
\center 
\includegraphics[height=0.7\textwidth, width =0.35\textwidth]
{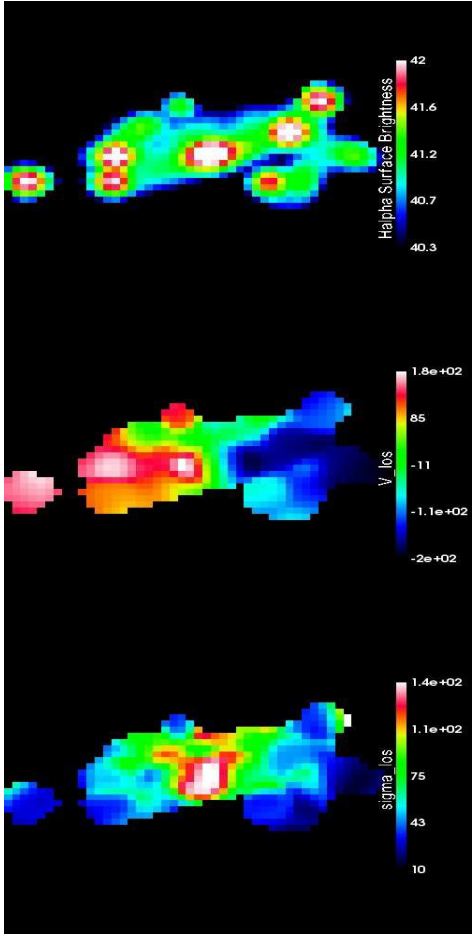} 
\caption{ 
Galaxy A at $z=2.3$ ``observed" in \Halpha\ at $70^\circ$ inclination, 
with the major axis horizontal.  
Top: \Halpha\ surface brightness (erg s$^{-1}$ \kpc$^{-2}$). 
Middle: line-of-sight mean velocity (km s$^{-1}$). 
Bottom: line-of-sight velocity dispersion (km s$^{-1}$). 
The velocities are weighted by \Halpha\ emissivity. 
The box size is $10\kpc$ and the pixel size is $200\pc$. 
All maps are convolved with a Gaussian filter of FWHM$=0.5 \kpc$,  
corresponding to $\sim 0.06$ arcsec.  
The irregular \Halpha\ morphology dominated by giant clumps 
has only little effect on the kinematics, which is dominated by a regular  
pattern of a rotating disc with a central bulge of high velocity dispersion.} 
\label{fig:obs_mw3_disk} 
\end{figure}  
 
\begin{figure}  
\vskip 0.1cm 
\center 
\includegraphics[width =0.48\textwidth]{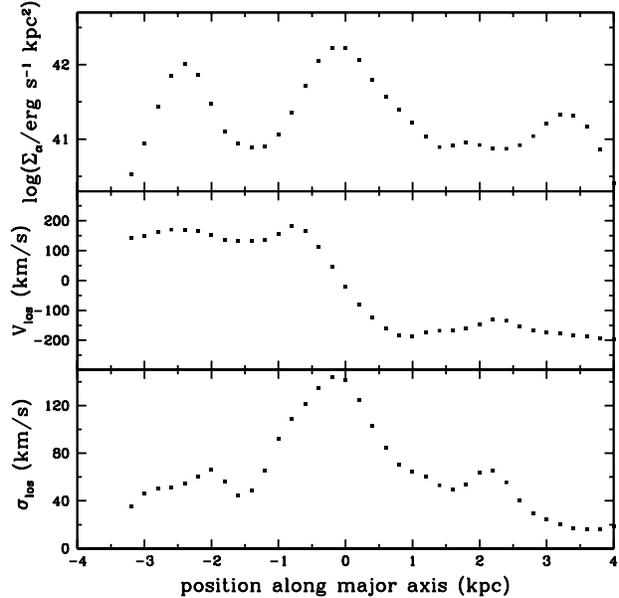} 
\caption{ 
Simulated long-slit ``observations" along the major axis of the Galaxy A disc 
at $z=2.3$.  
Top: \Halpha\ surface brightness. 
Middle: line-of-sight mean velocity.  
Bottom: line-of-sight velocity dispersion. 
After correcting for inclination and beam smearing, 
the rotation curve is rather flat at a level of $180 \kms$,  
and the intrinsic central velocity dispersion is $\sim 90 \kms$.}  
\label{fig:obs_major_axis} 
\end{figure} 
 
We have learned that the internal kinematics of most giant clumps in our  
simulated galaxies at $z \sim 2$ is dominated by rotation, with  
rotation velocities that could be as high as $120 \kms$ or more in the biggest 
clumps, and with a median rotation-support parameter of $\Rot =0.78$.
One might have assumed that this rotation signal should be detectable. 
On the other hand, the internal velocity dispersion is not very different from 
that of the surrounding disc, and is therefore expected to be hardly noticeable. 
Here we make a quick attempt to learn about the potential observability of the  
clump rotation signal. 
We refer in particular to observations in \Halpha, tracing the 
ionized gas in the star-forming regions, which highlight the giant clumps.  
With a clump diameter $\lsim 1 \kpc$, the main observational obstacle is the  
beam smearing, which is currently at the level of FWHM $\sim 0.2$ arcsec 
\citep{Genzel11},  
corresponding to  $\sim 1.6\kpc$. 
Other effects that may reduce the rotation signal are (a) contamination  
by foreground and background gas in the perturbed disc, that may have high  
velocities driven by supernova feedback, and (b) the unknown inclination of  
the clump spin axis relative to the line of sight due to the common tilts of  
the clumps with respect to the disc. 
In this preliminary study we perform mock observations of the 2D kinematics  
of a few disc clumps in our simulated galaxies, exploring different levels  
of beam smearing.  
 
\subsection{\Halpha\ kinematics in the disc} 
\label{sec:kinematics}
 
In order to generate mock \Halpha\ observations, we first compute the star  
formation rate density, $\rho_{\rm SFR}$, using the distribution of stellar  
particles younger than 10 Myr.  
Then, we use standard conversion factors to compute the \Halpha\ emissivity  
$\epsilon_{{\rm H}\alpha}$,  
based on the \citet{Kennicutt98} conversion and adjusted to a  
\citet{Chabrier03} IMF, 
\begin{equation} 
\log \epsilon_{\rm H\alpha} = \log \rho_{\rm SFR} + 41.33 \, , 
\end{equation} 
where $\epsilon_{\rm H\alpha}$ is in $\ergs \kpc^{-3}$  
and $\rho_{\rm SFR}$ is in $\sy \kpc^{-3}$. 
This equation holds as long as the \Halpha\ emissivity traces the underlying  
star-formation law, which is the case on scales of a few hundred parsecs  
\citep{Kennicutt07}. 
The \Halpha\ surface brightness is obtained by projecting the 
emissivity along a line-of-sight (los), {\bf n}, characterized by an inclination 
angle $i$ with respect to the galaxy rotation axis 
(variations in the surface brightness due to variations of dust opacity are not included).

We then convolve it with a Gaussian filter, $h_{\rm G}$, of a given FWHM, 
\begin{equation} 
{\rm S}_{\rm H\alpha}=\left(  \int  \epsilon_{\rm H\alpha} \ dl \right)  
\ast h_{\rm G} \, .  
\end{equation} 
We compute the mean velocity along the line-of-sight,  
weighted by the \Halpha\ emissivity, and convolve it with the same Gaussian,  
\begin{equation} 
\bar{v}_{\rm los}(x,y) = \frac {1}{\rm S_{\rm H\alpha}}  
\left[  \left( \int  \epsilon_{\rm H\alpha} {\bf v} \cdot {\bf n} \ dl \right)
\ast h_{\rm G} \right] \, ,  
\end{equation} 
where the integrals are along the line-of-sight {\bf n} through the point 
(x,y). 
The line-of-sight velocity dispersion is computed via, 
\begin{equation} 
\sigma^2_{\rm los}(x,y)= \overline{v^2}_{\rm los}(x,y) 
- \bar{v}^2_{\rm los}(x,y) 
\, . 
\end{equation} 
where, $\overline{v^2}_{\rm los}$ is the \Halpha-weighted variance of the 
velocity  along the line-of-sight, convolved with the Gaussian filter, 
\begin{equation} 
\overline{v^2}_{\rm los}(x,y) = \frac{1}{\rm S_{\rm H\alpha}}  
\left[ \left( \int  \epsilon_{\rm H\alpha} ({\bf v} \cdot {\bf n})^2 \ dl 
\right)    
\ast h_{\rm G} \right] \, .  
\end{equation} 
 
\Fig{obs_mw3_disk} shows \Halpha\ maps of galaxy A at $z=2.3$, observed at an 
inclination $i=70^\circ$. This inclination, with $\sin i \simeq 0.94$, 
is close enough to edge-on for capturing most of the rotation signal, 
but inclined enough to make all the giant clumps visible individually. 
The figure shows the \Halpha\ surface brightness, ${\rm S}_{\rm H\alpha}$,  
the line-of-sight mean velocity, $\bar{v}_{\rm los}$,  
and the line-of-sight velocity dispersion, $\sigma_{\rm los}$. 
We imposed a threshold in star formation surface density,  
$\Sigma_{\rm SFR}>0.1 \sy \kpc^{-2}$ or ${\rm log \ S}_{\rm H\alpha} > 40.33$,  
a pixel size of $0.2 \kpc$, and a Gaussian smoothing with FWHM$=0.5 \kpc$,  
corresponding to $\sim 0.06$ arcsec at $z \sim 2$.   
 
The \Halpha\ density image shows an extended, near edge on, thick 
and highly perturbed disc. 
The \Halpha\ luminosity highlights the central bulge and  
several large clumps of $\sim 1 \kpc$ in size.  The clumps   
account for about half the total star formation rate in the galaxy. 
In spite of this irregular clumpy morphology, the kinematics is dominated 
by a systematic pattern of a rotating disc with  
$V_{\rm rot,max} \simeq 180 \kms$. 
This overall rotation pattern is disturbed and shows non-axisymmetric features 
that reflect significant local non-circular motions, 
which could be driven by the incoming streams or by the disc instability 
itself. 
For example, the clump with a high positive velocity near the positive minor 
axis is an off-disc satellite in an orbit not related to the disc rotation. 
However, in general, one sees no obvious correlation between the kinematic 
features and the giant clumps as observed in the surface-brightness map. 
The velocity dispersion is dominated by the bulge  
with a global fall off as a function of radius,  
and the clumps are not associated with significant peaks in velocity  
dispersion --- if anything, some of them are associated with local minima. 

\Fig{obs_major_axis} shows a simulated long-slit ``observation" of  
${\rm V}_{\rm los}$ and $\sigma_{\rm los}$ along the major axis of the disc  
shown in \fig{obs_mw3_disk}. 
The surface brightness shows the large star-forming region at the central  
bulge and two clumps that lie along the major axis. 
Overall, the \Halpha\ surface brightness does not decrease with radius, 
with the inter-clump medium at roughly a constant level of \Halpha\ 
surface brightness out to the disc edge. 
The line-of-sight 
velocity along the major axis shows a roughly flat rotation curve of  
${\rm V_{\rm rot}} \simeq {\rm V}_{\rm los} /  \sin i = 180 \kms$, 
outside the central 1-kpc region.  
The beam smearing has only a 10\% effect on the value of the maximum disc 
rotation velocity. 
Inside that region, the rotation pattern shows a near solid-body rotation,  
which is not an artifact of low gas density in the bulge,  
but the beam smearing of FWHM$= 0.5\kpc$ is 
responsible for a reduction of 70\% in the ``observed" velocity gradient  
compared to the unsmoothed case.  
The line-of-sight 
velocity dispersion along the major axis shows a high central peak, 
and a weak decline with radius outside the central region.   
The beam smearing converts rotation signal into velocity dispersion 
at the disc centre, increasing the central velocity dispersion by 60\% 
compared to the unsmoothed value of  
$\sigma_{\rm intrinsic}= \sigma_{\rm los}(r=0)/  \sin i =90 \kms$. 
This central velocity dispersion is a signature of a massive spheroidal bulge 
that dominates the dynamics in the central 1-kpc region.  
 
The mock observations presented here are in qualitative agreement with  
observed massive discs at $z \sim 2$ in the SINFONI survey  
\citep{Genzel06,Genzel08,Genzel11}.  They both show a perturbed, thick 
rotating disc, a central mostly stellar bulge, and a few giant clumps 
with no obvious kinematic signal. 
Galaxy A is less massive than the most massive galaxies observed 
by a factor of a few, and the giant clumps are less massive than the observed 
clumps in proportion, but this does not seem to make a qualitative difference. 
Galaxy E is closer in mass to the massive observed galaxies, and it shows 
similar kinematics, with clumps of a few times $10^9\msun$. 
However, the current sample of simulated galaxies does not contain 
clumps as massive as the few extreme clumps observed with $\sim 10^{10}\msun$.

\subsection{Rotating clumps in \Halpha\ observations} 

\begin{figure*} 
\includegraphics[width =0.8\textwidth]{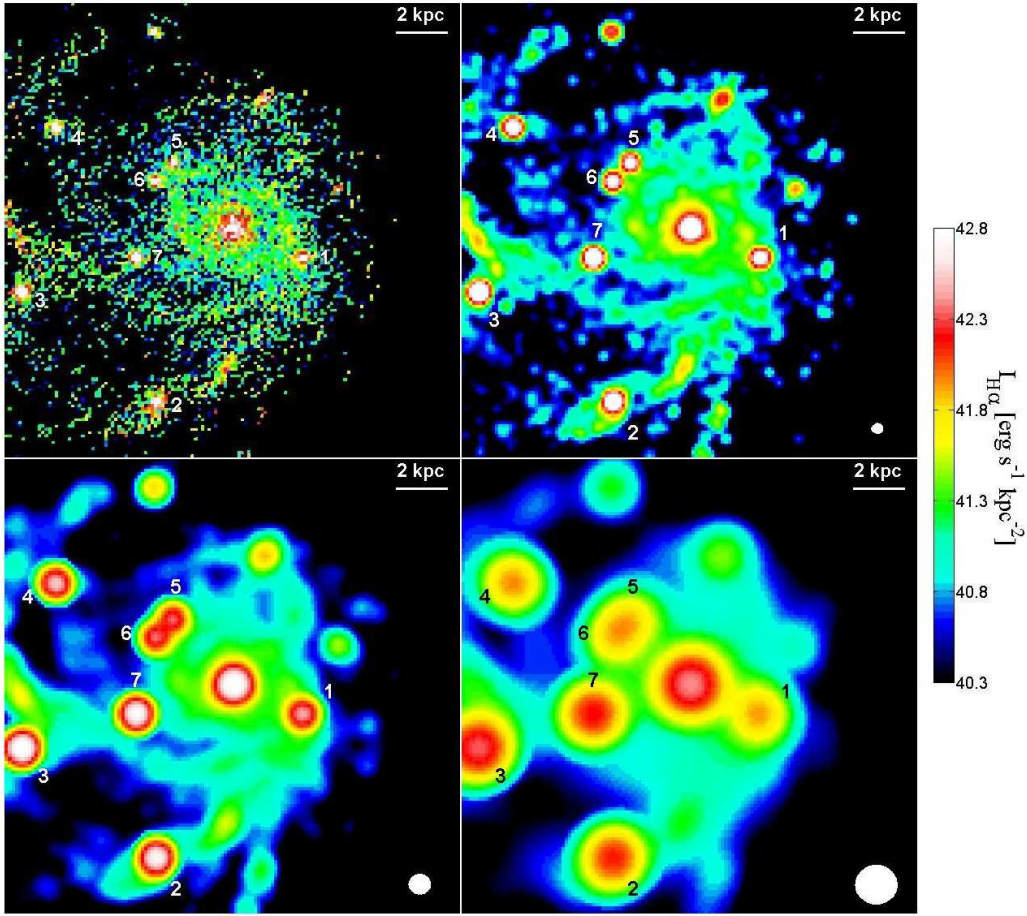}
\caption{
The effect of Gaussian smoothing on the morphology of galaxy E at $z=2.3$.
Shown is the \Halpha\ surface brightness.
The orientation and projection depth are the same as in \fig{sfg1},
namely the disc is face on and the depth is $5\kpc$.
The 4 panels employ Gaussian smoothing with FWHM$=0,0.4,0.8,1.6 \kpc$.
A white circle in the bottom right corner of each panel has a diameter equal
to the FWHM in that panel.
The seven clumps of this snapshot are marked in each panel, though with
FWHM$=1.6\kpc$, clumps 5 and 6 are seen as one clump.}
\label{fig:smoothing}
\end{figure*}
 
The beam smearing has an important effect on the images of  
the high-$z$ clumpy discs. 
\Fig{smoothing} shows a face-on \Halpha\ view of galaxy E at $z=2.3$  
for different Gaussian smoothings, ranging from no smoothing to FWHM$=1.6\kpc$, 
corresponding to $\sim 0.2$ arcsec. 
The \Halpha\ surface brightness was computed much the same way as described 
in \se{kinematics}, except that the star formation 
rate density, $\rho_{\rm SFR}$, was computed using the distribution of 
stellar particles younger than 100 Myr. 
The unsmoothed \Halpha\ image roughly follows the high density in the gas 
map shown in  \fig{sfg1}, with the giant clumps particularly pronounced.   
At a beam smearing of FWHM$=0.4\kpc$ (0.05 arcsec), the clumps are more than 
doubled in size and the transient structures in the disc are still clearly 
seen.  At larger smoothing scales, the clumps are not resolved --
they get gradually bigger and their contrast relative to the background disc 
diminishes.  With FWHM$=1.6\kpc$ (0.2 arcsec), 
clumps 5 and 6 are confused to be one clump.
We learn that typical Toomre clumps in discs of $\sim 10^{11}\msun$
are not expected to be resolved with beam smearing of FWHM $\sim 0.2$
arcsec.\footnote{
HST imaging of stellar light in clumps could reach comparable or better 
resolution, FWHM$\leq 1.2\kpc$ at $z=2$ \citep{forster11b}. This provides
complementary information on clump stellar mass, radius, and age 
\citep[][,CANDELS]{Grogin11}, but it is not useful in the quest for
rotation signal.}

\begin{figure*} 
\includegraphics[width =0.8\textwidth]{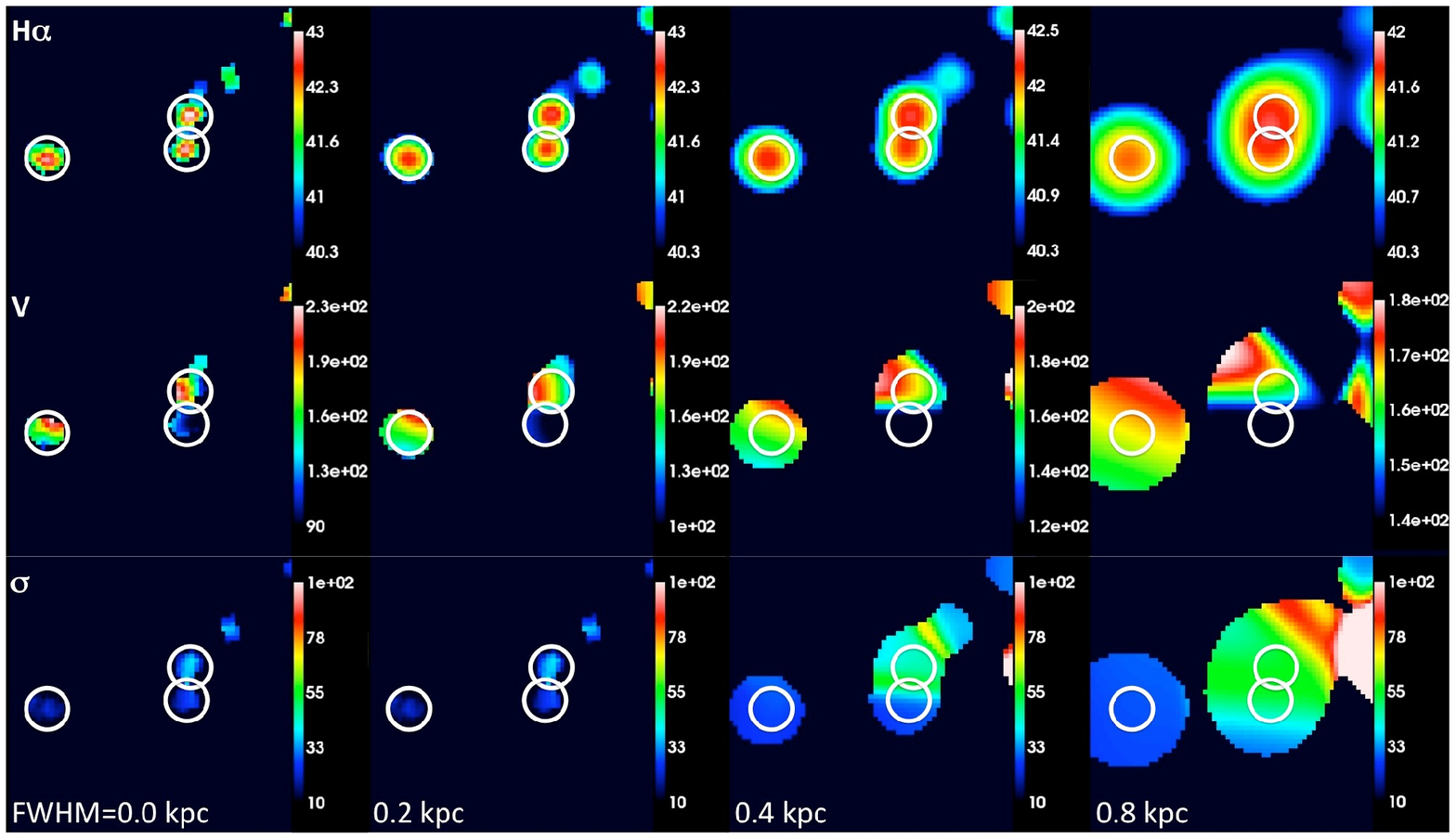} 
\includegraphics[width =0.8\textwidth]{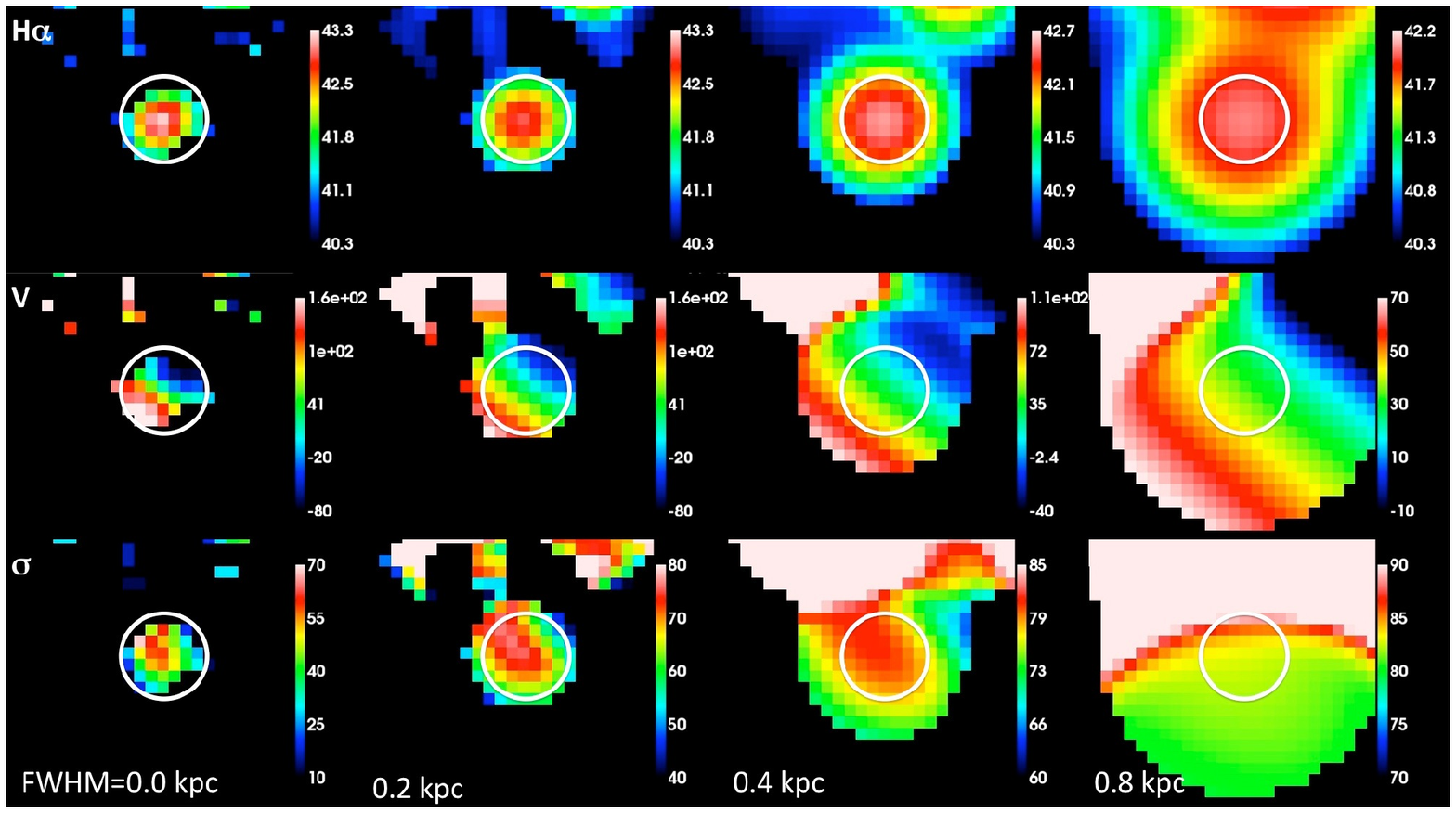} 
\caption{ 
``Observing" rotating giant clumps with different beam smearing. 
Zoom in on clumps 1 and 4 of galaxy A at $z=2.3$ (top panels, the clump at the 
top and the clump on the left)  
and on clump 2 of galaxy E at $z=2.2$ (bottom panels). 
Shown are \Halpha\ surface brightness (top), \Halpha-weighted 
line-of-sight velocity (middle), and velocity dispersion (bottom). 
The clumps are viewed nearly edge on, and almost naked, excluding the  
foreground and background, to maximize the rotation signal. 
The beam smearing ranges from zero on the left to FWHM$=0.8\kpc$   
($\sim 0.1$ arcsec) on the right. 
The white circles mark radii of $0.3\kpc$ and $0.42\kpc$ for the clumps of 
galaxy A and E respectively, 
marking the extent of the unsmoothed clump in \Halpha. 
}
\label{fig:summaryraw} 
\end{figure*} 
 
\Fig{summaryraw} shows mock observations zooming-in on  
three rotationally supported clumps from our cosmological simulations: 
clumps 1 and 4 of galaxy A at $z=2.3$ and clump 2 of galaxy E at $z=2.2$. 
In galaxy A we see one side of the disc, with the clumps along the 
major axis given the chosen line of sight and the bulge at the right of the 
frame. In galaxy E the bulge is above the top of the frame, and the 
clump is along the minor axis given the chosen line of sight.  
The clump masses are $(0.4,\,0.2,\,1.2)\times 10^9\msun$, 
their radii determined from the gas density profiles
are $(0.45,\,0.32,\,0.43)\kpc$, 
their maximum rotation velocities are $(68,\,60,\,129)\kms$,
and the rotation parameters are $\Rot=0.86,\,0.86,\,0.88$. 
The maximum velocity gradients across the whole clump are  
$\gv = 2\Vc/(2\Rc) = (150,\,190,\,280)\kms\kpc^{-1}$,
but it is typically higher over the range that emits \Halpha. 
The galactic discs are chosen to be observed at high inclinations of  
$\sin i=0.98,\,0.94$, nearly edge on, 
and the effective inclinations of the clump spins and the lines of sight are 
also large, $\sin i_{\rm c} = 0.98,\,0.89,\,0.94$.  
The observed rotation signal is quantified by the maximum gradient across  
the clump, $\gv = \max \{\Delta V/ (2 \Rc \sin i)\}$.   
 
\Fig{summaryraw} shows \Halpha\ surface density, and \Halpha-weighted 
line-of-sight velocity and velocity dispersion. 
The clumps are exposed naked, with no background or foreground contamination 
from the turbulent and rotating gas in the disc outside the clumps. 
This is achieved by excluding cells with emissivity below a threshold of 
$10^{42.2} \ergs \kpc^{-3}$ for galaxy A and 
$10^{42.5} \ergs \kpc^{-3}$ for galaxy E. 
Each clump is ``observed" with 4 different beam smearings, from no smearing, 
where the signal is smoothed at the $100\pc$ pixel scale, through Gaussian 
smoothings of FWHM$=0.2$, $0.4$, and $0.8\kpc$. The latter is comparable to 
the true clump diameter, and it corresponds to $0.1$ arcsec, which is about
half the beam smearing in current observations \citep{Genzel11}. 
 
Clump 3 of galaxy E is close in mass to the typical observed clumps 
and can thus serve for a direct comparison,  
but it is significantly smaller than the most massive  
clumps observed, with $\sim 10^{10}\msun$.  
Galaxy A and its clumps are even smaller, and significantly  
less massive than the big observed galaxies and their clumps, 
so a quantitative comparison between them requires scaling. 
Based on Toomre instability (see DSC09), 
we expect the clump quantities to scale with the galaxy quantities, 
and $V \sim R \sim M^{1/3}$. This implies that a   
clump $\sim 8$ times more massive is expected to have size and  
velocities twice as big. 
In particular, when ``observing" a clump of mass $\Mc$ with a given 
beam smearing, say FWHM=0.1 arcsec, we mimic the observation of a clump of 
mass $8\Mc$ with twice the beam smearing, FWHM=0.2 arcsec. 
Once observed with the proper beam smearing,   
the velocity gradient, which scales like $V/R$, is expected to be independent  
of clump mass. 
 
Clump 1 of galaxy A has been analyzed in some detail in \fig{disc_mw3}, 
\fig{clump1_mw3}, and \fig{prof_cos}. 
Its rotation curve is rather flat out to $\Rc$ and it is 
gradually declining outside the clump edge.  
For this clump, with no beam smearing and naked, we read across the clump  
$\Delta V = 125\kms$, which recovers much of the maximum rotation  
velocity of $2 \times 63\kms$.   
Over a scale of $2R=0.5\kpc$, 
the diameter of the clump in \Halpha, unsmoothed,
we obtain a gradient of $\gv = 250\kms\kpc^{-1}$.
With beam smearing of FWHM$=0.8\kpc$, we measure 
$\Delta V=28\kms$ across $2R=0.8\kpc$, namely $\gv=35\kms\kpc^{-1}$. 
The velocity gradient is reduced by a factor of 7 by beam smearing 
of 0.1 arcsec alone.  
When including the foreground and background contamination, 
the gradient is reduced further to $\gv = 30\kms\kpc^{-1}$, another 20\% 
reduction for this clump that resides in the middle of the disc. 
We measure at the clump position an un-smoothed velocity dispersion of 
$\sigma =42 \kms$, and after beam smearing of FWHM$=0.8\kpc$ it grows 
to $\sim 60 \kms$, but at such smoothing there is hardly any dispersion 
signal associated with the clump itself ---  
we basically see a continuous large-scale shallow gradient of dispersion 
from the bulge (on the left of the picture) outward. 
 
 
 
Clump 4 of galaxy A is extremely tilted relative to the disc,  
$\cos({\rm tilt})=-0.13$, 
but it happens to still be not far from edge-on relative  
to the line of sight, $\sin i_{\rm c}=0.89$. 
The unsmoothed gradient across the \Halpha\ clump 
is $\gv = 250\kms\kpc^{-1}$, 
and after smoothing with FWHM$=0.8\kpc$ it is drastically reduced to 
$\gv = 15\kms\kpc^{-1}$.  
As in clump 1, there is no noticeable dispersion signal associated with the 
clump. 
The contamination by foreground and background in this clump is negligible, 
because it resides at the outskirts of the disc. 
Clump 3 of galaxy A, seen just below clump 1, is not seen in the velocity map 
because its line-of-sight 
velocity is just outside the range shown by the color scheme. 
 
The massive clump 2 of galaxy E at $z=2.2$, unsmoothed and naked, shows  
$\Delta V = 240\kms$, which recovers much of the maximum rotation 
velocity of $2 \times 129\kms$. 
With $2R = 0.65 \kpc$, the unsmoothed \Halpha\ diameter,
we obtain a gradient of $\gv = 375 \kms\kpc^{-1}$.
With beam smearings of FWHM$=0.2,0.4,0.8\kpc$, we measure 
$\Delta V=240, 150, 80\kms$, and $\gv=300, 125, 40\kms\kpc^{-1}$ respectively. 
The velocity gradient is reduced by a factor of 9 by beam smearing 
of 0.1 arcsec alone.  
We learn that even this massive clump, which is highly rotation supported 
and observed nearly edge on, shows only a small rotation signal. 
 
 
The three clumps analyzed here were all selected to be ``observed"  
with high inclinations relative to the line of sight in order to maximize the  
rotation signal. 
Given the frequent occurrence of big tilts between clump and disc,  
even if the disc is observed nearly edge on,  
a typical clump is likely not to be edge on, and thus to show an even weaker  
rotation signal.  
 
In summary, given the current beam smearing, 
our simulations predict that the typical clumps, which are rotation supported, 
should show only a weak observable rotation signal,  
with $\Delta V \sim 10-40\kms$ and $\gv \sim 15-30\kms\kpc^{-1}$. 
The actual signal in individual cases could be even weaker because of  
(a) low rotation support, (b) low (unknown) inclination of the clump relative  
to the line of sight,  
and (c) contamination by gas in the disc outside the clump. 
The typical clumps, with internal velocity dispersion comparable to  
the velocity dispersion in the disc, are not expected to show a noticeable  
signal in the smoothed velocity dispersion field. 
These predictions for the typical rotation-supported \insitu disc  
clumps are consistent with the marginal evidence for weak systematic  
prograde clump rotation, at a level of $\gv \sim 10-40 \kms\kpc^{-1}$,  
and the no noticeable dispersion signal associated with the clumps,  
as observed for the typical clumps of $\sim 10^9\msun$ \citep{Genzel11}. 
This does not apply to the extreme massive clumps observed, which 
do not have obvious counterparts among the \insitu Toomre clumps in our  
simulated discs (see \se{exsitu}.

\section{Ex-situ clumps} 
\label{sec:exsitu} 

Among the 86 clumps detected in our simulated discs,  
we identify 9 (i.e. $\sim 10\%$) as \exsitu clumps,  
which came in with the inflowing streams as minor-merging galaxies. 
They are listed in \tab{clumps} with a proper comment, 
and appear in some of the statistics figures of \se{stat} 
marked with a special symbol. 
Three of these \exsitu clumps happen to appear in one snapshot, 
clumps 1, 2 and 4 in
galaxy E at $z=3$, shown in \fig{sfg1}.  
This is a rare event, immediately following an episode of multiple
minor mergers. 
We defer to another paper a detailed study of the \exsitu clumps, 
including their origin, evolution, structure, and kinematics, refering to the
three components of gas, stars and dark-matter. Here we limit the discussion 
to some of the features that distinguish them from the \insitu clumps.

\begin{figure} 
\includegraphics[width =0.48\textwidth]{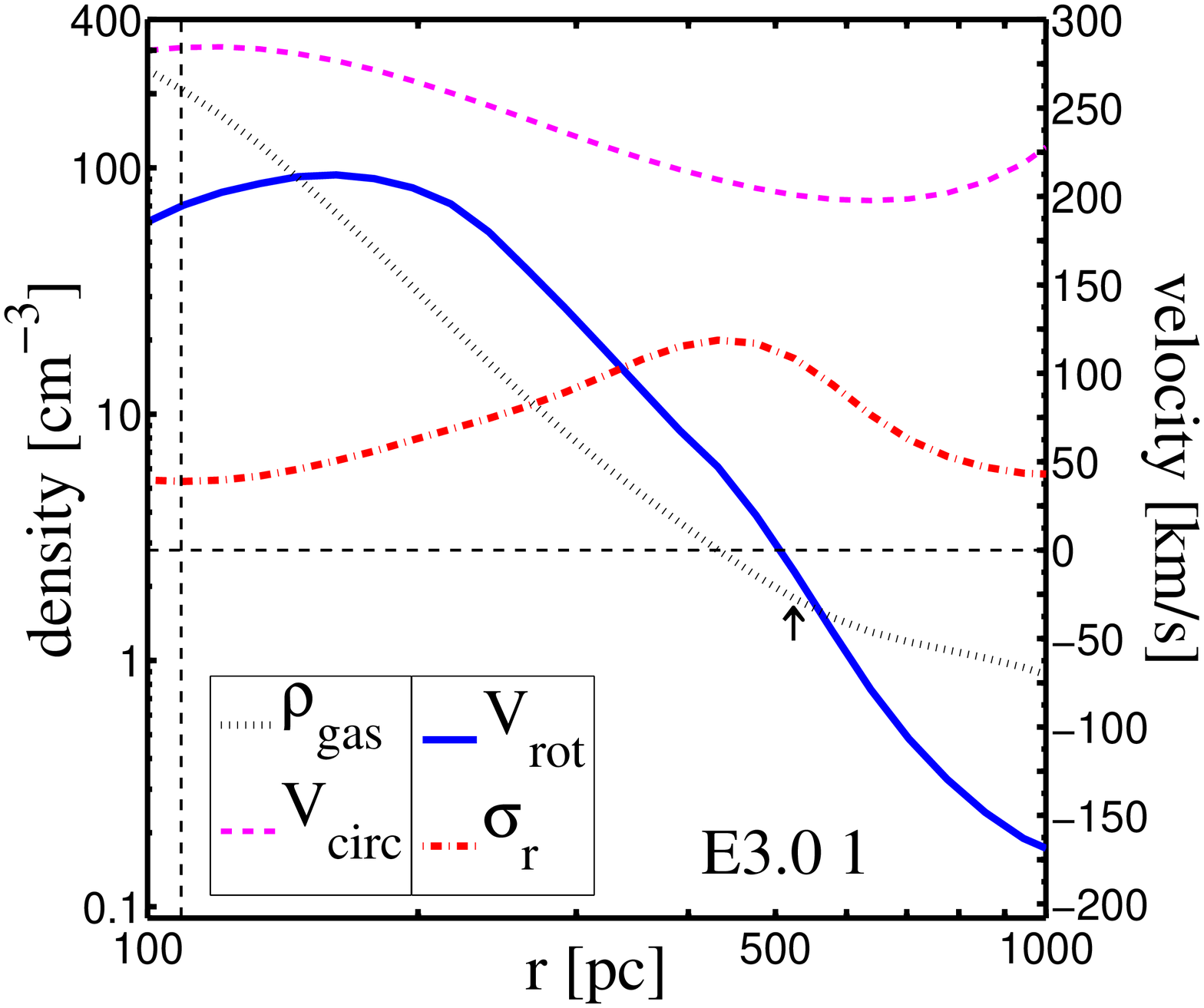}
\caption{
Profiles of the gas in the equatorial plane of \exsitu clump 1 of galaxy E
at $z=3.0$, similar to \fig{prof_cos}.
The rotation curve rises to a maximum of $\sim 210\kms$ but then declines
quickly toward the clump edge, beyond which it has a negative value,
reflecting the retrograde nature of the clump rotation relative to the disc.
}
\label{fig:sfg1_dm_vprofile}
\end{figure}

The \exsitu clumps are similar in their gas appearance to the \insitu clumps,  
and are positioned in or near the disc at different radii, so  
they cannot be easily distinguished by their morphological appearance.  
However, in these clumps the dark-matter fraction within the clump radius 
ranges from 0.1 to 0.4 (compared to less than a few percent in the \insitu 
clumps), their mean stellar ages are typically $400-1000\Myr$  
(compared to $20-400\Myr$ in the \insitu clumps), 
and more than $80\%$ of their stars were formed at distances more than $10\kpc$
from the disc plane. 
The \exsitu clumps mark the massive tail of the clump mass distribution, 
with clump 1 of galaxy E at $z=3$ being the most massive clump in the 
sample, and with 6 of the 9 \exsitu clumps in the mass range 
$(2-5) \times 10^ 9 \ \msun$, compared to only 4 \insitu clumps in that range,  
all involved in clump interactions or mergers within the disc.
Similarly, 4 of the \exsitu clumps have a circular velocity $\Vcc$ in the  
range $210-250\kms$, compared to the 3 \insitu clumps with the highest $\Vcc$ 
being in the range $180-200\kms$, all involved in mergers,  
and all other \insitu clumps having $\Vcc < 140\kms$. 
 
Many of the \exsitu clumps share the overall disc rotation pattern, 
but in some cases they have significant velocity components along $r$ or $z$. 
However, 3 \exsitu clumps actually share the overall disc kinematics 
with deviations much smaller than the dispersion in each component. 
One of these, clump 2 of galaxy E at $z=3$, is of mass  
$\Mc\sim 3\times 10^9\msun$ in a disc of $\Md \sim 3.3\times 10^{10}\msun$ 
and a bulge twice as massive, thus representing a non-negligible minor merger. 
This means that
minor mergers below a mass ratio of 1:10 
may end up joining the disc spatially and kinematically, and  
could be easily confused with the \insitu clumps, though they account for 
only $\sim 10\%$ of the disc clumps.  
 
In terms of their internal kinematics, 5 of the \exsitu clumps are rotation 
supported with $\Rot > 0.88$ and 5 are with a small tilt. 
They all seem to be in Jeans equilibrium. 
However, 4 of the clumps are severely tilted by more than 90 degrees,  
and 3 of the clumps are not supported by rotation, $\Rot \leq 0.35$. 
Therefore, 
the \exsitu clumps can provide interesting examples of massive clumps that  
show only little rotation even when marginally resolved. 
This is because they are not expected to obey the toy model of \se{theory}, 
and the spins of the incoming galaxies could in principle be oriented at a  
random direction compared to the disc rotation axis. 
Imagine an \exsitu clump with a retrograde spin compared to the host disc. 
The rotation curve in the clump frame will flip from rotation in one sense  
in the inner radii to rotation in the opposite sense in the outer regions 
of the clump, where it blends with the overall disc rotation.  
In this case, it would be hard to detect any rotation  
signal unless the beam smearing is much smaller than the clump size. 
There are four examples for this among the \exsitu clumps in our simulations. 
Two of them are clumps 1 and 4 of galaxy E at $z=3$, which have the largest 
tilts in our whole sample, $\cos{\rm tilt}=-0.61$ and $-0.40$, respectively. 
Their inner rotation corresponds to a rotation support factor 
$\Rot = 0.11$ and $1.02$, but in both cases it vanishes and then 
flips to the opposite sense near $r \sim 500\pc$, 
as demonstrated for clump 1 in \fig{sfg1_dm_vprofile}.  
These clumps will show no detectable rotation signal when observed 
with beam smearing of $FWHM \geq 0.1$ arcsec.  
An example of similar nature is clump 5 of galaxy E at $z=2.4$, 
with $\cos{\rm tilt}=-0.39$ and $\Rot = 0.53$, which is a closely 
interacting clump. 
These cases may provide a clue for the possible origin of some of the 
few observed extremely massive clumps that do not show significant rotation 
despite being marginally resolved \citep{Genzel11}. 
Indeed, these clumps have masses $\sim 10^{10}\msun$, 
at the level of $\sim 10\%$ of their host disc mass. 
This exceeds the Toomre clump mass of only $\sim 2\%$ of the disc mass, 
as expected from straightforward disc instability analysis (DSC09), and 
as revealed for the \insitu clumps in our simulations.  
Being an \exsitu clump, or a merger remnant of \insitu clumps, helps 
achieving a clump mass more massive than the Toomre mass, though 
our simulations have not revealed so far clumps as massive as the extreme 
observed clumps. 
 

\section{Gradients of age and gas fraction} 
\label{sec:age} 
 
\begin{figure}  
\includegraphics[width =0.48\textwidth] {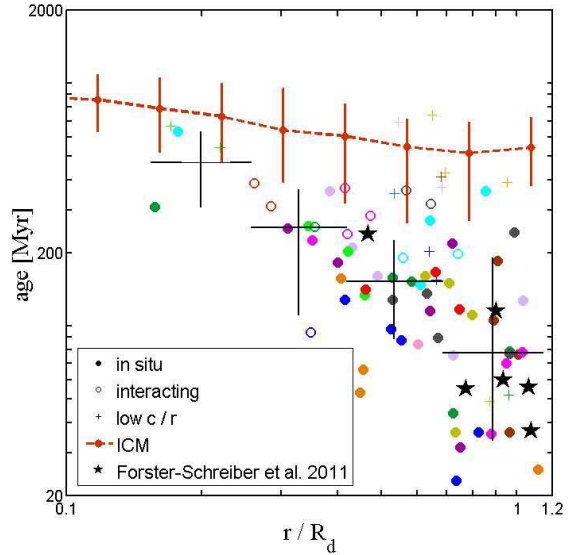} 
\caption{ 
Mean stellar age in clumps versus distance from the disc center relative to
disc radius.
Symbols and color code for the individual clumps are as in \fig{R} to \fig{c}. 
The black crosses mark the median and standard deviation of log(age) in
distance bins, and they are best fitted by the power law age$\prop r^{-1.2}$.
The analysis excludes the low contrast clumps.
The dashed line shows the mean stellar age of the inter-clump stellar material
in cylindrical shells, best fit by a power law age$\propto r^{-0.26}$.
The black star symbols correspond to estimates based on observations by 
\citet{forster11b}.
There is a significant gradient of clump age with radius, steeper than 
the gradient of the inter-clump material, and consistent with
observations.
} 
\label{fig:age} 
\end{figure} 

\begin{figure} 
\includegraphics[width =0.48\textwidth] {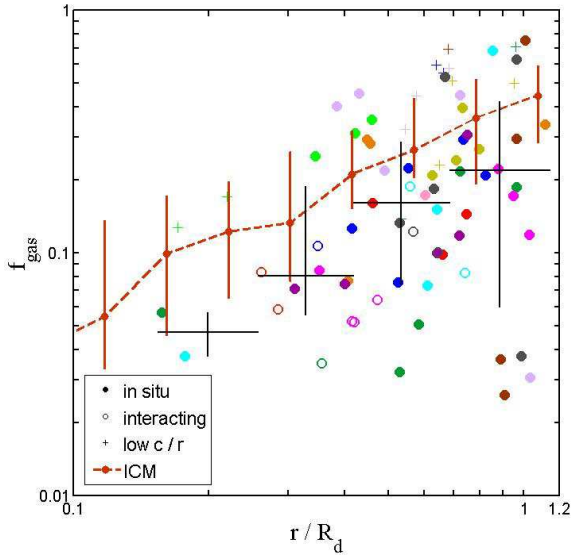}
\caption{
Gas fraction in clumps versus distance from the disc center relative to
disc radius. Symbols and lines are as in in \fig{age}.
The linear regression analysis yields a slope of 0.82.
The two outliers with low gas fraction near $\Rd$ lie somewhat off plane
with non-negligible $V_z$ --- the slope becomes unity when these two outliers
are excluded.
There is a significant gradient of gas fraction with radius.
}
\label{fig:fgas}
\end{figure}

In order to validate our confidence in our predictions for the 
giant-clump support, it is worthwhile to work out testable predictions for the 
general scenario of high-$z$ disc instability with long-lived 
giant clumps as simulated here. 
Predictions of this sort are associated, for example, with 
the preferred formation of giant clumps in the outer parts of the disc 
and their inward migration toward a bulge in the disc center, 
which imply an age gradient throughout the disc. 
 
\Fig{age} shows for all the \insitu clumps in our simulated sample 
the mass weighted mean stellar age within each clump radius as a function 
of its distance from the disc center, normalized to the outer disc radius. 
One can see a clear trend, where the age is roughly inversely proportional 
to the radius, fit by a power law $\propto r^{-1.2}$,
declining from a median of $\simeq 350 \Myr$ ($300-600\Myr$) 
at $r<0.3\Rd$ to a median $\simeq 70 \Myr$ ($20-400\Myr$) toward $r \sim \Rd$. 
This trend is steeper than the corresponding gradient in the disc stars
between the clumps, which is flatter than $r^{-0.3}$ 
(Mandelker et al., in preparation).
The steeper gradient in the clumps reflects the 
clump migration inwards on top of the inside-out growth of the disc. 
This difference in the age 
gradients between the clumps and the inter-clump material
 thus provides a tool for distinguishing between 
the scenario simulated here, in which the clumps survive intact during their 
migration to the bulge, and the scenario where the clumps are disrupted by 
feedback in $\lsim 100\Myr$, mixing the stellar populations of clumps and disc
\citep{Genel10}. There are preliminary observational indications for a steep
age gradient similar to the gradient predicted here for long-lived clumps
\citep{forster11b}.

\Fig{fgas} similarly shows the gas fraction within clumps as a function
of distance from the disc center.
One can see a clear trend, where the gas fraction is roughly proportional
to the radius, increasing from 5-10\% in the inner disc
to $\sim 30\%$ in the outer disc.
Two outliers with low gas fraction near $\Rd$ are clumps that
lie somewhat off plane with non-negligible $V_z$.
Recall that the overall gas fraction in our simulations at $z \sim 2$
is on the low side compared to observations, associated with a slight
overproduction of stars at higher redshifts. This implies that the absolute
values of the gas fraction may be unreliable to within a factor of 2,
but the trend with radius is likely to be indicative.  
 
The 10\% population of \exsitu clumps would have contaminated these trends, 
as 8 of the 9 \exsitu clumps have mean ages older than $400\Myr$,
and 6 out of the 9 have gas fractions less than 0.05,
while they span the whole radius range (\tab{clumps}).

\section{GMCs at low redshift} 
\label{sec:gmc} 
 

Our simulations of gas-rich discs at high redshift 
reveal that the typical giant clumps that form \insitu due to gravitational 
instability are largely supported by rotation, and indicate that this is not 
an effect of limited resolution. 
While this is yet to be confronted with high-redshift observations, 
the low-redshift analogs, the giant molecular clouds (GMCs), are 
observed not to be supported by rotation and not in equilibrium 
\citep{Blitz80,Alves98,Blitz04,Phillips99}. 
As a first step in trying to approach this issue, 
we study the clump support in an isolated $z=0$ disc galaxy 
from a RAMSES simulation \citep[][hereafter B10]{Bournaud10}, technically 
similar to the simulations that served our convergence test at high  
redshift. 
The simulation technique is the same as described in \se{subs}, 
except that the initial gas fraction is significantly lower,
11\% within the stellar disc radius, 
the initial stellar mass is $3.3 \times 10^9 \msun$, one fifth of the stellar 
mass in the high-z case, and the stellar disc scale length is 1.5 kpc, 
a factor of 3.7 smaller than in the high-z disc. The model also includes a 
stabilizing stellar bulge of one tenth of the stellar disc mass. 
The spatial resolution is $\epsilon_\mathrm{AMR}=0.8$~pc and the mass 
resolution is $m_\mathrm{res} = 5 \times 10^3\msun$, 
with the Jeans-length floor at 3.2~pc. 
The simulated GMCs at $z=0$ are found to be $\sim 10^6\msun$ compared to the 
$\sim 10^9\msun$ high-redshift clumps. They are typically 10 times smaller 
in size, with similar central gas volume densities. 
Thus, the spatial resolution relative to the clump size is slightly better 
than the MR resolution level in our high-$z$ isolated simulations. 
With $\sim 100$ spatial elements per clump in the MR simulation, 
the rotation parameter was found to be only about 15\% higher than in the HR 
cases, so we expect the resolution effects to be small both in  
the high-$z$ clumps and the low-$z$ GMCs. 
 
\begin{figure}  
\includegraphics[width=0.5\textwidth]{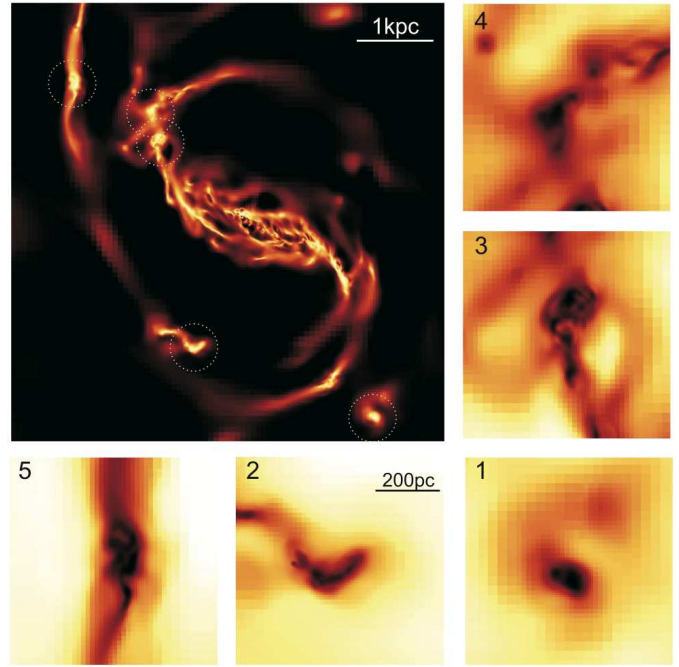} 
\caption{ 
Face-on view of the gas in the low-redshift disc simulation (B10). 
Surface densities above 50~M$_{\odot}$~pc$^{-2}$ are shown in log-scale. 
The insets zoom in on the five GMCs analyzed in \fig{FBappen2}. 
The GMCs are irregular in shape and dominated by gas.} 
\label{fig:FBappen1} 
\end{figure} 
 
\begin{figure*} 
\includegraphics[width=0.9\textwidth]{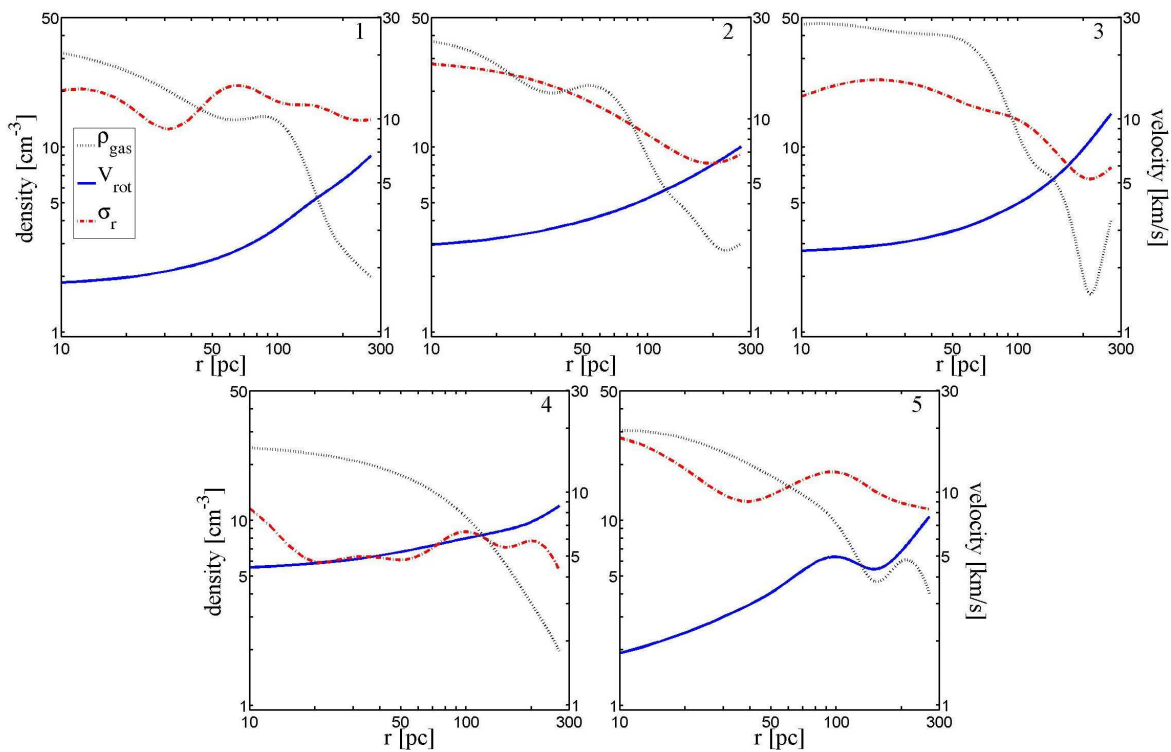} 
\caption{ 
Profiles of the gas in the equatorial plane for the
5 GMCs from the low-$z$ isolated disc simulation (B10). 
Each panel shows the profiles of density (left axis), rotation velocity and
radial velocity dispersion (right axis).
The profiles are smoothed as described in Appendix \se{clumps}.
Within the clump, the rotation velocity is smaller than the velocity disperion. 
} 
\label{fig:FBappen2} 
\end{figure*} 
 
 
\Fig{FBappen1} shows a face-on view of the gas surface density in 
the simulated disc at $t=268$~Myr after the start, once the turbulent speed  
and ISM density distribution reached a steady state.  
Also shown are five main clouds, which are indeed well resolved.  
These clouds contain mostly gas. 
They have typical masses and radii $\sim 10^6$~M$_{\odot}$ and 
$\sim 50-100$~pc as expected from a Toomre-instability estimate, 
and their typical internal circular velocity is $\Vcc \sim 7.5\kms$. 
The five clouds are analyzed in the very same way as the high-$z$ giant 
clumps. The mass density, rotation velocity and radial velocity dispersion 
profiles are shown in \fig{FBappen2}. 
We see that in 4 out of the 5 GMCs, the rotation velocity within the clump is 
$\Vc \sim 2.5\kms$, so the typical rotation parameter is only $\Rot \sim 0.1$. 
The rotation velocity is rising only near and outside the clump edge, 
where it reflects the overall disc rotation. 
The velocity dispersion is $\sigma \sim 10 \kms$ with an uncertainty  
$< 1\kms$ (B10), so $2\sigma^2$ is significantly larger than $\Vcc^2$,  
indicating that these dispersion-dominated clumps are transients and  
not in Jeans equilibrium. 
Indeed, the clump lifetimes are $\sim 20-60\Myr$.  
It is encouraging to find that our simulated low-$z$ GMCs are {\it not\,} 
rotation supported, indicating that the rotation support of  
the high-$z$ clumps is not a numerical artifact. 
 
 
Fully understanding the origin of the difference between the level of  
rotation support in high-$z$ and low-$z$ clumps is beyond the scope of the  
current paper, which focuses on the high-$z$ clumps.  
We had a quick look at the low-$z$ clumps in order to verify our  
evaluation that the rotation support at high redshift is not a numerical  
artifact of resolution, and to demonstrate that having the high-$z$ clumps  
supported mainly by rotation is not contradictory to having the low-$z$  
clumps not supported by rotation (as observed). 
However, we can highlight important differences and similarities between 
the high-$z$ and low-$z$ clumps, which may provide hints for the origin 
of the difference in rotation support.  
 
Assuming that the high-$z$ clumps and the low-$z$ clumps both 
form entirely by Toomre gravitational instability,  
a major difference between them is expected to be 
that the high-$z$ clumps are big and composed of gas and stars, 
while the low-$z$ clumps are small and made of gas only. 
At high redshift, the gas fraction is high, and the stars are relatively young, 
with their velocity dispersion not much higher than that of the gas, 
$\sigs \sim \sigg \sim 50\kms$. In this case 
both the stars and the gas participate in the gravitational instability 
almost as equals, so the clumps are made of the two components. 
The clump mass, being proportional to the square of the 
surface density of the baryons, 
is big, a few percent of the disc mass ($\Mc \sim 10^9\msun$ with  
a circular velocity $\Vcc \sim 50-100\kms$).  
At low redshift, the gas fraction in the disc is low, and the stars keep 
their high velocity dispersion $\sim 50 \kms$ from the time of their formation. 
In order to maintain a $Q\sim 1$ Toomre instability under such conditions, 
the gas has to cool to $\sim 10 \kms$ (Caciatto et al. in prep.).  
Under the condition $\sigg \ll \sigs$, only the gas participates in the 
instability and the clumps contain gas only.  
Since the gas surface density is low, the forming clumps are small, 
$\Mc \sim 10^{-4} \Md$. 
  
The density contrast between surface density in the clumps and in the 
background disc, measured at high $z$ for all the baryons and at low $z$ 
for the gas only,  
is found in our isolated simulations to be $\sim 15$ in both cases. 
Using the simple model of \se{theory}, if angular momentum is conserved, 
the contraction factor $c \sim 4$ implies rotation support with $\Rot \sim 0.8$ 
both at high and low redshift. 
The fact that low-$z$ clumps are not rotation supported, $\Rot \sim 0.1$, 
indicates that they must have lost a significant fraction 
of their angular momentum, by a mechanism that is yet to be understood. 
This angular-momentum loss may be associated with the escape of a fraction 
of the few sub-clumps. 
Given the lower circular velocity in the low-$z$ clumps, they are also 
expected to be more susceptible to the effects of supernovae feedback  
\citep{DekelSilk86} or radiation-driven feedback  
\citep{murray10,KrumholzDekel} 
in producing thermal pressure or driving outflows. 
  
As an alternative, it is possible that the trigger for instability and 
clump formation is somewhat different at low $z$. 
While the formation of giant clumps in our simulated high-$z$ discs is 
initiated by linear gravitational instability in regions where $Q<1$ 
\citep{Bournaud07,CDB,Genzel11}, 
the self-gravitating collapse of the low-$z$ GMCs could be triggered 
by the formation of non-linear perturbations due to gas compression 
associated with spiral density waves \citep[see also][]{Dobbs11}  
or supersonic turbulent flows. 
Indeed, some of our low-$z$ clouds seem to form by compression in local shocks, 
as can be seen, for example, in cloud~2, shown here at $t=268$~Myr. 
This cloud started forming just a bit earlier, at $t=254$~Myr,  
where an in-plane small-scale velocity component analysis (Fig.~9 in B10)  
shows a compression flow into a filament, with a local velocity gradient  
of $27\kms$ over 200~pc.  
This supersonic velocity also exceeds the circular velocity of 
this clump, which has a mass of $7\times 10^5$~M$_{\odot}$ and a half-mass 
radius of 88~pc, indicating that the compression velocity is higher than 
expected from self-gravitating collapse (though the compression velocity 
is large only in one direction). 
This cloud at $t=268$~Myr has an elongated shape, with 5-6 smaller clumps 
spread along the filament, suggesting that gas compression by large-scale 
flows created the initial filament, which subsequently became unstable and 
fragmented into a few clumps. 
If low-$z$ GMCs form primarily by turbulent gas compression rather than by 
pure gravitational instability, then their initial rotation pattern 
may differ from the global disc rotation assumed in \se{theory}, and they may 
end up rotating with lower velocity than the clumps formed by pure 
gravitational instability. 
 
\section{Summary and Discussion} 
\label{sec:conc} 
 
We addressed the internal support of the \insitu giant clumps in  
gravitationally unstable galactic discs at high redshift,  
using an analytic model and hydro AMR simulations. 
Zoom-in cosmological simulations with maximum resolution of $35-70\pc$ 
were complemented by isolated-disc simulations reaching $\sim 1\pc$ resolution. 
 
The simple analytic model predicts that the protoclumps should be  
rotating in the sense of the disc rotation. For a disc with 
a flat rotation curve, if angular momentum is conserved during the collapse 
of the clump, the relative rotation support of the clump,  
$\Rot=\Vc^2/\Vcc^2$, is predicted to be $\Rot \sim 0.2\,c$, 
where $c$ is the 2D collapse factor of the clump, roughly the square root of  
the surface-density contrast of the clump with respect to the disc.  
 
We have analyzed 77 \insitu disc clumps from several snapshots of five zoom-in 
cosmological simulations of galaxies in the redshift range $z=2-3$, 
eliminating the few clumps (10\%) that contain a significant peak in 
dark-matter density and are  
suspect of being external mergers. We also identified $\sim 10\%$ of the \insitu 
clumps as associated with a close clump interaction or a recent clump merger, 
but have not eliminated them from the analysis.  
The typical clumps are found to be rotation supported, with a median 
value of $\Rot = 0.78$, and where in 82\% of the \insitu clumps the  
rotation provides most of the support, $\Rot > 0.5$.  
The clumps are to a good approximation in Jeans equilibrium, obeying the Jean 
equation for an isotropic rotator, with the centrifugal force and turbulent 
pressure force balancing the gravitational attraction.  
While the typical clump rotation is prograde as expected,  
except for a few outliers, 
the clump spin and disc axis show a median tilt of $40^\circ$, with a  
significant tail toward larger tilts. These tilts are associated in many  
cases with deviations of the local plane from the global disc plane. 
In general, the clump rotation and contraction factor are consistent 
with the toy-model prediction and with conservation of angular momentum in the clumps
within a factor of a few.
There are clumps that indicate significant exchange 
of angular-momentum, loss by up to $50\%$ or gain by up to a factor of two. 
 
The isolated-galaxy simulations show clumps with similar properties. 
They are typically rotation supported, but with somewhat smaller tilts. 
At a resolution comparable to the resolution of the cosmological simulations, 
they show a similar level of rotation support. 
Once the clump substructure and internal turbulence are fully resolved, 
the rotation support parameter is reduced by $\sim 30\%$, at the expense of 
a comparable increase in the support by velocity dispersion. The median  
is reduced to $\Rot \sim 0.6$, confirming the general result that 
rotation is the dominant source of support in most clumps. 
 
We made a first attempt to address the detectability of clump rotation given  
the non-negligible beam smearing and other observational limitations at high 
redshifts. For this purpose, we mimicked \Halpha\ observations of three clumps 
from our simulated discs, selected to be highly rotation supported and  
optimally inclined for a maximum rotational signal. 
The beam-smearing FWHM was varied from zero to $0.8 \kpc$, 
the latter corresponding to $\sim 0.1$ arcsec at $z \sim 2$, 
which is about half the beam smearing in current observations. 
A simple scaling argument implies that an ``observation" of a simulated  
clump of a few $\times 10^8\msun$ with a beam smearing of 0.1 arcsec  
should have a similar relative smearing effect to the case where a  
clump 8 times more massive is observed with a beam smearing twice as broad.  
It also implies that the observed velocity gradient across the clump,  
$\gv = \Delta V/(2\Rc \sin i)$, should be independent of mass. 
We find that with a beam smearing of 0.1 arcsec, the rotation signal is reduced  
by a factor of 5-10, to the level of $\gv \sim 15-30\kms\kpc^{-1}$. 
When the whole clump population is considered, the typical rotation  
signal is expected to be even weaker because many clumps would be less 
favorably tilted and with an unknown inclination relative to the line of  
sight, and because of foreground and background contamination by the  
turbulent gas in the disc outside the clumps.  
The rotation signal is also reduced by observational noise, and when   
the overall disc rotation is subtracted. 
The velocity dispersion within the clumps is comparable to and sometimes 
smaller than the dispersion in the disc at the clump vicinity. 
This leaves no noticeable signal in the observed kinematics for any beam  
smearing, and the spurious dispersion due to smoothed rotation signal is  
not enough to make it detectable. 
These findings from ``observing" the simulations are consistent with the  
marginal detection of a very weak rotation signal, and the lack of a 
significant  dispersion signal in the typical observed clumps \citep{Genzel11}. 
 
The extreme observed clumps, with $\sim 10^{10}\msun$ and no significant 
rotation signal, are not reproduced as Toomre \insitu clumps in our 
current simulations, so they pose an interesting open question.  
These super-giant clumps may represent merging non-rotating bulges,
counter-rotating \exsitu clumps that joined the host disc (\se{exsitu}),  
mergers of two or more \insitu clumps, 
or, perhaps, big clumps out of equilibrium due to massive outflows  
\citep{murray10,Genel10}.  
 
If the high-$z$ clumps indeed tend to be supported by rotation, 
the fact that the low-$z$ GMCs are observed not to be supported by  
rotation poses a very interesting open question concerning the clump 
formation mechanisms in the two cases. 
We are encouraged by the finding in isolated-galaxy simulations   
that the low-$z$ clumps indeed tend not to be supported by rotation,  
while the high-$z$ clumps are mainly rotation supported. 
We verified that this is not a numerical artifact of resolution (\se{gmc}). 
The difference may be due to a different trigger for clump formation. 
While the high-$z$ clumps are born in a disc that is globally unstable ($Q<1$) 
to begin with, it is possible that the low-$z$ GMCs originate from turbulent 
compression in an otherwise stable disc ($Q>1$)  
before they collapse gravitationally.  
The difference may lie in the fact that the low-$z$ GMCs are less massive  
by 3 orders of magnitude.  
For example, if the sub-clumps are of similar masses at low and high redshift  
clumps, the timescale for two-body relaxation is shorter in the low-$z$ GMCs,  
which may be associated with escape of subclumps with high angular momentum. 
Another difference is that supernova feedback and momentum-driven stellar  
feedback are expected to be 
more effective in disrupting the less massive low-$z$ GMCs \citep{murray10}. 
The GMCs may therefore be short lived, either collapsing under gravity or  
expanding due to feedback, but never settled in a long phase of equilibrium. 
This implies that the clump contraction factor, and the associated surface  
density contrast, should be small, less than $\sim 3$ and $\sim 10$  
respectively. This is not the case for the typical clumps in our  
low-$z$ simulations.  However, the short-lived clumps in the simulations with 
enhanced outflows \citep{Genel10} may be of this nature. 
 
Supersonic turbulent motions can dissipate over a dynamical timescale, 
which is rather short in the giant clumps, $\sim 10 \Myr$, much shorter than 
the disc orbital time, $\sim 250 \Myr$, or the comparable 
time for clump migration to the disc centre. 
The turbulence dissipation time is therefore much shorter than the clump  
lifetime (unless the clump disrupts on a dynamical timescale).  
If turbulence was an important source of clump support against gravitational 
collapse, the turbulence should have been driven continuously  
by a mechanism that also operates on a dynamical timescale.  
This could be gravitational interactions between sub-clumps, and between the  
clump and the rest of the clumpy disc, but it requires further exploration. 
Our finding that most of the support is actually provided by rotation,  
which does not dissipate as quickly, alleviates the need for such an  
efficient driving mechanism for internal turbulence.  
 
If, contrary to the situation in our current simulations, 
the high-$z$ giant clumps are disrupted on a dynamical timescale, 
the clumps might never complete their collapse to rotation-supported  
equilibrium.  
This is the case in the simulations of \citet{Genel10}, where  
disruptive outflows are assumed by implementing an enhanced-outflow
version of the phenomenological
model of \citet{OppenDave06,OppenDave08}. 
While this model pushes the effect of outflows to the extreme, 
as it predicts clump lifetimes shorter than $50\Myr$ in disagreement with  
observational estimates, it is certainly possible that the simulations
discussed in our paper, which currently implement only energy-driven thermal 
feedback \citep{Ceverino09}, underestimate the role of outflows.  
Recent observational indications for massive outflows from massive high-$z$ 
galaxies \citep{Weiner09, Steidel10}, and from the giant clumps themselves 
\citep{Genzel11}, motivate an implementation of more efficient outflow  
driving mechanisms in our cosmological simulations. 
It should be done in a physically motivated way, and should allow 
most clumps to survive in equilibrium for at least $\sim 200 \Myr$, i.e., 
several clump free-fall times (Dekel \& Krumholz, in preparation).   
But such outflows may keep the clumps somewhat more extended, which 
may reduce the level of rotation support in them. 
 
Another missing physical mechanism in our current simulations is the effect 
of magnetic fields.  
In a different regime, magnetic braking is expected to reduce the angular  
momentum during the collapse of proto-stellar cores inside molecular clouds  
\citep{MagneticB79, MagneticB94}. 
Magnetic effects are expected to become important on a free-fall timescale 
when the ratio of clump mass to magnetic flux across the clump surface 
is $\sim (4 \pi^2 G \mu_0)^{-1/2}$, where $G$ is Newton's constant  
and $\mu_0$ is the permeability of free space \citep{NakanoNakamura78}.  
For a clump of mass $\Mc=10^9 \Msun$ and a proto-clump radius 
$\Rci =1\kpc$, this condition requires a magnetic field  
$B \sim (1/2)(\Mc/\Rci^2)(G \mu_0)^(0.5)$, which is $\sim 100 \mu G$.
This is one to two orders of magnitude larger than the 
ordered magnetic field in the Milky Way, which is a few $\mu G$  
on the large scales relevant for exerting torques on the giant clumps. 
The magnetic fields in $z \sim 2$ galaxies have not been measured yet,  
but if they are similarly produced by dynamo effects due to galactic rotation, 
they are likely to be comparable to the Milky-Way magnetic field, and thus 
to have only a small effect on any clump angular-momentum loss. 
On the other hand, the tangled magnetic fields on scales smaller than the 
giant clumps may be in equipartition with the turbulent motions in the ISM  
and reach $\sim 100 \mu G$ \citep{Birnboim09}, 
thus providing an additional term of pressure support  
comparable to the turbulent pressure present in the simulations. 

 
The above discussion leads to the conclusion that future cosmological 
simulations for the study of giant clumps should aim at  
(1) resolving the clump substructure on scales of a few pc,
(2) incorporating realistic momentum-driven outflows via radiative transport, 
and eventually 
(3) including magnetic fields via an MHD treatment.

\section*{Acknowledgments} 
 
We acknowledge stimulating discussions with Y. Birnboim, 
M. Cacciato,  
A. Klypin, T. Naab, and R. Teyssier. 
The simulations were performed in the astro cluster at HU, 
at the National Energy Research Scientific Computing Center (NERSC),  
Lawrence Berkeley National Laboratory, 
at NASA Advanced Supercomputing (NAS) at NASA Ames Research Center,
and at CCRT and TGCC under GENCI allocation 2011-GEN2192.
This work was partially supported by ISF grant 6/08, 
by GIF grant G-1052-104.7/2009, 
by a DIP grant, by NSF grant AST-1010033, 
and by an ERC grant ERC-StG-257720 (FB).

\bibliographystyle{mn2e} 
\bibliography{ClumpsRot16} 
 
\appendix 
 
\section{Cosmological simulations with the ART code} 
\label{sec:art} 
 
The cosmological simulations utilize the ART code 
\citep{Kravtsov97,Kravtsov03}, which accurately follows the evolution of a 
gravitating N-body system and the Eulerian gas dynamics using an adaptive mesh 
refinement approach. Beyond gravity and hydrodynamics, the code incorporates 
many of the physical processes relevant for galaxy formation, as described in 
\citet{Ceverino09} and in CDB10.  These processes, representing subgrid 
physics, include gas cooling by atomic hydrogen and helium, metal and molecular 
hydrogen cooling, and photoionization heating by a UV background with partial 
self-shielding. 
Cooling and heating rates are tabulated for a given gas 
density, temperature, metallicity and UV background based on the CLOUDY code 
\citep{Ferland98}, assuming a slab of thickness 1 kpc. A uniform UV background 
based on the redshift-dependent \citet{HaardtMadau96} model is assumed, 
except at gas densities higher than $0.1\cmc$, where a substantially 
suppressed UV background is used
($5.9\times 10^{26}{\rm erg}{\rm s}^{-1}{\rm cm}^{-2}{\rm Hz}^{-1}$) 
in order to mimic the partial self-shielding of dense gas. 
This allows the dense gas to cool down to temperatures of $\sim 300$K. 
The assumed equation of state is that of an ideal mono-atomic gas. 
Artificial fragmentation on the cell size is prevented by introducing 
a pressure floor, which ensures that the Jeans scale is resolved by at least 
7 cells (see CDB10). 
 
Star formation is assumed to occur at densities above a threshold of $1\cmc$ 
and at temperatures below $10^4$K. More than 90\% of the stars form at 
temperatures well below $10^3$K, and more than half the stars form at 300~K 
in cells where the gas density higher than $10\cmc$. 
The code implements a stochastic star-formation model that yields a 
star-formation efficiency per free-fall time of 5\%. At the given resolution, 
this efficiency roughly mimics the empirical Kennicutt law \citep{Kennicutt98}. 
The code incorporates a thermal stellar feedback model, in which the combined 
energy from stellar winds and supernova explosions is released as a constant 
heating rate over $40\Myr$ following star formation, the typical age of the 
lightest star that explodes as a type-II supernova. 
The heating rate due to feedback may or may not overcome the cooling 
rate, depending on the gas conditions in the star-forming regions 
\citep{DekelSilk86,Ceverino09}. No shutdown of cooling is implemented. 
We also include the effect of runaway stars by assigning a velocity kick of 
$\sim 10 \kms$ to 30\% of the newly formed stellar particles. 
The code also includes the later effects of type-Ia supernova and 
stellar mass loss, and it follows the metal enrichment of the ISM. 
 
The five selected halos were drawn from an N-body simulation. 
The initial conditions corresponding to each of the selected haloes 
were filled with gas and refined to a much higher resolution on an adaptive 
mesh within a zoom-in Lagrangian volume that encompasses the mass within 
twice the virial radius at $z =1$, roughly a sphere of comoving radius $1\Mpc$. 
This was embedded in a comoving cosmological box of side $20$ and $40\hmpc$ 
for galaxies A-C and D-E respectively. 
A standard $\Lambda$CDM cosmology has been assumed, with the 
WMAP5 cosmological parameters 
$\omm=0.27$, $\oml=0.73$, $\omb= 0.045$, $h=0.7$ and $\sigma_8=0.82$ 
\citep{WMAP5}. 
The zoom-in regions have been simulated with 
$\sim (4-11)\times 10^6$ dark-matter particles of mass 
$6.6\times 10^5\msun$ each, and the particles representing stars 
have a minimum mass of $10^4\msun$. 
Each galaxy has been evolved forward in time with the full hydro ART and 
subgrid physics on a adaptive comoving mesh refined in the dense regions 
to cells of minimum size 70 pc or smaller in physical units at all times. 
Each AMR cell is refined to 8 cells once it contains a mass in stars and 
dark matter higher than $2\times 10^6\msun$, equivalent to 3 dark-matter 
particles, or it contains a gas mass higher than $1.5\times 10^6\msun$. This 
quasi-Lagrangian strategy ends at the highest level of refinement 
that marks the minimum cell size at each redshift. 
In particular, the minimum cell size is set to 35 pc in physical units 
at expansion factor $a=0.16$ ($z=5.25$), but due to the expansion of the 
whole mesh while the refinement level remains fixed, 
the minimum cell size grows in physical units and becomes 70 pc by 
$a=0.32$ ($z=2.125$). 
At this time we add a new level to the comoving mesh, so the minimum cell 
size becomes 35 pc again, and so on. 
This maximum resolution is valid in 
particular throughout the cold discs and dense clumps, allowing cooling to 
$\sim 300$K and gas densities of $\sim 10^3\cmc$. 
 
\section{Isolated galaxy simulations with the RAMSES code} 
\label{sec:ramses} 
 
To investigate the effects of numerical resolution and small-scale 
substructure, we use simulations of idealized, isolated, gas-rich disc galaxies 
representative of $z \sim 2$ star-forming galaxies. 
The absence of continuous cosmological 
gas supply limits the duration of disc evolution that can be followed, 
but it permits a much higher resolution than the cosmological simulations. 
Overall, the initial conditions and evolution into $z=2$-like clumpy discs 
are as described in \citet[][BEE07]{Bournaud07}, but the current simulations 
were performed with much higher resolution using the AMR code RAMSES 
\citep{Teyssier02}, comparable in many ways to the ART code used in the 
cosmological simulations utilized in the rest of this paper. 
 
The technique relies on a quasi-Lagrangian refinement of the AMR grid 
and a barotropic equation of state for the heterogeneous ISM 
\citep{Teyssier10,Bournaud10}. Each AMR cell is refined once 
(i) it contains a gas mass higher than $m_\mathrm{res}$, 
or (ii) it contains more than 20 particles, 
or (iii) the local thermal Jeans length is resolved 
by less than four cells. The refinement continues up to a minimum  
cell size $\epsilon_\mathrm{AMR}$. To avoid artificial fragmentation, 
a pressure floor ensures that the thermal Jeans length is at least 
$4 \epsilon_\mathrm{AMR}$.   
Star formation occurs above a volume density threshold $n_\mathrm{SF}$, 
with an efficiency 4\% per free-fall time \citep{Teyssier10}. 
Supernovae feedback is implemented based on the blast-wave model of 
\citet{Dubois}, 
where a fraction $\eta_{\mathrm SN}=50\%$ of the supernova energy is 
deposited as bulk motion in a gas bubble of radius   
$3\times \epsilon_\mathrm{AMR}$, and the rest being radiated away. 
Gas cooling down to low temperatures (below 100~K at the 
highest resolution) is modeled using a barotropic equation of state, 
which assumes thermal equilibrium between heating by a standard UV 
background and cooling by fine-structure and metal lines. While a possible 
multi-phase instability is neglected, a cloudy, heterogeneous medium naturally 
results. This technique reproduces a realistic phase-space ISM 
structure for nearby disc galaxies \citep{Bournaud10}, their molecular clouds 
\citep{bournaud-iau270}, and basic properties of 
merging galaxies \citep{Teyssier10,bournaud-iau271}. 
  
We performed three simulations of a system that represents  
a massive disc at $z \sim 1-3$, at a high resolution (HR), medium resolution 
(MR) and low resolution (LR), as listed in Table~\ref{table-FB}. 
The resolution in the LR simulation is similar to the resolution 
in the cosmological simulations, with slightly lower spatial resolution 
but somewhat higher mass resolution. 
The HR simulation resolves gas densities up to $10^{7}\cmc$ and sizes of a 
few parsecs, comparable to today's molecular clouds. The resulting 
density power spectrum for the HR model (Bournaud et al. 2010, section 3.4 and 
Figure~15) is characteristic of a fully developed three-dimensional turbulence 
cascade, implying that the HR resolution is sufficient for 
convergence on the internal properties of kpc-size clumps. 
 
The simulations start with a pre-formed disc of half gas and half stars, 
and a stellar mass of $\sim 1.5 \times 10^{10}\msun$. 
The initial stellar disc is exponential with radial scale-length 
$R_{\mathrm d}/4$, 
truncated at radius $R_{\mathrm d}=16 kpc$. The gas disc is exponential with 
scale-length $R_{\mathrm d}/1.5$. The ratio of dark matter to baryons within 
$R_{\mathrm d}$ is 0.33, and the spherical dark matter halo has a 
\citet{Burkert95} profile with radial-scale length $R_{\mathrm d}/3$, 
matching high-redshift rotation curves. 


\section{Measuring clump properties in the simulations} 
\label{sec:clumps} 
 
\subsection{Clump finding} 
 
The disc plane is determined by the angular momentum of cold gas ($T<10^4$K)
within a sphere of radius comparable to the disc radius, as estimated by
eye.
Clump candidates are identified upon visual inspection of the face-on,
gas surface density map of the disc, searching for compact gas structures with an estimated
surface density contrast of $\sim 3$ or more compared to the local background.
Therefore, we exclude low-contrast, elongated features from our analysis.

A cube of side $\pm 1.2 \kpc$ is examined about the estimated centre of each
clump. For a smoother calculation, the AMR cells in this volume 
are split to create a uniform grid, with a cell size half the size 
of the smallest AMR cell. 
A friends-of-friends algorithm is then employed, linking together 
neighboring cells with gas temperature $T<10^4$K.  
Of all the identified groups with more than 80 cells, 
the one closest to the initial guess for the center is selected as 
the clump. In most cases, this is also the largest identified group
(excluding the central bulge). 
Otherwise, the group with the highest central density is selected.
The center of mass of this group is calculated, $\vec r_{\rm cm, temp}$,
and is further refined as follows. 
All the cells in the clump within 500 pc of $\vec r_{\rm cm, temp}$ 
are searched for the point which maximizes the mass in a sphere of 
radius $r=250\pc$ centered on that point. The clump center  
$\vec r_{\rm cm}$ is then taken to be the center of that sphere. 
The center of mass velocity of the clump, $v_{\rm cm}$, is the center of  
mass velocity inside that sphere. 
 
\subsection{Clump minor axis} 
 
The spherical density profile of the gas about 
the clump center is calculated in 80 equal bins of $\log r$, 
of width 0.15 dex each, in the 
interval 1.0 to 3.4 (where $r$ is in pc), namely spacing of 0.03 dex between
the bin centers.  
The profile is smoothed such that the 
density in every bin is replaced by a Gaussian weighted average of the 
densities in the 5 neigboring bins, with a FWHM of 0.06 dex. 
The radius of the spherical clump, $\Rs$, is then determined by searching the 
smoothed density profile at $r_i>178\pc$ for the smallest radius that meets  
one of the following three conditions: 
\begin{enumerate} 
\item $\rho'_i >-2$ 
\item $\rho'_i/\rho'_{i-1}<0.8$ 
\item $\vert\rho'_i\vert<\min\{\vert \rho'_{i-1}\vert,\vert\rho'_{i+1}\vert\}$ 
\end{enumerate} 
where $\rho'\equiv d\log \rho/d\log r$. 
If $\rho'_i>-2$ for all bins, the first condition is replaced by $\rho'_i>-1$.  
This happened in 4 out of the 86 clumps. 
The above definition of clump radius is very similar to the standard definition of subhalo radius generally used in dark-matter halo-finders \citep{Knebe11}. The subhalo radius is defined as the truncation radius or saddle point, where the density profile has an inflexion or upturn point due to the embedding of the subhalo within a background host.
The angular momentum is computed for the cold gas ($T<10^4$K) in a spherical 
shell between $(2/3)\Rs$ and $\Rs$ about $\vec r_{\rm cm}$.  
This determines the minor axis of the clump, $\hat z$. 
 
\subsection{Internal clump structure and kinematics} 
 
The profiles of density and velocity are computed in the major, equatorial 
plane of the clump, that is perpendicular to the minor axis. 
For this, we rotate the uniform cubic grid to match the clump frame.  
In an equatorial slice of thickness 100 pc along the minor axis,  
we use 80 cylindrical radial bins of width 0.15 dex spaced 0.03 dex as before. 
Using these bins, we compute the gas density profile, and the mass 
weighted average value of the rotation velocity $v_\phi$,  
as well as the three components of velocity dispersion  
$\sigmar$, $\sigma_\phi$ and $\sigma_z$. 
The profiles in the equatorial plane are smoothed in the same manner described  
in the previous section for the spherical profiles. The 
clump radius, $\Rc$, is determined from the density profile in the equatorial 
plane in the same manner as $\Rs$ was determined from the spherical density  
profile. 
The average rotation velocity $\Vc$ and radial velocity dispersion $\sigmar$  
are computed mass-weighted within the cylindrical bin $0.5\Rc-Rc$  
in the equatorial plane. 
For clumps where $0.5\Rc < 150$ pc, namely smaller than the force resolution, 
the inner radius $0.5\Rc$ is replaced by 150 pc. 
Our adopted definition of clump radius agrees well with the visual impression
from the face on images of the gas surface density in the clumps and the local
disc environment.  We tested several alternative definitions for the clump
radius and the average internal velocities, and found that our qualitative
conclusions are insensitive to the exact algorithm adopted.
 
\subsection{Global clump properties} 
 
The total clump mass $\Mc$, including gas, stars and dark matter, 
is computed within $\Rc$. The gas mass, stellar mass, and dark-matter mass, 
are computed accordingly. 
For the purpose of computing the effective circular velocity of the clump, 
we define the mass-weighted average radius $\Rsh$ in the shell $0.5\Rc-Rc$, 
compute the total mass within a sphere of that radius, $M(\Rsh)$, 
and then compute $\Vcc^2=GM(\Rsh)/\Rsh$.  
The rotation parameter is $\Rot = (\Vc/\Vcc)^2$. 
 
In order to determine a contraction factor $c$, we compute the 
clump baryonic surface density  
$\Sigma_{\rm c} = M_{\rm bar}(\Rsh)/(\pi \Rsh^2)$, 
and write $c^2 = \Sigma_{\rm c}/\Sigma_{\rm d}$ where $\Sigma_{\rm d}$ is the 
relevant background density in the disc. 
$\Sigma_{\rm d}$ is computed in a cylindrical shell from $R_{\rm bulge}$ 
to $\Rd$, and with height $h$.  
The values of these three numbers are determined by visual inspection as
follows. 
The disc radius is set where the gas surface density profile averaged over
circular rings drops below $10^{21}\cms$.
The bulge radius is taken to be where the stellar density profile
drops below $10^4\msun\pc^{-2}$.
The scale height is determined where the edge-on surface density
drops below $10^{21}\cms$, but the mean disc surface density is 
insensitive to the choice of scale height.
If the disc surface density severely deviates from axial symmetry,
e.g., in galaxy E at $z=2.2$-$2.4$, 
then the center of the cylindrical shell is taken to be at the center of 
mass in the disc.
We include all the gas in the disc. 
A star particle is assigned to the disc only if 
the $z$-component of its angular momentum $j_z$ 
(parallel to the total galaxy angular momentum) 
is higher than a fraction $f_J$ of the maximum angular momentum for the 
same orbital energy, $j_{max}=|v|r$, 
where $|v|$ is the magnitude of the particle velocity and $r$ is its distance 
from the center. We adopt as default $f_J=0.7$ (see CDB10). 
We note that
the relevant surface densities for quantifying a contraction 
factor as well as the clump radii
are quite uncertain, so the values of $c$ should be taken with a 
grain of salt. 

When comparing the clump masses and radii, 
we find a general trend $\Rc \propto \Mc^{1/3}$, as expected. 
The relation between these quantities do not reveal an obvious sign of 
resultion effects limiting the clump properties at small clumps, but 
naturally clumps of $\Rc < 300\pc$ or $\Mc < 3 \times 10^8\msun$ should be
considered less certain. 

The mean stellar age of each clump is determined by the mass-weighted average 
of the stellar ages inside $\Rc$. 

The profiles of gas density, rotation velocity and velocity dispersion
shown in \fig{FB3}, \fig{FB3a} and \fig{FBappen2} for
the clumps from the high-resolution isolated-galaxy simulations were 
obtained in radius bins and then smoothed using a cubic spline algorithm.
This is obtained by fitting a piecewise 3rd degree polynomial,
such that the sums in quadrature of both the residuals and the
second derivatives are minimized.
Tolerrance for the sum in quadrature of residuals
is taken to be equal to the total variance of the data for the velocity curves 
and 0.15 of the variance for the density profiles.

\begin{table} 
\caption{Virial properties of the selected galaxies. 
All virial properties are defined at the virial radius that
encompasses a mean mass density of 180 times the universal density. 
The radius $\Rv$ is expressed in proper kpc,
the mass $\Mv$ is in units of $10^{12}\msun$,
and the velocity $\Vv$ is in $\kms$.}
 \begin{center} 
 \begin{tabular}{cccc} \hline 
\multicolumn{2}{c} {Gal z}\ \ \ R$_{\rm v}$\ \ \ & M$_{\rm v}$ & V$_{\rm v}$  \\
\hline 
A 2.3 &  70 & 0.40 & 150 \\ 
A 2.1 &  88 & 0.60 & 171 \\ 
B 2.3 &  68 & 0.35 & 140 \\ 
C 2.3 &  83 & 0.61 & 180 \\ 
C 2.1 &  97 & 0.81 & 188 \\ 
C 1.9 & 109 & 0.94 & 192 \\ 
D 2.3 &  96 & 0.94 & 205 \\ 
E 3.0 &  91 & 1.35 & 252 \\ 
E 2.8 &  95 & 1.37 & 249 \\ 
E 2.7 & 100 & 1.43 & 248 \\ 
E 2.6 & 105 & 1.47 & 245 \\ 
E 2.4 & 109 & 1.52 & 244 \\ 
E 2.3 & 114 & 1.54 & 241 \\ 
E 2.2 & 118 & 1.55 & 237 \\ 
E 2.1 & 122 & 1.57 & 235 \\ 
E 2.0 & 127 & 1.63 & 234 \\ 
\hline 
 \end{tabular} 
 \end{center} 
\label{tab:0a} 
 \end{table} 
 
\begin{table} 
\caption{Kinematic disc properties of the selected galaxies. 
Rotation velocities and radial velocity dispersions for gas and stars
are in $\kms$.}
 \begin{center} 
 \begin{tabular}{ccccccccc} \hline 
\multicolumn{2}{c} {Gal z}\ \ \  $V_{\rm gas}$ & $\sigma_{\rm gas}$ & $(\sigma/V)_{\rm gas}$  & $V_{\rm star}$ & $\sigma_{\rm star}$ & $(\sigma/V)_{\rm star}$ \\
\hline 
A 2.3 & 180 & 20 & 0.11  & 160 & 33 & 0.22 \\ 
A 2.1 & 203 & 22 & 0.11 &  178 & 37 & 0.21 \\ 
B 2.3 & 180 & 25 & 0.14 &  157 & 34 & 0.26 \\ 
C 2.3 & 380 & 60 & 0.16  & 356 & 107 & 0.30 \\ 
C 2.1 & 427 & 52 & 0.12  & 331 &103  & 0.31  \\ 
C 1.9 & 465 & 41 & 0.09 & 378 & 128  &  0.34 \\ 
D 2.3 & 264 & 24 & 0.09 & 222 & 53    &  0.24 \\ 
E 3.0 & 370 & 29 & 0.08  & 311 & 88   & 0.28 \\ 
E 2.8 & 397 & 22 & 0.05  & 295 & 81   & 0.27  \\ 
E 2.7 & 382 & 25 & 0.06  & 309 & 81 & 0.26 \\ 
E 2.6 & 384 & 24 & 0.06 & 342 & 92 & 0.27  \\ 
E 2.4 & 400 & 23 & 0.06 & 322 & 92 & 0.28 \\ 
E 2.3 & 407 & 27 & 0.07 & 346 & 98 & 0.28 \\ 
E 2.2 & 419 & 17 & 0.04 & 345 & 102 & 0.29 \\ 
E 2.1 & 433 & 17 & 0.04 & 368 & 109 & 0.30 \\ 
E 2.0 & 424 &   9 & 0.02 & 346 & 107 & 0.31 \\ 
\hline 
 \end{tabular} 
 \end{center} 
\label{tab:0b} 
 \end{table} 
 
\begin{table*} 
\caption{Masses and Radii of the selected galaxies. 
The radii are in kpc and the masses are in $10^{10}\msun$.} 
 \begin{center} 
 \begin{tabular}{ccccccccccccc} \hline 
\multicolumn{2}{c} {Gal z}\  R$_{\rm disc}$ & R$_{\rm bulge}$ & M$_{\rm gas}$ & M$_{\rm star}$ & M$_{\rm DM}$ & M$_{\rm bulge}$ & M$_{\rm disc}$ & M$_{\rm disc }/$M$_{\rm total}$ & M$_{\rm gas}/$M$_{\rm disc}$ & M$_{\rm bulge}$/M$_{\rm disc }$ & M$_{\rm clumps}/$M$_{\rm disc}$\\\hline 
A 2.3 & 5.8 & 2.0 & 0.37 & 1.7 & 2.1 & 1.0  & 1.1  & 0.26 & 0.35 & 0.93  & 0.19 \\ 
A 2.1 & 5.5 & 2.0 & 0.36 & 2.4 & 2.1 & 1.2  & 1.6  & 0.32 & 0.23 & 0.75 & 0.14 \\ 
B 2.3 & 3.4 & 1.0 & 0.20 & 1.3 & 0.92   & 0.81 & 0.66  & 0.27 & 0.30 & 1.22 & 0.21 \\ 
C 2.3 & 4.6 & 1.5 & 0.46 & 11.0 & 3.4 & 6.6 & 5.3 & 0.35 & 0.09 & 1.25 & 0.008 \\ 
C 2.1 & 5.0 & 1.5 & 0.75 & 13.2 & 3.9 & 9.7 & 4.2 & 0.24 & 0.18 & 2.30 & 0.05 \\ 
C 1.9 & 5.0 & 1.5 & 0.97 & 17.9 & 4.8 & 13 & 5.9 & 0.25 & 0.17 & 2.22 & 0.06 \\ 
D 2.3 & 5.0 & 2.0 & 0.39 & 3.9 & 2.7 & 2.2 & 2.0 & 0.29 & 0.19 & 1.09 & 0.08 \\ 
E 3.0 & 6.0 & 2.0 & 0.75 & 9.7 & 7.9 & 7.1 & 3.3 & 0.18 & 0.23 & 2.16 & 0.08 \\ 
E 2.8 & 8.0 & 2.0 & 0.89 & 11.3 & 11.9 & 7.9 & 4.2 & 0.17 & 0.21 & 1.89 & 0.04 \\ 
E 2.7 &  8.0 & 2.0 & 0.93&11.8 & 11.5 & 7.9 & 4.8 & 0.20  &0.19&1.65& 0.07\\ 
E 2.6& 8.0 & 2.0 & 0.85&12.4 & 11.9 & 7.9 & 5.3 & 0.21&0.16&1.49& 0.09\\ 
E 2.4 & 8.0 & 2.0 & 0.83&13.3 & 12.1 & 8.2 & 5.8 & 0.22&0.14&1.41& 0.08\\ 
E 2.3 & 8.5 & 2.0 & 0.73&14.1 & 13.3 & 8.4 & 6.4 & 0.23 & 0.11&1.32& 0.11\\ 
E 2.2 & 8.5 & 2.0 & 0.80&14.9 & 13.6 & 8.7 & 6.9 & 0.24 & 0.12&1.26& 0.10\\ 
E 2.1 & 8.5 & 2.0 & 0.81&15.5 & 13.9 & 9.1 & 7.1 & 0.24&0.11&1.28& 0.08\\ 
E 2.0 & 9.0 & 2.0 & 0.90&15.9 & 14.8 & 9.4 & 7.4 & 0.23&0.12&1.28& 0.005\\ 
\hline 
 \end{tabular} 
 \end{center} 
\label{tab:0c} 
 \end{table*} 
 
%
 

\begin{table*} 
\caption{Clump properties.
``cl" marks the arbitrary clump number.
$\Rc$ is the clump radius in pc.
$\Mc$ is the clump mass in $10^8\msun$.
Velocities are in $\kms$.
``tilt" is $\cos({\rm tilt})$.
$\Sigma_{\rm c}$ is the baryonic surface density within the clump
in $10^8\msun \kpc^{-2}$. 
$f_{\rm DM}$ and $f_{\rm g}$ are the fractions of dark matter and gas within
the clump radius.
``age" is the mean stellar age within the clump in Myr.
$r$ and $z$ are the disc polar coordinates of the clump center in kpc.
} 
 \begin{center} 
 \begin{tabular}{cccccccccccccccccccc} \hline 
Gal $z$ & cl & $\Rc$ & $\Mc$ & $\Vcc$ & $\Vphi$ & $\sigma_{\rm r}$ & $\Rot$ & tilt & $\Sigma_{\rm c}$ & $c$ & $f_{\rm DM}$ & $f_{\rm g}$ & age & $r$ & $z$ & $V_r$ & $V_z$ & $\Vc/\Vd$ & comm. \\ 
\hline 
A2.3 & 1 & 454 & 3.9 & 68 & 63 & 24 & 0.86 & 0.94 & 1.1 & 4.2 & 0.07 & 0.12 & 128 & 2.4 & -0.20 & 13 &  20 & 190/170 & \\ 
      & 2 & 278 & 3.8 & 81 & 37 & 26 & 0.21 & 0.53 & 2.4 & 6.3 & 0.02 & 0.10 & 94 & 2.0 & 0.37 & 13 & 11 & 177/170 & merger  \\ 
      & 3 & 454 & 2.5 & 52 & 37 & 17 & 0.51 & 0.84 & 0.6 & 3.2 & 0.05 & 0.21 & 87 & 3.2 & -0.14 & -35 & 11 & 157/170 & \\ 
      & 4 & 322 & 1.9 & 61 & 57 & 9.2 & 0.86 & -0.13 & 1.3 & 4.6 & 0.01 & 0.21 & 36 & 4.8 & -0.34 & -1.4 & 13 & 166/170 & \\ 
      & 5 & 292 & 1.8 & 58 & 54 & 10 & 0.87 & 0.43 & 1.2 & 4.4 & 0.02 & 0.07 & 97 & 3.1 & -0.04 & -12 & 60 & 191/170 & \\ 
      & 6 & 411 & 2.6 & 57 & 51 & 13 & 0.80 & 0.34 & 0.9 & 3.8 & 0.02 & 0.29 & 23 & 4.2 & 0.58 & 33 & 44 & 180/170 & \\ 
      & 7 & 816 & 2.4 & 32 & 33 & 12 & 1.1 & 0.67 & 0.1 & 1.3 & 0.29 & 0.42 & 203 & 3.7 & 0.03 & -5.4 & 12 & 202/170 & low c\\ 
      & 8 & 609 & 1.9 & 32 & 26 & 27 & 0.67 & 0.80 & 0.2 & 1.6 & 0.18 & 0.46 & 155 & 3.8 & 0.24 & -37 & 14 & 172/170 & low c\\ 
      & 9 & 816 & 4.2 & 37 & 36 & 20 & 0.94 & 0.85 & 0.1 & 1.4 & 0.42 & 0.17 & 412 & 2.4 & -0.12 & -1.3 & -18 & 164/170 & \exsitu\\ 
      & 10 & 338 & 3.2 & 71 & 59 & 26 & 0.69 & 0.74 & 1.2 & 4.5 & 0.30 & 0.05 & 393 & 2.8 & 0.74 & -204 & -236 & 93/170 & \exsitu\\ 
\hline 
A2.1 & 1 & 229 & 3.2 & 87 & 85 & 18 & 0.95 & -0.02 & 3.1 & 5.9 & 0.00 & 0.03 & 127 & 5.7 & -0.62 & -61 & -93 & 137/150 & \\ 
      & 2 & 355 & 1.8 & 48 & 31 & 12 & 0.42 & 0.47 & 0.7 & 2.8 & 0.03 & 0.43 & 76 & 4.0 & -0.98 & -69 & -8.1 & 208/200 & \\ 
      & 3 & 900 & 4.9 & 40 & 32 & 18 & 0.64 & 0.52 & 0.2 & 1.3 & 0.21 & 0.45 & 374 & 3.7 & -0.63 & -37 & -29 & 222/200 & low c \\ 
      & 4 & 431 & 3.1 & 59 & 50 & 24 & 0.71 & 0.86 & 0.8 & 3.0 & 0.04 & 0.21 & 161 & 2.7 & 0.52 & -1.2 & -19 & 203/210 & \\ 
      & 5 & 552 & 3.9 & 51 & 44 & 28 & 0.74 & 0.84 & 0.4 & 2.2 & 0.12 & 0.40 & 211 & 2.4 & -0.07 & 1.3 & 2.0 & 190/210 & \\ 
      & 6 & 671 & 5.4 & 50 & 40 & 41 & 0.62 & 0.99 & 0.3 & 1.9 & 0.18 & 0.33 & 360 & 2.1 & 0.07 & 23 & -19 & 220/220 & \\ 
\hline 
B2.3 & 1 & 373 & 4.6 & 52 & 73 & 58 & 1.9 & 0.98 & 0.6 & 2.4 & 0.18 & 0.10 & 667 & 0.6 & 0.08 & 22 & 93 & 203/220 & low $r$ \\ 
      & 2 & 391 & 3.0 & 43 & 81 & 33 & 3.5 & 0.93 & 0.4 & 1.8 & 0.23 & 0.13 & 547 & 0.8 & -0.06 & -34 & 23 & 240/220 & low $r$  \\ 
      & 3 & 454 & 2.9 & 44 & 41 & 25 & 0.88 & 0.99 & 0.4 & 1.8 & 0.18 & 0.20 & 258 & 1.2 & 0.08 & 23 & 1.8 & 178/180 & \\ 
      & 4 & 454 & 2.6 & 42 & 28 & 30 & 0.46 & 0.94 & 0.3 & 1.7 & 0.16 & 0.26 & 203 & 1.4 & 0.01 & -1.9 & -0.3 & 174/180 & \\ 
      & 5 & 292 & 1.0 & 33 & 26 & 13 & 0.60 & 0.84 & 0.4 & 1.8 & 0.10 & 0.32 & 134 & 1.6 & -0.09 & 6.3 & -9.8 & 167/180 & \\ 
      & 6 & 338 & 7.5 & 110 & 55 & 47 & 0.25 & -0.11 & 3.3 & 5.4 & 0.18 & 0.11 & 218 & 2.6 & -0.07 & -94 & 169 & -318/-100 & \exsitu \\ 
\hline 
C2.3 & 1 & 278 & 24 & 211 & 121 & 101 & 0.33 & 0.96 & 15 & 5.6 & 0.09 & 0.01 & 940 & 4.9 & 0.54 & -166 & -166 & 300/260 & \exsitu \\ 
      & 2 & 454 & 4.2 & 47 & 34 & 27 & 0.53 & 0.97 & 0.5 & 1.0 & 0.11 & 0.12 & 352 & 2.5 & -0.38 & 22 & -104 & 431/440 & low c \\ 
\hline 
C2.1 & 1 & 776 & 7.4 & 56 & 89 & 84 & 2.5 & 0.35 & 0.4 & 1.2 & 0.16 & 0.37 & 534 & 2.9 & -2.0 & 41 & 62 & 437/410 & low c \\ 
      & 2 & 900 & 14 & 72 & 22 & 32 & 0.10 & 0.44 & 0.5 & 1.4 & 0.17 & 0.27 & 695 & 2.7 & 1.0 & 34 & -52 & 443/420 & low c \\ 
\hline 
C1.9 & 1 & 373 & 4.0 & 73 & 54 & 25 & 0.55 & 0.80 & 1.5 & 2.0 & 0.03 & 0.26 & 111 & 4.0 & 1.3 & -40 & -53 & 464/450 & \\ 
      & 2 & 292 & 2.4 & 63 & 56 & 24 & 0.79 & 0.43 & 1.4 & 1.9 & 0.03 & 0.20 & 160 & 3.1 & 0.55 & -8.6 & -105 & 544/470 & \\ 
      & 3 & 391 & 4.3 & 72 & 55 & 25 & 0.59 & 0.85 & 1.3 & 1.8 & 0.06 & 0.22 & 151 & 3.5 & 0.39 & -69 & 6.1 & 483/460 & \\ 
      & 4 & 229 & 3.1 & 78 & 65 & 42 & 0.70 & 0.81 & 2.4 & 2.5 & 0.00 & 0.40 & 36 & 3.6 & 0.00 & 28 & -8.6 & 453.6/460 & \\ 
      & 5 & 252 & 1.3 & 44 & 23 & 16 & 0.27 & 0.85 & 0.7 & 1.4 & 0.03 & 0.22 & 49 & 4.4 & 0.2 & -42 & 41 & 452/450 & low c \\ 
      & 6 & 671 & 3.0 & 42 & 54 & 27 & 1.6 & 0.91 & 0.3 & 1.4 & 0.16 & 0.42 & 390 & 4.8 & 0.06 & 42 & 85 & 404/440 & low c \\ 
      & 7 & 740 & 12 & 67 & 82 & 35 & 1.5 & 0.88 & 0.6 & 1.3 & 0.12 & 0.20 & 740 & 3.2 & -0.24 & 90 & -50 & 550/470 & low c \\ 
      & 8 & 476 & 2.3 & 41 & 41 & 16 & 1.0 & 0.96 & 0.3 & 0.9 & 0.15 & 0.43 & 427 & 3.5 & 1.1 & 116 & 46 & 371/460 & low c \\ 
\hline 
D2.3 & 1 & 338 & 2.7 & 67 & 54 & 18 & 0.66 & 0.95 & 1.4 & 3.5 & 0.05 & 0.15 & 141 & 2.3 & -0.48 & -58 & -4.6 & 286/300 & \\ 
      & 2 & 432 & 4.6 & 78 & 54 & 11 & 0.49 & 0.57 & 1.5 & 3.6 & 0.03 & 0.14 & 118 & 3.7 & -0.49 & 45 & 36 & 285/260 & \\ 
      & 3 & 373 & 8.3 & 114 & 108 & 32 & 0.90 & 0.01 & 3.8 & 5.8 & 0.02 & 0.10 & 167 & 3.3 & 0.60 & 31 & -12 & 222/270 & \\ 
\hline 
 \end{tabular} 
 \end{center} 
\label{tab:clumps} 
 \end{table*} 
 
\begin{table*} 
\caption{Continuation of \tab{clumps}} 
 \begin{center} 
 \begin{tabular}{cccccccccccccccccccc} \hline 
Gal $z$ & cl & $\Rc$ & $\Mc$ & $\Vcc$ & $\Vphi$ & $\sigma_{\rm r}$ & $\Rot$ & tilt & $\Sigma_{\rm c}$ & $c$ & $f_{\rm DM}$ & $f_{\rm g}$ & age & $r$ & $z$ & $V_r$ & $V_z$ & $\Vc/\Vd$ & comm. \\ 
\hline 
E3.0 & 1 & 525 & 49 & 219 & 73 & 111 & 0.11 & -0.61 & 8.3 & 8.0 & 0.17 & 0.04 & 727 & 1.3 & -0.47 & -148 & 76 & 469/450 & \exsitu \\ 
      & 2 & 265 & 31 & 250 & 246 & 40 & 0.97 & 0.99 & 21 & 13 & 0.11 & 0.07 & 636 & 4.3 & -0.04 & -6.1 & -0.7  & 368/390 & \exsitu \\ 
      & 3 & 391 & 5.2 & 88 & 78 & 13 & 0.78 & 0.09 & 2.2 & 4.1 & 0.02 & 0.29 & 36 &  5.8 & -0.48 & 135 & 40 & 319/350 & \\ 
      & 4 & 292 & 31 & 236 & 239 & 27 & 1.02 & -0.40 & 18 & 12 & 0.15 & 0.01 & 831 & 7.4 & 1.1 & 59 & 465 & 351/320 & \exsitu \\ 
      & 5 & 816 & 7.0 & 56 & 80 & 31 & 2.00 & 0.98 & 0.4 & 1.7 & 0.27 & 0.50 & 414 & 4.1 & -0.19 & 42 & 64 & 370/390 & low c \\ 
      & 6 & 373 & 1.8 & 41 & 21 & 16 & 0.27 & 0.66 & 0.5 & 1.9 & 0.05 & 0.72 & 76 & 6.0 & 0.16 & -164 & -5.5 & 347/350 & \\ 
      & 7 & 391 & 4.4 & 80 & 79 & 37 & 0.97 & -0.26 & 1.8 & 3.7 & 0.04 & 0.02 & 185 & 5.4 & -1.8 & -5.7 & 181 & 461/370 & \\ 
      & 8 & 579 & 8.5 & 86 & 114 & 114 & 1.76 & -0.05 & 1.3 & 3.1 & 0.07 & 0.03 & 106 & 5.3 & -1.1 & 17 & 101 & 472/370 & \\ 
\hline 
E2.8 & 1 & 338 & 5.0 & 87 & 38 & 52 & 0.19 & 0.35 & 2.3 & 4.1 & 0.04 & 0.07 & 157 & 3.3 & 0.33 & -51 & 110 & 432/410 & 3 \\ 
      & 2 & 355 & 3.9 & 75 & 54 & 16 & 0.51 & 0.68 & 1.7 & 3.5 & 0.05 & 0.28 & 53 & 3.6 & 0.67 & -20 & -4.5 & 374/420 & \\ 
      & 3 & 292 & 2.8 & 65 & 41 & 31 & 0.39 & 0.04 & 1.5 & 3.2 & 0.05 & 0.27 & 66 & 3.6 & 0.26 & -39 & -37 & 455/420 & \\ 
      & 4 & 671 & 6.3 & 71 & 59 & 22 & 0.71 & -0.01 & 0.8 & 2.3 & 0.06 & 0.32 & 25 & 8.9 & 0.42 & -19 & -83 & 226/280 & \\ 
\hline 
E2.7 & 1 & 432 & 5.0 & 80 & 84 & 28 & 1.09 & 0.72 & 1.6 & 3.0 & 0.07 & 0.07 & 183 & 3.2 & -1.0 & 97 & -33 & 380/410 & 3 \\ 
      & 2 & 307 & 6.2 & 108 & 98 & 24 & 0.83 & 0.15 & 4.2 & 4.9 & 0.03 & 0.07 & 253 & 2.5 & 0.59 & 108 & -116 & 442/440 & 1 \\ 
      & 3 & 355 & 3.2 & 71 & 32 & 8.4 & 0.78 & 0.69 & 1.5 & 3.0 & 0.02 & 0.30 & 32 & 6.0 & 0.80 & -98 & 26 & 383/370 & \\ 
      & 4 & 431 & 14 & 136 & 118 & 34 & 0.76 & 0.69 & 4.4 & 5.0 & 0.01 & 0.12 & 220 & 5.8 & 0.89 & -127 & 6.6 & 355/370 & \\ 
      & 5 & 373 & 6.6 & 100 & 103 & 36 & 1.06 & 0.43 & 2.8 & 4.0 & 0.02 & 0.10 & 115 & 5.1 & -0.59 & -3.1 & 19 & 335/380 & 4 \\ 
      & 6 & 500 & 1.8 & 150 & 158 & 38 & 1.1 & -0.03 & 4.5 & 5.1 & 0.18 & 0.01 & 988 & 8.0 & 1.4 & 73 & -3.3 & 165/360 & \exsitu \\ 
\hline 
E2.6 & 1 & 338 & 5.5 & 96 & 66 & 30 & 0.46 & 0.87 & 3.0 & 3.9 & 0.03 & 0.21 & 44 & 5.8 & 0.33 & -173 & 38 & 411/380 & \\ 
      & 2 & 342 & 24 & 200 & 199 & 43 & 0.99 & 0.67 & 12 & 8.0 & 0.01 & 0.03 & 257 & 2.9 & 0.12 & 1.3 & -46 & 438/440 & mer. 2,4 \\ 
      & 3 & 476 & 21 & 158 & 148 & 46 & 0.88 & 0.98 & 4.7 & 4.9 & 0.16 & 0.05 & 965 & 2.6 & 0.23 & 14 & 7.0 & 451/440 & \exsitu 6 \\ 
      & 4 & 252 & 8.1 & 125 & 109 & 53 & 0.77 & 0.97 & 5.8 & 5.4 & 0.05 & 0.05 & 309 & 1.3 & 0.02 & 79 & 11 & 569/520 & 5 \\ 
      & 5 & 338 & 3.2 & 75 & 57 & 23 & 0.58 & 0.29 & 1.8 & 3.0 & 0.01 & 0.19 & 79 &  7.7 & -1.1 & -35 &  84 & 305/350 & \\ 
      & 6 & 391 & 1.8 & 41 & 28 & 25 & 0.48 & 0.17 & 0.4 & 1.5 & 0.05 & 0.68 & 52 & 7.7 & 0.28 & -34 & -44 & 309/350 & low c \\ 
      & 7 & 454 & 4.2 & 73 & 64 & 41 & 0.75 & 0.19 & 1.3 & 2.6 & 0.08 & 0.05 & 152 & 4.7 & -0.19 & 53 & -129 & 410/410 & \\ 
      & 8 & 252 & 2.1 & 66 & 29 & 22 & 0.20 & -0.34 & 1.7 & 3.0 & 0.01 & 0.03 & 158 & 4.2 & -0.47 & -3.8 & -75 & 306/420 & \\ 
\hline 
E2.4 & 1 & 609 & 10 & 94 & 91 & 81 & 0.92 & 0.95 & 1.5 & 3.3 & 0.06 & 0.17 & 136 & 5.1 & -0.20 & -61 & -43 & 393/410 & 6 \\ 
      & 2 & 432 & 4.1 & 69 & 33 & 18 & 0.23 & -0.2 & 1.2 & 2.9 & 0.04 & 0.51 & 89 & 5.3 & 0.01 & -164 & -25 & 363/410 & \\ 
      & 3 & 552 & 5.4 & 64 & 24 & 14 & 0.14 & 0.40 & 0.8 & 2.4 & 0.06 & 0.59 & 77 & 7.7 & -1.0 & -1.9 & -2.4 & 321/360 & \\ 
      & 4 & 411 & 9.7 & 110 & 91 & 30 & 0.69 & 0.61 & 0.4 & 4.8 & 0.05 & 0.10 & 319 & 5.1 & 0.05 & 84 & 86 & 426/410 & interact 1 \\ 
      & 5 & 391 & 5.6 & 91 & 66 & 32 & 0.53 & -0.39 & 2.3 & 4.1 & 0.03 & 0.12 & 363 & 4.5 & 0.17 & 66 & 2.7 & 301/410 & interact \\ 
      & 6 & 432 & 9.6 & 108 & 105 & 25 & 0.94 & 0.68 & 2.8 & 4.5 & 0.03 & 0.13 & 128 & 4.2 & 0.14 & 42 & -3.8 & 470/410 &  5 \\ 
      & 7 & 307 & 3.8 & 87 & 89 & 30 & 1.03 & -0.03 & 2.8 & 4.5 & 0.02 & 0.04 & 244 & 7.9 & -0.95 & -29 & -50 & 318/360 & 7 \\ 
\hline 
E2.3 & 1 & 411 & 13 & 135 & 136 & 41 & 1.01 & 0.94 & 4.7 & 7.6 & 0.03 & 0.08 & 226 & 3.0 & 0.27 & 34 & -20 & 523/480 &  1 \\ 
      & 2 & 307 & 5.3 & 98 & 55 & 24 & 0.31 & -0.01 & 3.3 & 6.5 & 0.01 & 0.22 & 36 & 7.5 & 0.34 & -84 & -24 & 369/330 & \\ 
      & 3 & 525 & 8.4 & 95 & 99 & 21 & 1.09 & 1.0 & 1.8 & 4.8 & 0.02 & 0.12 & 78 & 8.8 & 0.2 & -2.8 & 24 & 308/330 & \\ 
      & 4 & 307 & 3.9 & 84 & 77 & 56 & 0.84 & 0.69 & 2.4 & 5.5 & 0.01 & 0.17 & 70 & 8.1 & 0.00 & 61 & -20 & 319/330 & \\ 
      & 5 & 411 & 8.2 & 108 & 115 & 24 & 1.14 & 0.96 & 3.1 & 6.2 & 0.04 & 0.05 & 372 & 3.5 & -0.27 & 127 & 45 & 414/460 & interact  5\\ 
      & 6 & 411 & 12 & 132 & 123 & 52 & 0.86 & 0.59 & 4.8 & 7.8 & 0.03 & 0.05 & 240 & 3.6 & -0.04 & 69 & 29 & 414/460 & interact 6\\ 
      & 7 & 736 & 21 & 125 & 127 & 62 & 1.04 & 0.34 & 2.3 & 5.3 & 0.09 & 0.06 & 285 & 4.0 & -0.07 & 17 & 27 & 396/440 & mer. 4,7 \\ 
\hline 
E2.2 & 1 & 740 & 6.8 & 63 & 54 & 22 & 0.74 & 0.98 & 0.5 & 2.8 & 0.10 & 0.61 & 360 & 7.2 & 0.00 & -6.2 & -5.2 & 326/360 & \\ 
      & 2 & 432 & 12 & 129 & 121 & 31 & 0.88 & 0.97 & 4.1 & 7.5 & 0.02 & 0.07 & 148 & 5.2 & 0.35 & 32 & -16 & 420/420 &  2 \\ 
      & 3 & 638 & 12 & 104 & 111 & 58 & 1.14 & 0.64 & 1.8 & 5.0 & 0.05 & 0.08 & 198 & 6.3 & 0.37 & 14 & 1.0  & 386/380 & interact 3 \\ 
      & 4 & 525 & 12 & 113 & 118 & 22 & 1.10 & 0.99 & 2.5 & 5.9 & 0.05 & 0.18 & 192 & 4.8 & 0.12 & 101 & 29 & 393/420 & interact 4 \\ 
      & 5 & 350 & 15 & 130 & 128 & 72 & 0.97 & 0.98 & 4.2 & 7.7 & 0.10 & 0.03 & 634 & 1.5 & -0.08 & 115 & 6.5 & 534/510 & 5 \\ 
      & 6 & 638 & 9.2 & 87 & 88 & 29 & 1.02 & 0.89 & 1.2 & 4.1 & 0.07 & 0.14 & 273 & 5.4 & -0.08 & -51 & 5.5 & 397/400 & \\ 
\hline 
E2.1 & 1 & 552 & 34 & 179 & 194 & 83 & 1.18 & 0.98 & 6.1 & 7.4 & 0.06 & 0.08 & 389 &  2.2 & -0.08 & -45 & 4.9 & 537/490 & mer. 3,4 \\ 
      & 2 & 411 & 25 & 180 & 165 & 64 & 0.84 & 0.98 & 8.2 & 8.6 & 0.04 & 0.06 & 313 & 2.4 & 0.02 & -21 & 13 & 440/500 & mer. 2,6 \\ 
\hline 
E2.0 & 1 & 278 & 3.4 & 80 & 71 & 20 & 0.80 & 0.98 & 2.3 & 4.0 & 0.02 & 0.17 & 84 & 5.4 & -0.05 & 105 & 11 & 441/430 & \\ 
\hline 
 
 \end{tabular} 
 \end{center} 
 \end{table*} 
 
 
\bsp 
 
\label{lastpage} 
 
\end{document}